\documentclass[prb,aps,showpacs,groupedaddress,superscriptaddress,twocolumn,toc=flat]{revtex4-2}

\usepackage[table]{xcolor}

\usepackage[colorlinks=true,linkcolor=blue,urlcolor=blue,citecolor=blue]{hyperref}

\usepackage[utf8x]{inputenc}
\usepackage{color}
\usepackage{bbm} 

\usepackage{amsfonts,amsmath,amssymb,stmaryrd}

\usepackage{graphicx}
\usepackage{subfigure}  
\usepackage{bbm} 
\usepackage{hyperref}
\usepackage{epsfig}
\usepackage{mathrsfs}
\usepackage{verbatim}
\usepackage{centernot}
\usepackage{ulem}
\usepackage{longtable}
\usepackage{nicefrac}
\usepackage[framemethod=tikz]{mdframed}
 
\usepackage{booktabs}

\renewcommand{\l}{\left(}
\renewcommand{\r}{\right)}

\newcommand{\bra}[1]{\langle#1|}
\newcommand{\ket}[1]{|#1\rangle}

\renewcommand{\H}{\hat{\mathcal{H}}}

\renewcommand{\c}{\hat{c}}
\renewcommand{\a}{\hat{a}}
\newcommand{\s}{\hat{s}}
\newcommand{\f}{\hat{f}}
\newcommand{\cd}{\hat{c}^\dagger}

\newcommand{\ad}{\hat{a}^\dagger}
\renewcommand{\sd}{\hat{s}^\dagger}
\newcommand{\fd}{\hat{f}^\dagger}
\newcommand{\bd}{\hat{b}^\dagger}

\renewcommand{\b}{\hat{b}}
\newcommand{\hd}{\hat{h}^\dagger}
\newcommand{\h}{\hat{h}}
\newcommand{\dd}{\hat{d}^\dagger}
\renewcommand{\d}{\hat{d}}
\newcommand{\n}{\hat{n}}

\newcommand{\hc}{\text{h.c.}}

\newcommand{\U}{\hat{U}}

\newcommand{\Ud}{\hat{U}^\dagger}

\newcommand{\eff}{\text{eff}}
\newcommand{\tr}{\text{tr}}

\DeclareMathOperator{\sgn}{sgn}

\renewcommand{\P}{\hat{\mathcal{P}}}
\newcommand{\Sz}{\hat{S}^z}
\newcommand{\Sp}{\hat{S}^+}
\newcommand{\Sm}{\hat{S}^-}

\newcommand{\Svec}{\hat{\vec{S}}}
\newcommand{\comm}[2]{\left[#1,#2\right]}

\usepackage{array}

\usepackage{cancel,ifthen}
\newcommand{\cmnt}[2][NoInPuT]{\ifthenelse{\equal{#1}{NoInPuT}}{}{{\color{red}\sout{#1}}} {\color{blue} #2}}

\usepackage{bm}	
\renewcommand{\vec}[1]{\bm{#1}}

\LTcapwidth=\textwidth

\graphicspath{ {./figures/} }

\begin{document}
\normalem	

\title{Particle zoo in a doped spin chain: Correlated states of mesons and magnons}

\author{Petar \v{C}ubela}
\affiliation{Department of Physics and Arnold Sommerfeld Center for Theoretical Physics (ASC), Ludwig-Maximilians-Universit\"at M\"unchen, Theresienstr. 37, M\"unchen D-80333, Germany}
\affiliation{Munich Center for Quantum Science and Technology (MCQST), Schellingstr. 4, D-80799 M\"unchen, Germany}

\author{Annabelle Bohrdt}
\address{ITAMP, Harvard-Smithsonian Center for Astrophysics, Cambridge, MA 02138, USA}
\affiliation{Department of Physics, Harvard University, Cambridge, MA 02138, USA}

\author{Markus Greiner}
\affiliation{Department of Physics, Harvard University, Cambridge, MA 02138, USA}

\author{Fabian Grusdt}
\email[Corresponding author email: ]{fabian.grusdt@physik.uni-muenchen.de}
\affiliation{Department of Physics and Arnold Sommerfeld Center for Theoretical Physics (ASC), Ludwig-Maximilians-Universit\"at M\"unchen, Theresienstr. 37, M\"unchen D-80333, Germany}
\affiliation{Munich Center for Quantum Science and Technology (MCQST), Schellingstr. 4, D-80799 M\"unchen, Germany}


\date{\today}

\begin{abstract}
It is a widely accepted view that the interplay of spin- and charge-degrees of freedom in doped antiferromagnets (AFMs) gives rise to the rich physics of high-temperature superconductors. Nevertheless, it remains unclear how effective low-energy degrees of freedom and the corresponding field theories emerge from microscopic models, including the $t-J$ and Hubbard Hamiltonians. A promising view comprises that the charge carriers have a rich internal parton structure on intermediate scales, but the interplay of the emergent partons with collective magnon excitations of the surrounding AFM remains unexplored. Here we study a doped one-dimensional spin chain in a staggered magnetic field and demonstrate that it supports a zoo of various long-lived excitations. These include magnons; mesonic pairs of spinons and chargons, along with their ro-vibrational excitations; and tetra-parton bound states of mesons and magnons. We identify these types of quasiparticles in various spectra using DMRG simulations \cite{Hauschild2018,Hauschild2019}. Moreover, we introduce a strong-coupling theory describing the polaronic dressing and molecular binding of mesons to collective magnon excitations. The effective theory can be solved by standard tools developed for polaronic problems, and can be extended to study similar physics in two-dimensional doped AFMs in the future. Experimentally, the doped spin-chain in a staggered field can be directly realized in quantum gas microscopes.
\end{abstract}

\maketitle

\section{Introduction}
Field theoretic approaches to quantum spin models in lattices, such as the Heisenberg antiferromagnet (AFM), provide very successful descriptions of these paradigmatic quantum many-body systems \cite{Haldane1983a} and have led to a thorough understanding of their various quantum phase transitions in different dimensions \cite{Sachdev2011}. Key to their success is the underlying hypothesis that the coarse-grained fields on long length-scales feature similar behavior as the microscopic local magnetic moments underlying the spin model. More formally, a simple renormalization-group (RG) procedure yielding the effective low-energy field theory does not change the particle-content of the analyzed fields. However, in dimensions larger than one and with mobile dopants included, this approach has not been able to explain the rich phase diagram of high-temperature superconductors so far.

In this article, we take a different perspective and explore emergent structures, at low- to intermediate energies, in a doped quantum spin chain. The zoo of constituents we find defies a naive field-theoretic description: we identify emergent parton structures of spinons and chargons, forming mesonic bound states with a rich spectrum of ro-vibrational internal excitations. Moreover, these mesons interact with collective magnon excitations in the surrounding spin system, which leads to polaronic dressing on the one hand and, more exotically, to long-lived meson-magnon bound states. In a phenomenological field-theoretic model, each of these constituents should be described by a separate quantum field, with mutual interactions between all of them. Describing how these new fields emerge at intermediate length- or energy-scales in a thorough RG procedure is a challenging task, even for the simple toy model we consider. Hence we focus on a microscopic description of the individual emergent bound states and analyze their characterizing properties, such as their dispersion relations, zero-point energies, and mutual interactions. To this end, we apply the powerful theoretical tools developed for the description of Bose polarons \cite{Chevy2010,Devreese2020,Grusdt2015Varenna,Rath2013,Shi2018}.

\begin{figure}
    \centering
    \includegraphics[scale=0.5]{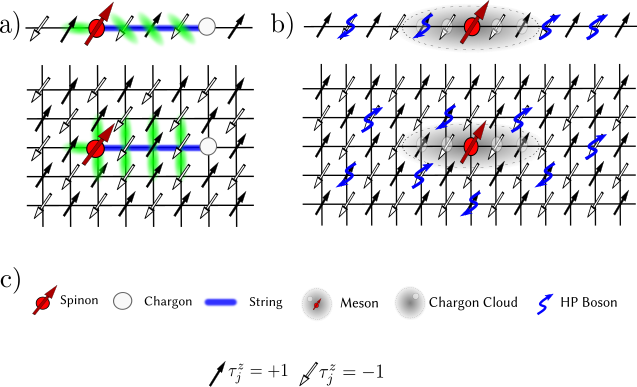}
    \caption{Particle zoo in a doped spin chain: We consider a doped mobile hole in a spin-chain subject to a staggered magnetic field. a) Ignoring transverse spin fluctuations, the staggered field leads to a confining force between spinons and chargons connected by a string of overturned spins (top row), which leads to meson formation. A similar situation is realized in a mixed-dimensional model, where a strong gradient prevents hole motion along the direction of the gradient (bottom row). b) Transverse spin couplings give rise to Holstein-Primakoff (HP) magnon fluctuations in the surrounding spin background. The latter interact with the spinon, which is surrounded by the strongly fluctuating but tightly bound chargon cloud. All constituents making up the zoo of excitations are summarized in c), where we also indicate the background Ising fields $\tau^z_j$, affected by the hole motion, around which we expand in the generalized $1/S$ approximation employed here.}
    \label{Fig1}
\end{figure}

Concretely, we consider doped one-dimensional spin chains. When featuring ${\rm SU}(2)$ invariance, these systems display spin-charge separation \cite{Kim1996,Sing2003,Ogata1990,Hilker2017,Vijayan2020}: the collective excitations of the spin-chain are fractionalized spinons, which co-exist with free chargons. In this limit, non-trivial bound states of the constituents are absent \cite{Ogata1990} and bosonization techniques provide a powerful field-theoretic description of the doped system in terms of Luttinger liquids \cite{Giamarchi2004}. As we demonstrate, the situation changes drastically when a staggered magnetic field is included, breaking the ${\rm SU}(2)$ symmetry, see Fig.~\ref{Fig1} a): Now spinons and chargons are confined \cite{Giamarchi2004,Borla2020,Kebric2021}, the undoped ground state has gapped collective magnon excitations, and spin-charge separation breaks down. Despite this confinement, the situation is far from trivial: As we will show, doped holes in this system host a zoo of excitations reflecting their rich internal structure, and their interaction with gapped magnons can lead to even more complicated multi-parton bound states, see Fig.~\ref{Fig2}.

In several regards, our 1D model is motivated by the physics of mobile holes in a ${\rm SU}(2)$-invariant 2D Hubbard model. In contrast to the 1D case with ${\rm SU}(2)$ symmetry, the ground state of the two-dimensional (2D) Heisenberg model has long-range magnetic order and gapless spin-$1$ magnon excitations. This effect is mimicked by the external staggered magnetic field we consider in our model, which introduces magnetic order and leads to similar spin-$1$ magnon modes, although with a non-vanishing gap. There is strong evidence that a doped hole in the 2D AFM features a rich internal meson structure \cite{Beran1996}, with discrete vibrational \cite{Brunner2000,Mishchenko2001,Bohrdt2020PRB} and rotational excitations \cite{Grusdt2018,Bohrdt2021}. In our 1D model, we reveal similar structures and develop an effective strong-coupling description.

Remarkably, the coupling of mesons to collective magnon excitations remains poorly understood in 1D and 2D, in particular around zero momentum where we show that the competition of mesons and magnons is most pronounced. In the present article we fill this gap and apply a powerful theoretical framework, the so-called generalized $1/S$ expansion \cite{Grusdt2018}, to describe the coupling of mesons to magnons in a systematic manner. As a result, we obtain an effective polaron Hamiltonian describing the dressing, or even binding, of a spinon-chargon meson with additional magnon excitations. Our work paves the way for similar studies in doped 2D Mott insulators, and may lead to a better understanding of the charge carriers and their interactions with magnons in underdoped copper oxides. In particular we expect that our formalism will be useful for understanding transport measurements involving magnetic polarons, such as the long-time spreading dynamics of a hole reported in~\cite{Hubig2020,Bohrdt2020Dyn,Ji2021,Nielsen2022arXiv}. 

Experimentally, the model we consider can be realized with ultracold atoms in optical lattices, which have recently made significant advances in studying doped quantum magnets \cite{Bohrdt2021review}. On a mean-field level, our model moreover maps to a doped mixed-dimensional $t-J$ model \cite{Grusdt2018SciPost}, which can be realized by subjecting a Fermi-Hubbard system to a strong tilt along one of the lattice directions \cite{Hirthe2022arXiv}. Ultracold atom realizations allow to measure spectra \cite{Stewart2008,Feld2011,Bohrdt2018,Brown2019} like the ones we calculate here to identify the emergent zoo of excitations; moreover they can directly visualize string patterns \cite{Endres2011,Hilker2017,Chiu2019} or the dressing cloud of magnetic polarons in configuration-space \cite{Koepsell2019}, making them ideal platforms to explore the emergent structures we predict on intermediate length scales.

\section{Model and main results}
In this article, we study a simple but rich one-dimensional model of a doped AFM. Our starting point is an ${\rm SU}(2)$-invariant Heisenberg spin chain. An additional staggered magnetic field of strength $\pm h$ on alternating sites along the $z$-direction breaks the ${\rm SU}(2)$ symmetry, introduces long-range magnetic correlations, and leads to collective magnon excitations with a tunable gap controlled by $|h|$. To describe mobile holes doped into this model, we use a $t-J$ Hamiltonian:
\begin{eqnarray}\label{model}
    \nonumber \H = -t \sum_{j,\sigma} \P \l \cd_{j+1,\sigma}\c_{j,\sigma} + \hc \r \P\\
    + J \sum_{j} \Svec_{j+1} \cdot \Svec_j - h \sum_j \l -1\r^{j} \Sz_j.
\end{eqnarray}
Since we will only consider a single doped hole in this article, we dropped the nearest-neighbor interaction $-J/4 ~ \hat{n}_{j+1} \hat{n}_j$ typically included in the $t-J$ model \cite{Auerbach1994}. A similar model, including phonons, has been studied in Ref.~\cite{Kogoj2014}.

\subsection{Lattice gauge Hamiltonian}
For later purposes, we find it convenient to write the Hamiltonian as a sum of two separate parts: (i) a $t-J_z$ part which conserves each individual spin in the so-called squeezed space \cite{Ogata1990,Kruis2004a,Hilker2017} obtained by removing holes from the chain:
\begin{multline}
    \H_{t-J_z} = -t \sum_{j,\sigma} \P \l \cd_{j+1,\sigma}\c_{j,\sigma} + \hc \r \P\\
    + J_z \sum_{j} \hat{S}^z_{j+1} \hat{S}^z_j - h \sum_j \l -1\r^{j} \Sz_j.
    \label{EqHIsing}
\end{multline}
To keep our analytical formalism later on general, we introduced the coupling $J_z$, which is simply $J_z=J$ for the original model in Eq.~\eqref{model}.

Remarkably, the Hamiltonian in Eq.~\eqref{EqHIsing} is exactly equivalent to a $\mathbb{Z}_2$ lattice gauge theory (LGT), as shown in Refs.~\cite{Grusdt2020PRL,Kebric2021}. In this mapping, chargons (i.e. spin-less holes) and spinons (i.e. Ising domain walls) carry $\mathbb{Z}_2$ gauge charges and are connected by a $\mathbb{Z}_2$ electric string $\tau^x_{\langle i,j \rangle}$. The staggered field $\pm h$ leads to a term $h \sum_{\langle i,j \rangle} \hat{\tau}^x_{\langle i,j \rangle}$ in the $\mathbb{Z}_2$ gauge invariant Hamiltonian. The latter has been shown to cause spinon-chargon confinement for any infinitesimal $h \neq 0$ \cite{Borla2020,Kebric2021}. This $\mathbb{Z}_2$ LGT formalism forms the basis for our mesonic description of a doped hole.

In addition, the full Hamiltonian in Eq.~\eqref{model} includes (ii) transverse spin fluctuations,
\begin{equation}
    \H = \H_{t-J_z} + \H_{J_\perp},
\end{equation}
where we find it most convenient to write
\begin{equation}
    \H_{J_\perp} = \frac{J_\perp}{2} \sum_{j} \l \hat{S}^+_{j+1} \hat{S}^-_{j} + \hc \r.
    \label{eqHJperp}
\end{equation}
Again we introduced the more general coupling strength $J_\perp$ in this term, although for our original model in Eq.~\eqref{model} $J_\perp = J$.
Later on, we will include such transverse spin fluctuations on top of a N\'eel ordered ground state distorted by the hole motion by introducing Holstein-Primakoff bosons (magnons), see Fig.~\ref{Fig1} b).

Finally, we note that in the limit $h / J_\perp \to \infty$, the transverse fluctuations $\H_{J_\perp}$ can always be treated perturbatively, independent of the ratios $J_z/J_\perp$ or $t/J_\perp$. To lowest order, only the $t-J_z$ part of the Hamiltonian, Eq.~\eqref{EqHIsing}, remains and it follows that the model has an emergent $\mathbb{Z}_2$ gauge structure for large values of $h$.

\subsection{Main results: particle zoo in the spin chain}
The separation of the Hamiltonian in two components lends a natural understanding of our results. Our main goal is to understand the ground and excited states of a mobile dopant in the spin chain. As described in detail below, we find that the $\mathbb{Z}_2$ gauge structure of the $t-J_z$ part of the Hamiltonian, or equivalently (in our model) the string-picture of magnetic polarons \cite{Bulaevski1968,Trugman1988,Grusdt2018}, introduces parton constituents, namely spinons and chargons, which are confined by the linear string tension generated by the staggered field $h$. The resulting mesonic spinon-chargon bound state has a rich internal structure constituted by inversion-even and inversion-odd vibrational modes of the $\mathbb{Z}_2$ electric string, or equivalently the string of overturned Ising spins, connecting the spinon and the chargon. We probe these states directly in spectra calculated by time-dependent matrix product states (td-MPS), see Sec.~\ref{secDMRGresults}, and compare to an effective strong-coupling description that we develop here, see Sec.~\ref{secStrongCplngThy}.

\begin{figure}
    \centering
    \includegraphics[width=0.5\textwidth]{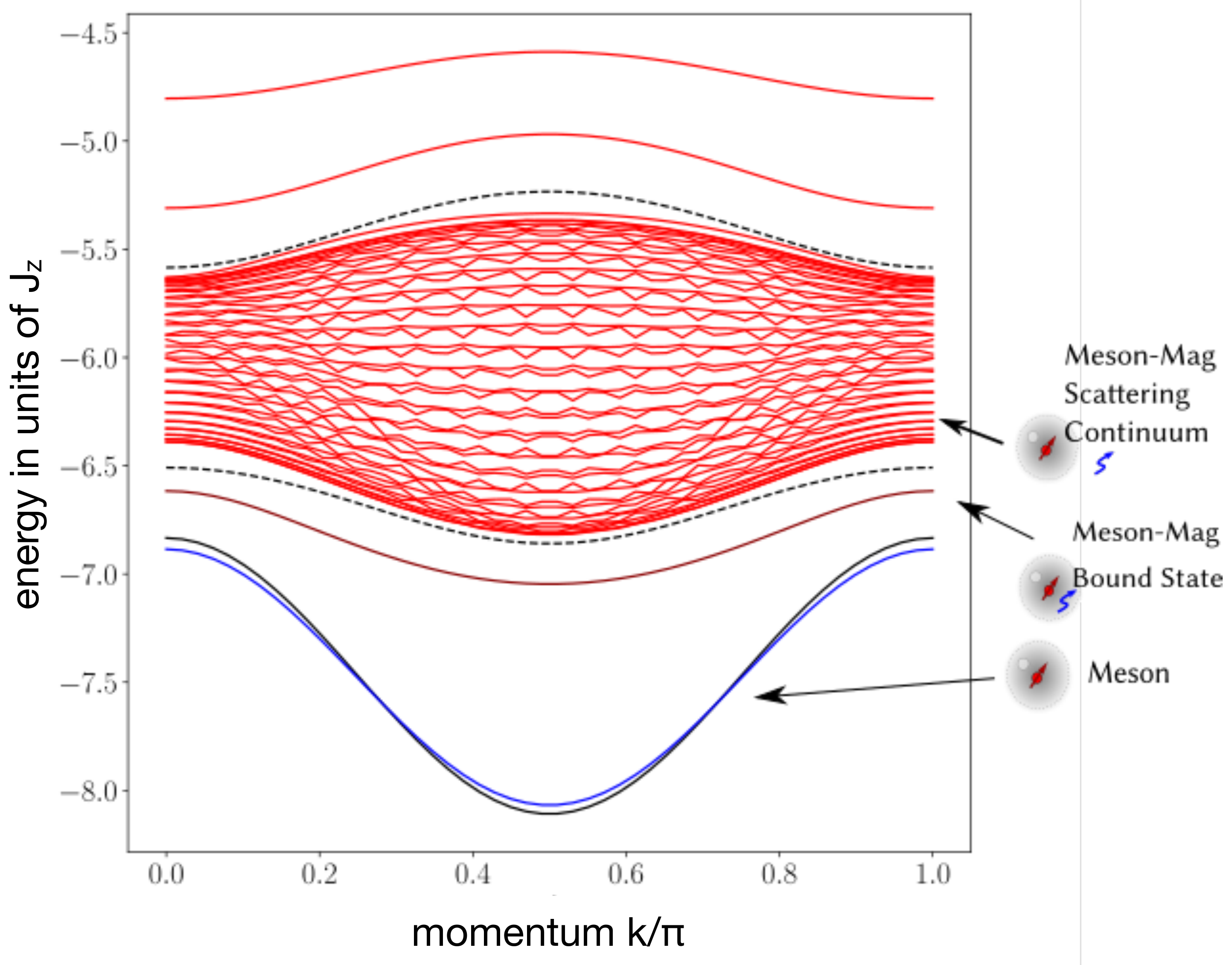}
    \caption{Polaronic bands in the presence of meson-magnon interactions at low energies: The overall ground state at momentum $k=\pi/2$ is realized by a weakly dressed meson (solid blue line). Before the broad meson-magnon continuum is reached at higher energies (wide red band), we predict a weakly dispersing meson-magnon bound state (dark red line), corresponding to a tetra-parton configuration. The black lines indicate the bare meson dispersion (solid) and edges of the meson-magnon continuum (dashed) in the absence of meson-magnon interactions, respectively. At higher energies (not shown) we find ro-vibrational internal meson excitations. Calculations were performed using the strong-coupling generalized $1/S$ approximation introduced in the text; we chose parameters $h=0.6 J$ and $t=5 J$.}
    \label{Fig2}
\end{figure}

The transverse couplings introduced by $\H_{J_\perp}$ lead to vacuum fluctuations of magnons in the absence of a doped hole. This effect can be captured by a simple linear spin-wave expansion around the classical N\'eel state, which we achieve by a Holstein-Primakoff approximation. In the vicinity of the meson, the distortion of the N\'eel background caused by the spinon-chargon pair introduces additional couplings to magnons which give rise to additional rich physics: On one hand, they lead to polaronic dressing and weak mass renormalization of the meson around the dispersion minimum at momentum $k=\pi/2$. This is shown for the lowest-energy mesonic state (solid blue line) in Fig.~\ref{Fig2}. 

More dramatically, the interactions with magnons can give rise to meson-magnon bound states. Since the magnon itself can be viewed as a bound state of two confined spinons, this state constitutes an emergent tetra-parton composite. As demonstrated in Fig.~\ref{Fig2}, for sufficiently large values of $h$ our effective model of the meson-magnon coupling predicts a low-lying meson-magnon bound state at relatively low energies below the meson-magnon scattering continuum. This should be contrasted with the higher excitation energies of ro-vibrational internal meson modes. We confirm our prediction of meson-magnon bound states in td-MPS calculations of one-hole spectra in a sector with total spin $S_z=3/2$, see Sec.~\ref{secDMRGresults}.

Finally, meson-magnon interactions can have a pronounced effect on the quasiparticle dispersion of the dressed meson around momentum $k=0$. In this region of momentum space, the bare meson dispersion approaches the meson-magnon scattering continuum most closely, as indicated by the dashed and solid black lines in Fig.~\ref{Fig2}. Without meson-magnon interactions and for sufficiently weak fields $h\ll t,J$, we find that they can even cross, leading to a decaying bare meson state inside the meson-magnon scattering continuum. However, in Sec.~\ref{SubSecAvdMgnDcy} we analyze our effective meson-magnon Hamiltonian and find indications that meson-magnon interactions in the 1D chain are strong enough to avoid such quasiparticle decay \cite{Verresen2018}. Namely, the meson and magnon bands repel and an isolated quasiparticle band of the mesonic magnetic polaron survives even around $k=0$. This prediction is further supported by td-MPS simulations at small fields where the effect is most pronounced.

Methodologically, we deviate from the standard approach typically used to describe magnetic polaron formation in an AFM \cite{Kane1989,Sachdev1989,Martinez1991}. As mentioned above, we first take into account how the mobile hole distorts the N\'eel background with pure Ising interactions. This allows us to make a direct connection to the $\mathbb{Z}_2$ LGT and identify the parton content of the meson. Moreover, we can relatively easily capture the competition between the tunneling term $t$ and the linear string tension $\propto h$, to all orders in $t/h$. This is achieved within a strong-coupling theory. Next we introduce generalized Holstein-Primakoff bosons (loosely speaking, magnons) by expanding around the \emph{already distorted} N\'eel state (we refer to this approach as the generalized $1/S$ approximation \cite{Grusdt2018}). This yields additional couplings of the meson to the magnons; importantly, the strength of these couplings is only of order $J$, and a fraction of $t$ for some further corrections we identify. Hence, perturbative or simple variational approaches are sufficient to capture the additional meson-magnon interactions. This should be contrasted with the traditional $1/S$ approximation \cite{Kane1989,Sachdev1989,Martinez1991} where the hole hopping $t$ itself leads to magnon creation: as a result, the effective Hamiltonian is strongly coupled when $t>h$ and direct analytical insights are harder to obtain.

\subsection{Possible experimental realizations}
Experimentally, the model in Eq.~\eqref{model} we study can be realized in different ultracold atom setups. We propose to use ultracold fermionic Lithium or Potassium atoms which have very successfully explored the ${\rm SU}(2)$-invariant 2D Fermi-Hubbard model \cite{Bohrdt2021review}. The main obstacle in these systems is to implement the staggered magnetic field, which requires local addressability on the scale of an optical wavelength, see e.g.~\cite{Weitenberg2011}, and sizable magnetic moments in order to distinguish different spin states, in a regime close to an atomic Feshbach resonance to realize super-exchange couplings. 

A first option is to use Potassium atoms in a quantum gas microscope \cite{Cheuk2015} which have a sizable magnetic moment \cite{Nichols2018}, allowing for a local modulation of the magnetic field. A second option is to work in a mixed-dimensional setting where tunneling is strongly suppressed by strong gradients along all but one lattice direction \cite{Grusdt2018SciPost}. Moreover, we assume that nearest-neighbor AFM Ising couplings between all spins are present, which dominate over the weak super-exchange couplings along the gradient directions. This can be realized in an optical lattice by adding Rydberg dressing \cite{GuardadoSanchez2017}. When doping only the central chain with one hole and keeping all neighboring chains at half filling, the surrounding spin chains can generate an effective staggered field term $\pm h$ if they are sufficiently cold. Here we assumed, in a mean-field spirit, that the wavefunctions of the different chains approximately factorize. Similarly, in mixed-dimensional settings with $SU(2)$ invariant spin-exchange interactions \cite{Grusdt2018SciPost,Hirthe2022arXiv} we expect a ground state with broken $SU(2)$ symmetry in qualitatively very similar physics. 

Finally, we note that the 1D model in Eq.~\eqref{model} can be equally realized with bosons as long as one ensures to have AFM Heisenberg couplings between the spins \cite{Trotzky2008,Dimitrova2019}. The statistics of the dopants is irrelevant, as can be shown by a Jordan-Wigner transformation. Hence the model in Eq.~\eqref{model} can also be simulated in qubit arrays or digital quantum computers \cite{Arute2020}, without the need to incorporate fermionic statistics.

\section{Numerical DMRG spectra}
\label{secDMRGresults}
In this section we present our numerical results, largely based on td-MPS simulations \cite{Paeckel2019}, which support our main findings about the structure and interactions of doped holes in the 1D spin chain with a staggered field. We already compare our numerical results to predictions by the semi-analytical strong-coupling meson-magnon theory introduced in the subsequent sections. This theory provides a unified understanding of all our key numerical observations.  

Detailed descriptions of the numerical td-MPS simulations we performed can be found in Refs.~\cite{Bohrdt2020PRB,Bohrdt2021}; our algorithm builds upon the earlier works \cite{Kjaell2013,Zaletel2015,Verresen2018spec}. To ensure proper convergence of the MPS calculations, we performed the same convergence checks, in time and bond-dimension, as described in \cite{Bohrdt2020PRB,Bohrdt2021}.

\subsection{Ground state: Dressed hole}
\label{subsecDMRGgs}
\begin{figure}
    \centering
    \includegraphics[width=0.5\textwidth]{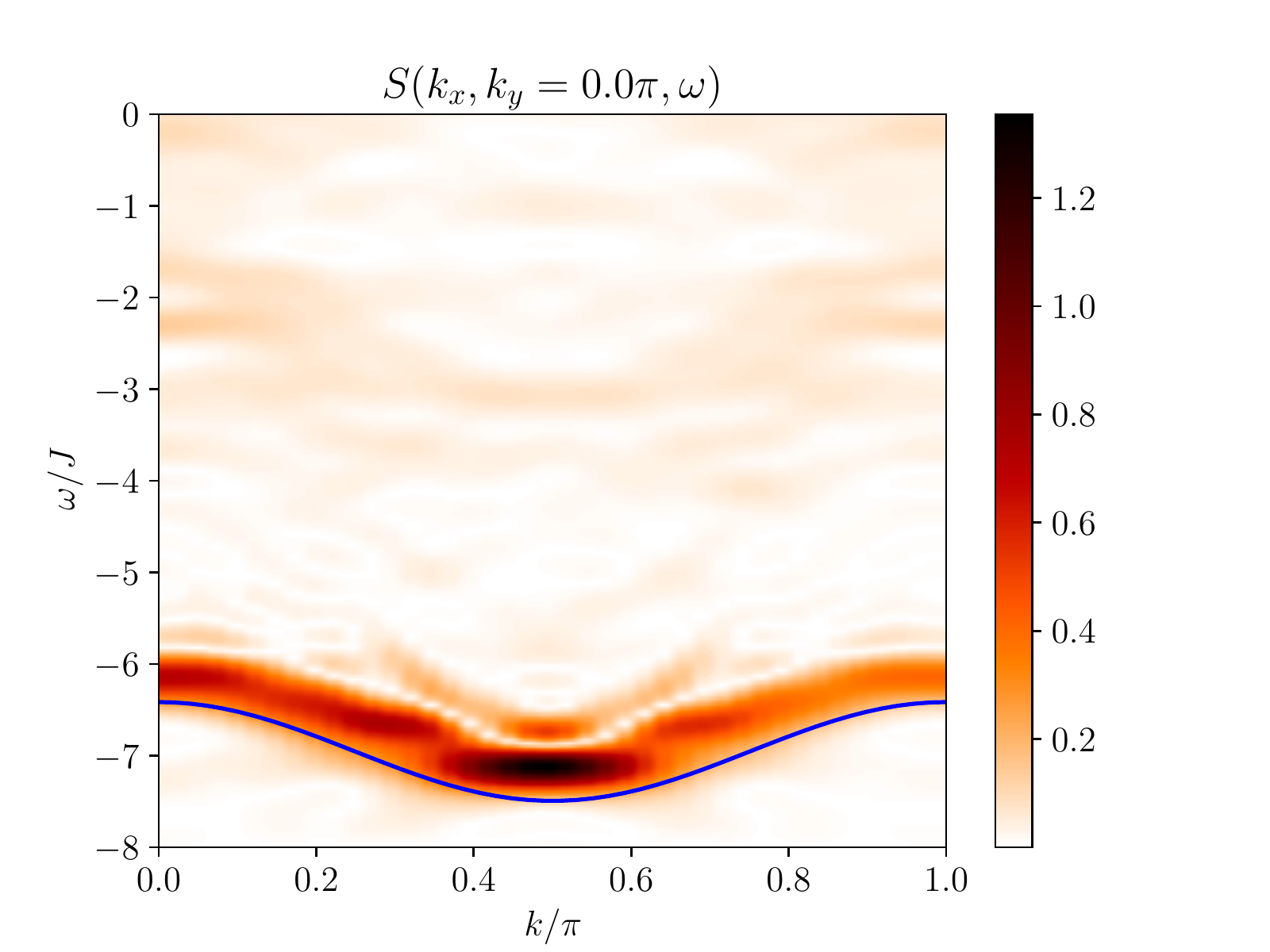}
    \caption{The standard one-hole ARPES spectrum reveals a pronounced quasiparticle peak at the lowest energy. The dispersion minimum is located at $k=\pi/2$, as predicted by our semi-analytical theory (solid blue line). Here we consider $h=1.0J$ and $t=5J$; the color scale is in a.u..}
    \label{StandARPES}
\end{figure}

In Fig.~\ref{StandARPES} we start by showing the standard one-hole angle-resolved photoemission spectrum (ARPES), defined by 
\begin{equation}
    S(k,\omega) = -\frac{1}{\pi} \text{Im}{\int_0^\infty dt \frac{e^{i (\omega-E_0) t}}{\sqrt{L}} \sum_j e^{i k j} \mathcal{G}_j(t)},
    \label{eqAkwDef}
\end{equation}
with the Green's function
\begin{equation}
    \mathcal{G}_j(t) = \sum_\sigma \bra{\Psi_0} \cd_{j,\sigma} e^{- i \H t} \c_{0,\sigma} \ket{\Psi_0}
    \label{eqGjStandard}
\end{equation}
where $\ket{\Psi_0}$ is the ground state with energy $E_0$ of $\H$ with zero holes. 

In the spectrum, we observe a pronounced quasiparticle peak at low energy which corresponds to the magnetic polaron. The comparison with our semi-analytical theory shows that it is located around the expected energy, and shows the same dispersion relation with a minimum at $k=\pi/2$. At higher energies the spectrum is relatively featureless for the considered value of $h/J=1.0$ in Fig.~\ref{StandARPES}. As we show next, additional features becomes visible for larger values of $h$.

\subsection{Ro-vibrational excitations: Mesonic states}
\label{subsecDMRGmeson}
Now we calculate a rotational variant of the ARPES spectrum, where spinon-chargon excitations with odd $(\xi=-1)$ and even ($\xi=+1$) inversion symmetry can be detected. It is defined as in Eq.~\eqref{eqAkwDef} but using the rotational Green's function \cite{Bohrdt2021}
\begin{equation}
    \mathcal{G}^{\rm rot}_{j,\xi}(t) = \sum_\sigma \bra{\Psi_0} \cd_{j,\sigma} \hat{X}_{j,\xi}^\dagger e^{- i \H t} \hat{X}_{0,\xi} \c_{0,\sigma} \ket{\Psi_0}
\end{equation}
where $\hat{X}_{j,\xi}=\sum_\sigma \cd_{j,\sigma} ( \c_{j+1,\sigma}+\xi \c_{j-1,\sigma})$ creates an additional excitation of the spinon-chargon pair.

\begin{figure}
    \centering
    \includegraphics[width=0.47\textwidth]{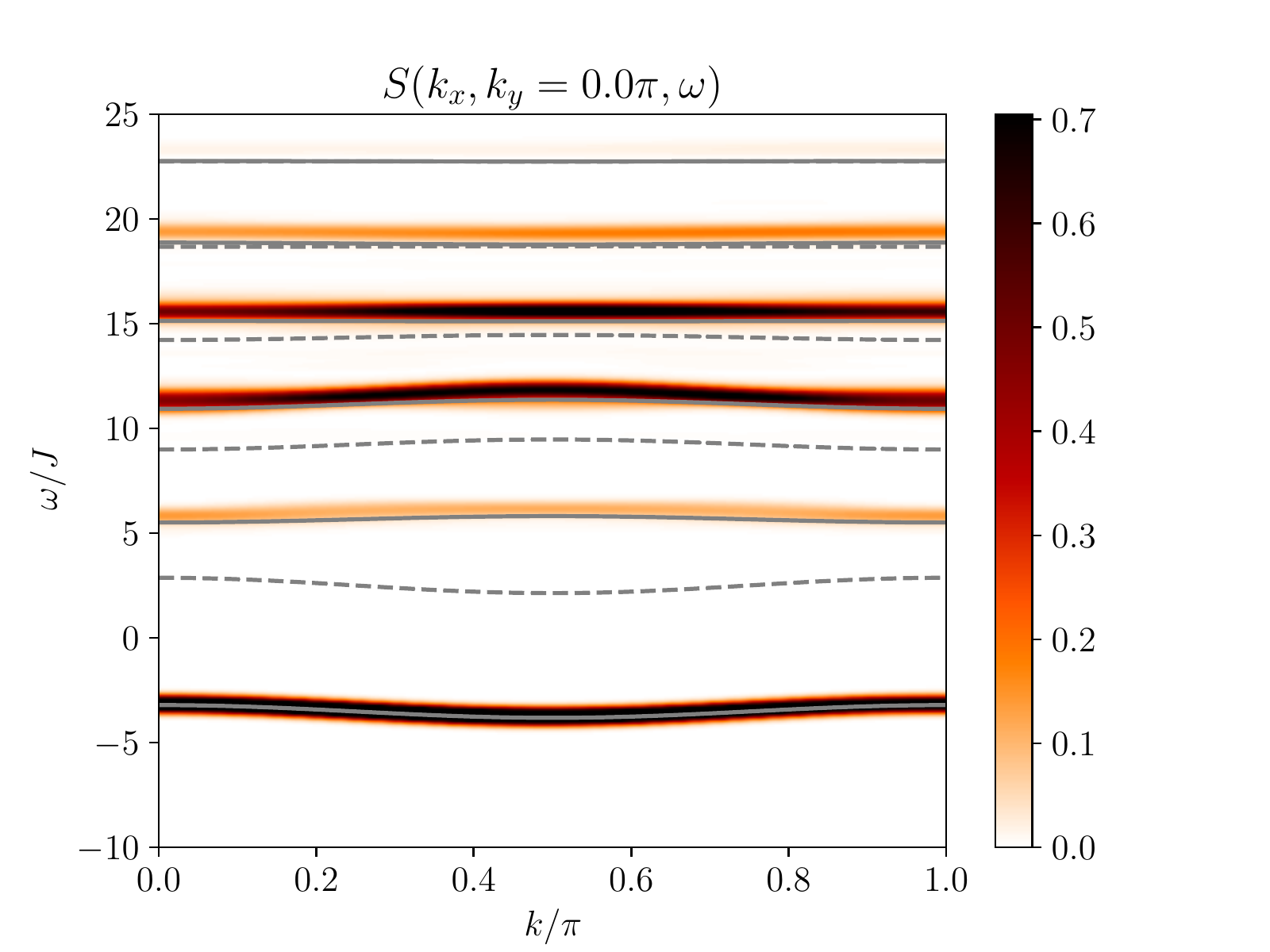}\\
    \includegraphics[width=0.47\textwidth]{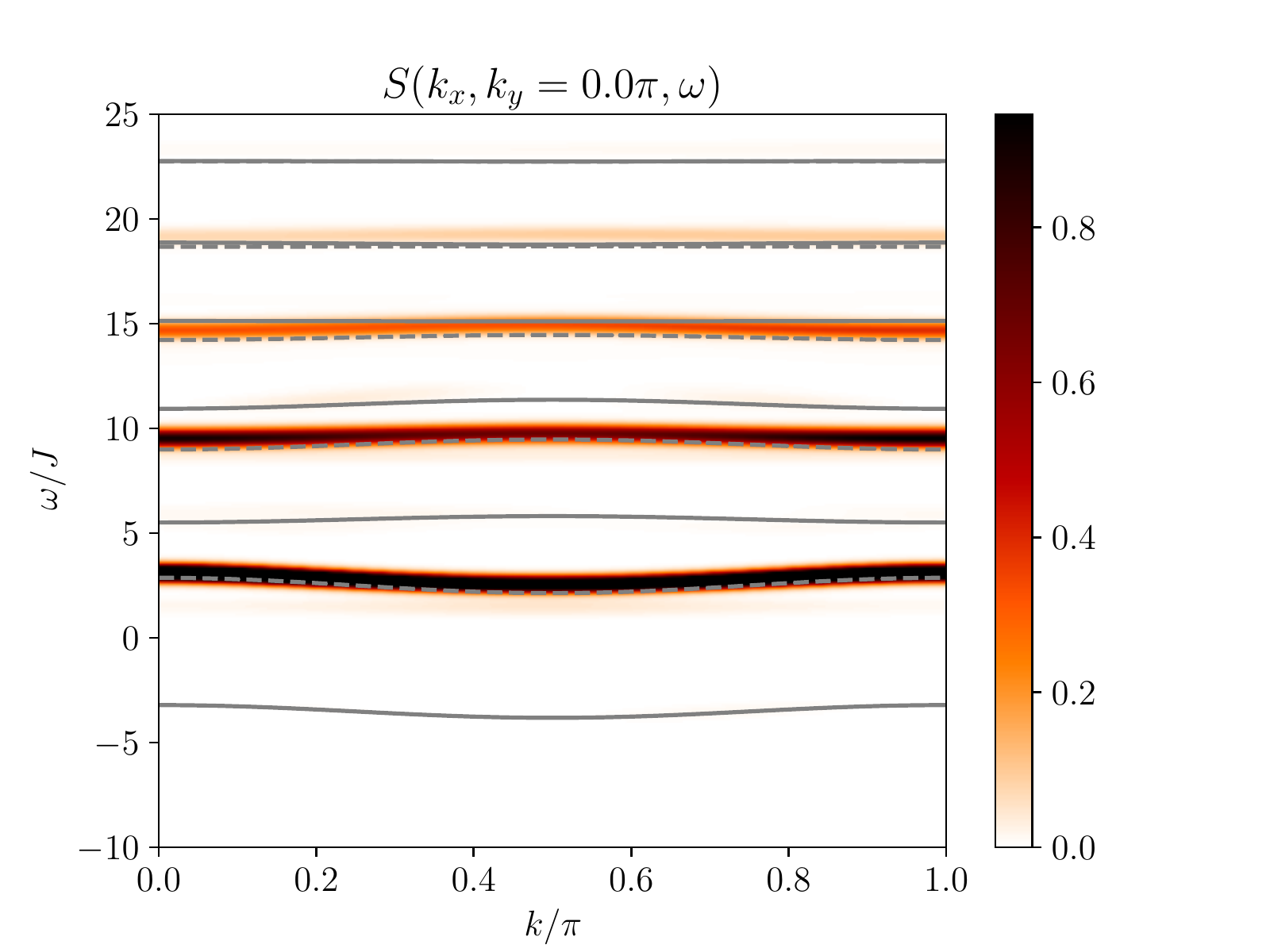}
    \caption{The rotational one-hole ARPES spectrum reveals a series of long-lived vibrational excitations with even ($\xi=+1$, top) and odd ($\xi=-1$, bottom) parity. We compare the td-MPS spectra with bare meson resonances calculated from our strong-coupling theory (gray solid lines: $\xi=+1$ even; gray dashed lines: $\xi=-1$ odd). Here we consider $h=4.0J$ and $t=5J$; the color scale is in a.u..}
    \label{RotARPES}
\end{figure}

In Fig.~\ref{RotARPES} we show our results for $h=4J$. In both parity sectors $\xi=\pm 1$ we observe pronounced vibrational peaks, which correspond to vibrational modes of the spinon-chargon string. The absence of even (odd) peaks in the odd (even) spectrum indicates that the parity $\xi$ is a good emergent quantum number at all momenta, not only at $k=0,\pi/2$ where the system is strictly inversion symmetric. This is a direct indication for the existence of an internal meson structure \cite{Bohrdt2021}.

In Fig.~\ref{RotARPES} we also compare the peak positions observed in td-MPS with predictions by our strong-coupling theory. The observed peaks in our full numerical spectra are in excellent agreement with our semi-analytical predictions. In the latter, for simplicity, we neglected corrections from magnon-dressing which are weak at large values of $h$. Nevertheless, note that significant charge fluctuations are present since we consider $t>h$ in the figure.

\subsection{Meson-magnon bound states}
\label{ARPESBound}
Next we show that even more complex excitations can arise when the mesonic hole interacts with its spin environment. Specifically, the meson can form a stable bound state with a magnon excitation. To demonstrate the existence of such bound states, we first consider an even more involved type of spectral function. To obtain spectral weight in the sector with one hole and one extra magnon, we create an excitation with total spin $S^z=3/2$ by flipping a spin next to the hole. This corresponds to working with the meson-magnon Green's function
\begin{equation}
    \mathcal{G}^{\rm mes-mag}_j(t) = \sum_\sigma \bra{\Psi_0} \cd_{j,\sigma} ~\hat{\Xi}^\dagger_{j}~ e^{- i \H t} ~ \hat{\Xi}_{0}~ \c_{0,\sigma} \ket{\Psi_0},
\end{equation}
where $\hat{\Xi}_{j} = \sum_{\delta=\pm 1}\cd_{j+\delta,\overline{\sigma}} \c_{j+\delta,\sigma}$ flips an additional spin and $\overline{\uparrow}=\downarrow$ ($\overline{\downarrow}=\uparrow$).

\begin{figure}
    \centering
    \includegraphics[width=0.5\textwidth]{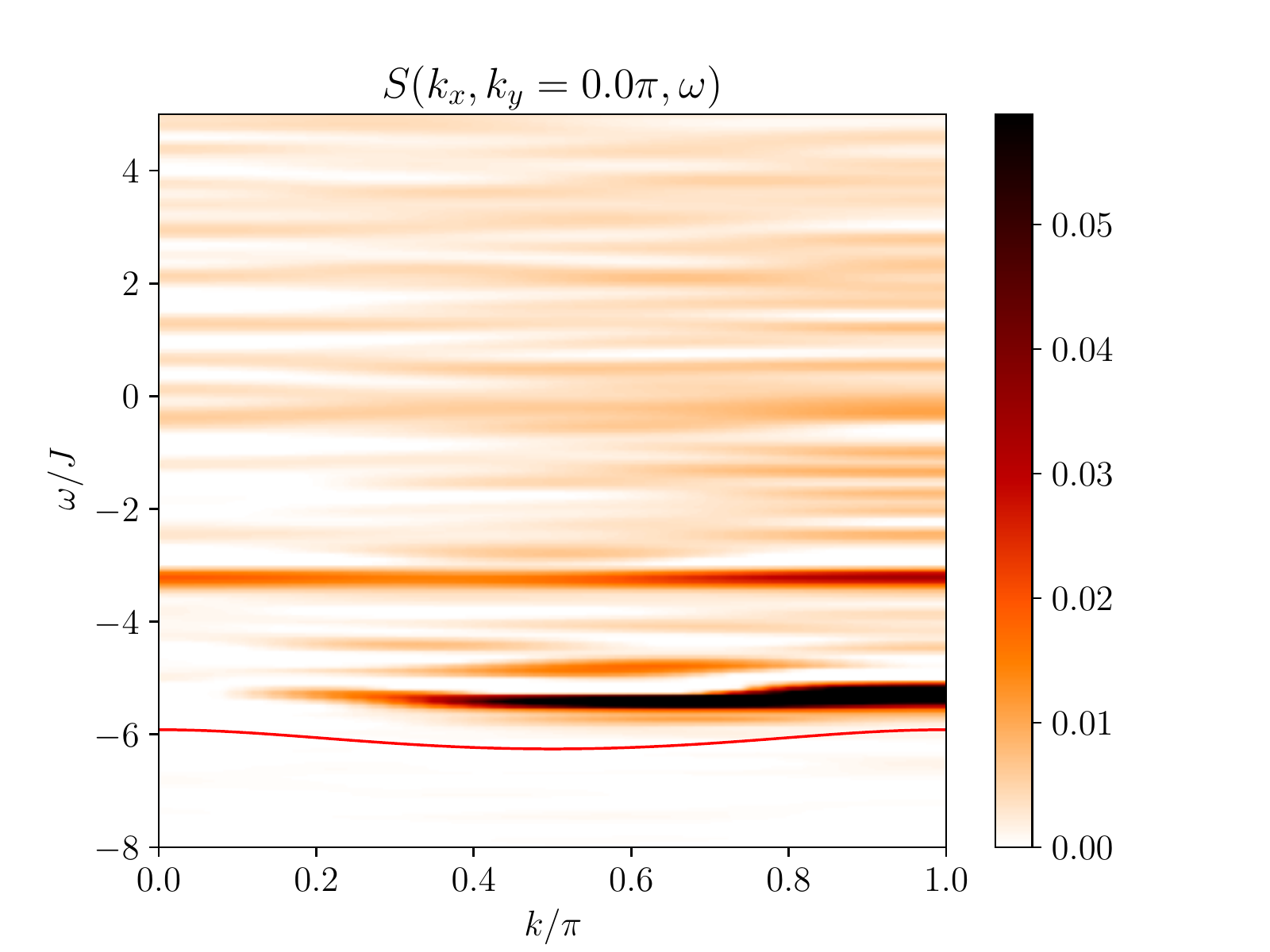}
    \caption{The spin-flip one-hole ARPES spectrum probes the sector with total spin $S^z=3/2$. It reveals a long-lived weakly dispersing meson-magnon bound state. We compare the td-MPS spectrum with our semi-analytical prediction for the meson-magnon bound state (solid red line). Parameters are $h=1.0J$ and $t=5J$; the color scale is in a.u..}
    \label{StagFlipARPES}
\end{figure}

The resulting spin-flip one-hole ARPES spectrum is shown in Fig.~\ref{StagFlipARPES}. There we observe a low-lying pronounced quasiparticle peak featuring a weakly dispersing band. Comparison to our semi-analytical prediction in the one-hole plus one magnon sector yields good qualitative agreement up to a small overall energy shift a fraction of $J$. Hence we interpret the observed feature as a stable meson-magnon bound state.

To further analyze the robustness of the meson-magnon bound state, we need to check whether it lies energetically below the meson-magnon scattering continuum. This is the case for the bound state predicted by our semi-analytical theory: Indeed, in Fig.~\ref{Fig2} we observe an isolated one-magnon excited state (lowest red band) between the mesonic ground state (blue) and the meson-magnon continuum (filled red band). To test this scenario in our fully numerical DMRG simulations, we calculate the meson-magnon binding energy, which is defined as follows:
\begin{equation}
    E_{\rm mm} = (E_{{\rm 1h},3/2} - E_{{\rm 0h},0}) - (E_{{\rm 1h},1/2} + E_{{\rm 0h},1} - 2 E_{{\rm 0h},0}).
\end{equation}
Here $E_{n{\rm h},s}$ denotes the ground state energy in the sector with $n$ holes and total spin $S^z=s$. If $E_{\rm mm}<0$, the meson-magnon state is located below the scattering continuum and forms a stable bound state.

\begin{figure}
    \centering
    \includegraphics[width=0.5\textwidth]{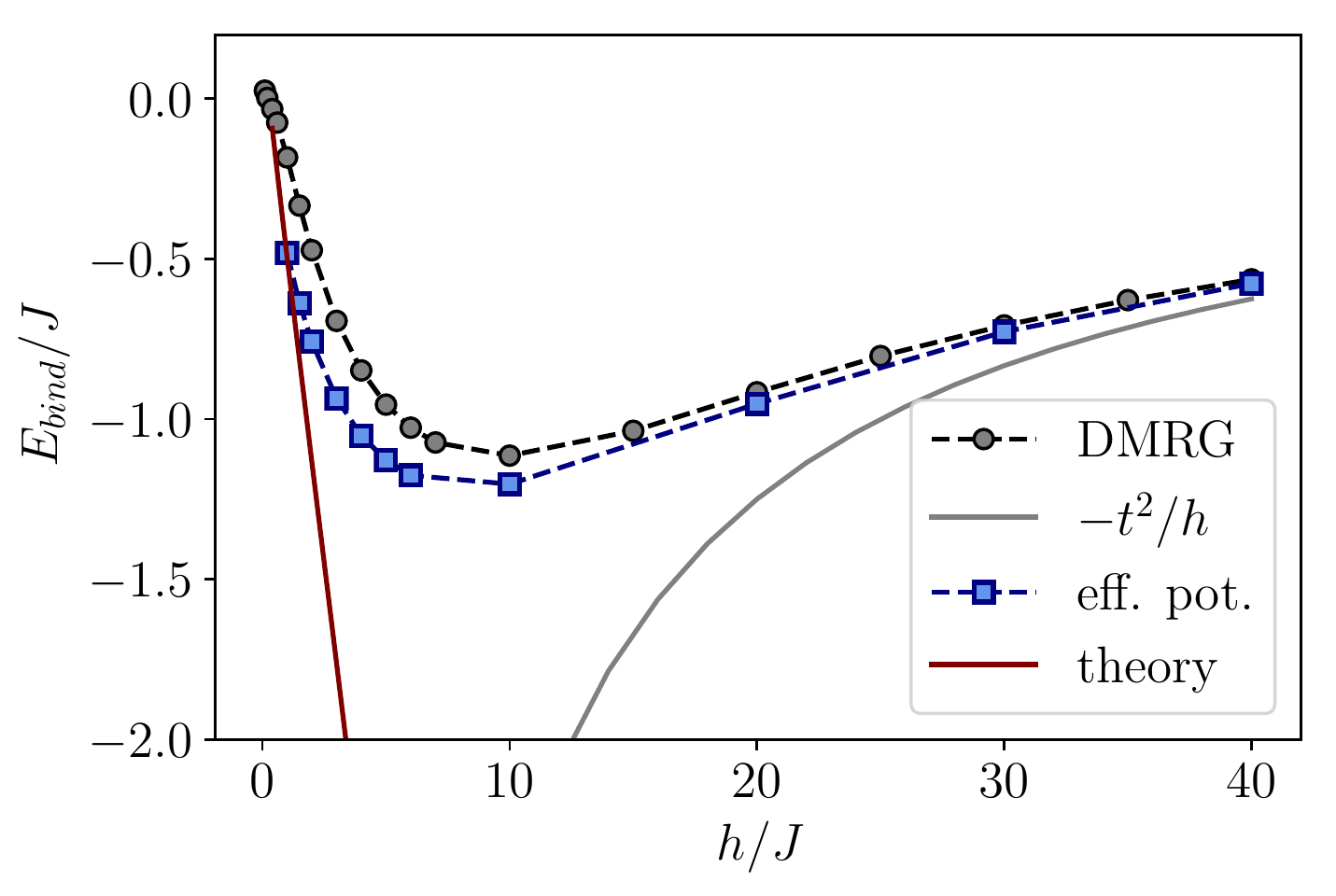}
    \caption{The meson-magnon binding energy evaluated from DMRG simulations of the ground state with and without an additional hole and magnon. We used an additional magnetic field of strength $h_{\rm edge}=2 J$ at the boundaries of the chain to avoid boundary effects.} 
    \label{MesMagBndgEngy}
\end{figure}

In Fig.~\ref{MesMagBndgEngy} we plot the numerically obtained meson-magnon binding energy for various field strengths $h/J$, at fixed $t/J=5$. We find consistently that  $E_{\rm mm}<0$ beyond a critical field strength $h_c = 0.3(1) J$, confirming the existence of a stable bound state as anticipated from the spin-flip ARPES spectrum. In the figure, we also compare our results to the semi-analytical theory (solid red) and an effective theory valid at large $h \gg J$ (see Appendix \ref{ApdxBoundStateLargeh}). For small $h$, our semi-analytical theory is in good agreement with the numerics. At larger values of $h$ we observe deviations, which can be attributed to some simplifying approximations we made, see Sec.~\ref{secSolSCthy} for a detailed discussion. The large-$h$ theory provides good qualitative agreement everywhere.

\begin{figure}
    \centering
    \includegraphics[width=0.5\textwidth]{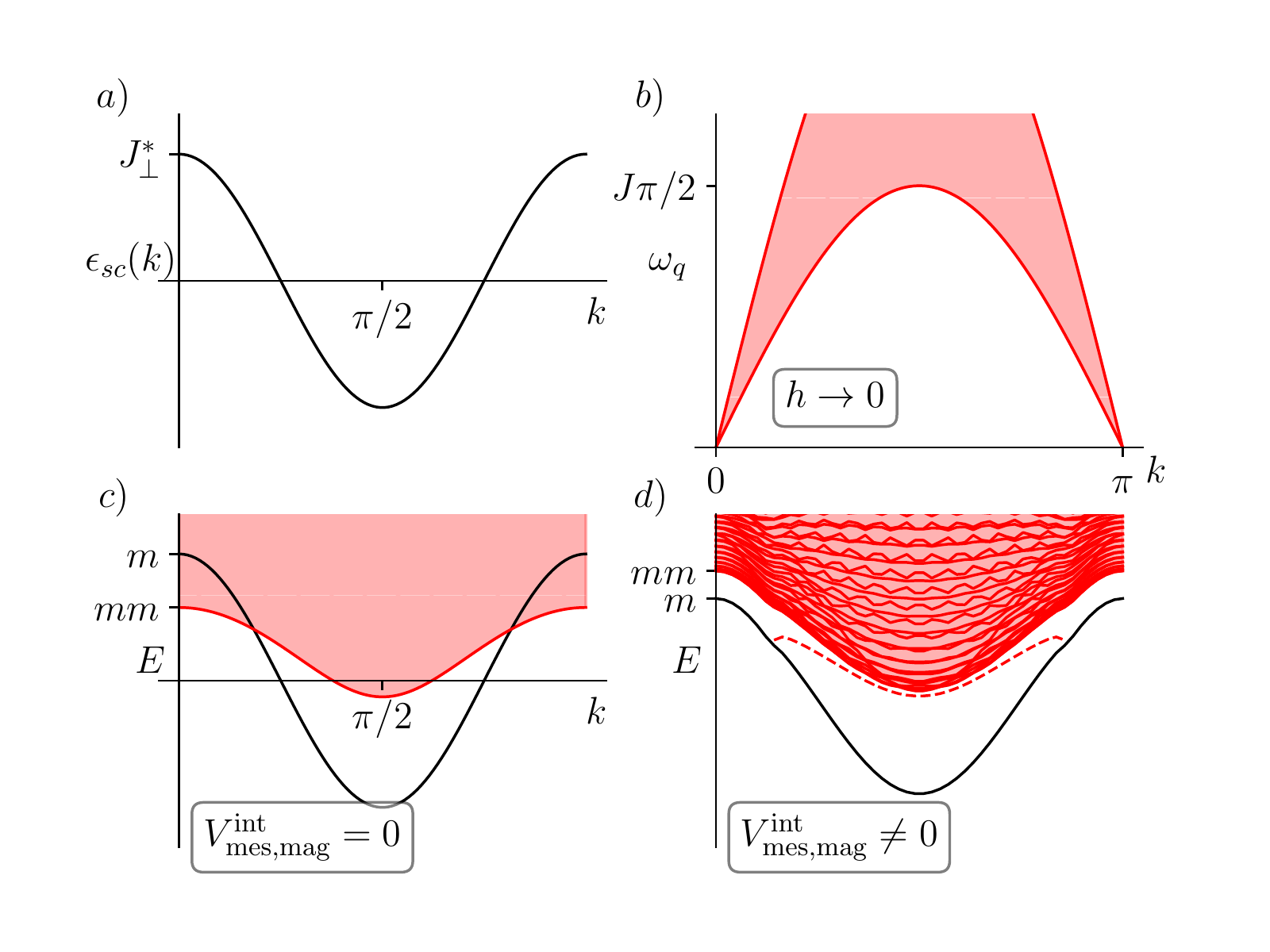}
    \caption{Avoided quasiparticle decay. a) The meson dispersion in the limit $t \gg h,J$ approaches its strong coupling shape $J_\perp^* \cos(2k)$. b) The magnon dispersion in the limit $h \ll J$ approaches the 1D spinon-continuum, with a lower edge at $(J \pi/2) |\sin(q)|$. c) Without interactions, the bare meson state ($m$) would enter the meson-magnon ($mm$) continuum when $t \gg J \gg h$. d) In the presence of sufficiently strong meson-magnon interactions, the mesonic quasiparticle band remains stable for all momenta.}
    \label{FigAvdQPdecay}
\end{figure}

\subsection{Avoided magnon decay in a weak field}
\label{secAvdMagDcySmry}
Finally, to study the effect of meson-magnon interactions around zero total system momentum $k=0$, we return to the standard ARPES spectrum, i.e. the Green's function in Eq.~\eqref{eqGjStandard}. However now we consider a parameter regime where $t \gg J > h$. Beyond a critical value $t>t_c(J,h)$ depending on $h$ and $J$, the bare spinon-chargon dispersion is predicted by our strong coupling theory to enter the magnon continuum in the absence of meson-magnon interactions. Let us begin by specifying what is meant by this.

We consider two states, both at the same total system momentum $k$. In the first -- the bare meson state -- all momentum is carried by the spinon-chargon pair. From the strong-coupling theory we predict its energy to be
\begin{equation}
    \varepsilon_{\rm sc}(k) = J_\perp^* \cos(2k),
\end{equation}
with a renormalized tunneling $J_\perp^*$, see Fig.~\ref{FigAvdQPdecay} a). In Sec.~\ref{subsecDMRGmeson} we already confirmed that the meson dispersion takes this general shape. The second, competing state we consider contains an additional magnon excitation with momentum $q$ and energy $\omega_q$, see Fig.~\ref{FigAvdQPdecay} b). To obtain the same total momentum $k$, the meson carries momentum $k-q$ and the total energy of the state is:
\begin{equation}
    E_{k,q}=\varepsilon_{\rm sc}(k-q) + \omega_q.
\end{equation}
This defines the meson-magnon continuum.

Since for $h \neq 0$ the magnon spectrum $\omega_q > 0$ is gapped, the lowest energy state at $k=\pi/2$ always corresponds to a single spinon-chargon pair. However, at $k=0$ the situation is much more interesting. In particular, the meson state with $k=0$ and energy $\varepsilon_{\rm m}=\varepsilon_{\rm sc}(k)=J_\perp^*$ is very competitive with the meson-magnon state at $q=\pi/2$ which has energy $\varepsilon_{\rm mm}=\omega_{\pi/2}-J_\perp^* = E_{k=0,q=\pi/2}$. Indeed, for very small values of $h \to 0$ we can show analytically using our strong-coupling theory that $J_\perp^* \to J$. Moreover, in this limit the magnon dispersion approaches the well-known spinon continuum which has a lower-edge at $\varepsilon_{\rm s}(q) = J \pi/2 |\sin(q)|$ \cite{Giamarchi2004}. Hence
\begin{equation}
    \varepsilon_{\rm mm} - \varepsilon_{\rm m} =  \frac{\pi}{2} J - 2 J = -0.429... J < 0,
\end{equation}
and we conclude that, in the absence of meson-magnon interactions, the meson-magnon state has lower energy; i.e. the bare meson enters the meson-magnon continuum in this limit, see Fig.~\ref{FigAvdQPdecay} c). 

\begin{figure}[t!]
    \centering
    \includegraphics[width=0.5\textwidth]{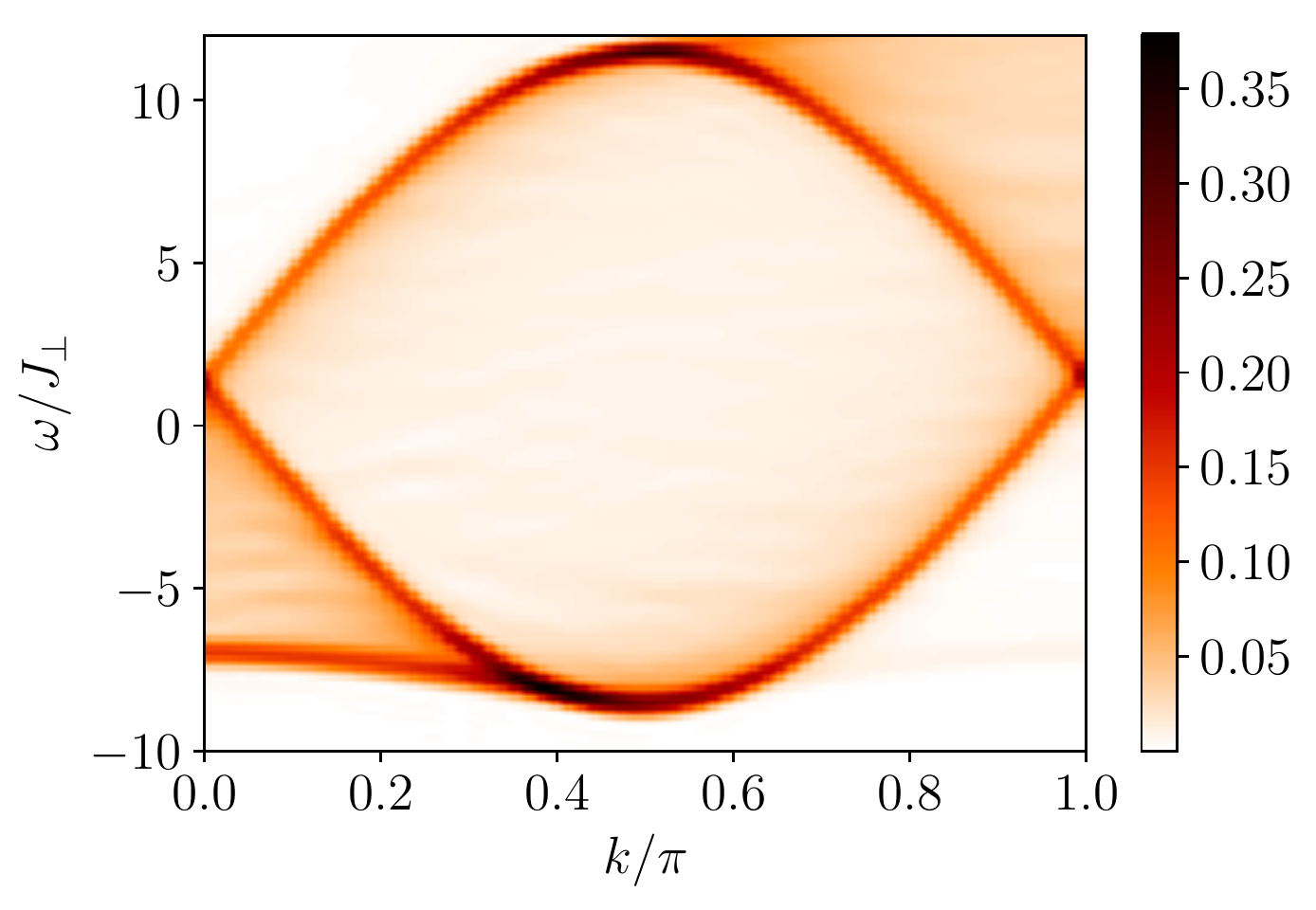}
    \caption{Avoided quasiparticle decay in the weak-field large-tunneling limit, seen in ARPES spectra obtained by td-MPS simulations. We observe a pronounced quasiparticle peak at the lowest energies for all momenta $k$, even around $k=0$ where the meson comes closest in energy to the meson-magnon continuum. We show td-MPS results for $t=5 J$ and $h=0.01 J$; the color scale is in a.u..}
    \label{FigAvoidedMagDcytdMPS}
\end{figure}

Before we proceed, we note that the same result is obtained when the meson plus two-magnon continuum is considered. This case becomes relevant if the meson can only couple to pairs of magnons. Making the same considerations as above does not change the outcome, as one can see by placing the second magnon in the $q=0$ state whose energy $\omega_{q=0} \to 0$ as $h \to 0$.

At first glance, the scenario we find appears reminiscent of supersonic polarons emitting Cherenkov phonons \cite{Seetharam2021}; But we completely ignored interactions between mesons and magnons so far, which can destroy the quasiparticle. However, recently it has been shown that sufficiently strong couplings of a quasiparticle to an excitation continuum can, on the contrary, stabilize the quasiparticle band and lead to a complete avoidance of quasiparticle decay \cite{Verresen2018}. This scenario is sketched in Fig.~\ref{FigAvdQPdecay} d). As we show below in Sec.~\ref{SubSecAvdMgnDcy}, our effective theory indicates that meson-magnon couplings in the doped spin chain Eq.~\eqref{model} become strong enough at long wavelengths to cause an avoided quasiparticle decay.

In Fig.~\ref{FigAvoidedMagDcytdMPS} we confirm this prediction by td-MPS simulations of the ARPES spectrum. In fact, all ARPES spectra we considered showed a clearly defined quasiparticle band at low energies, for all momenta. In Fig.~\ref{FigAvoidedMagDcytdMPS} we consider the most extreme regime where $t \gg J \gg h$ and our argument above predicts the meson quasiparticle band to enter the magnon continuum in the absence of interactions. 
Specifically we assumed $t=5 J$ and $h=0.01 J$.\\ 

\section{Strong-coupling theory\\ of doped holes}
\label{secStrongCplngThy}
In this section we discuss the generalized $1/S$ expansion for the 1D $t-J$ model in a staggered field, Eq.~\eqref{model}. It combines a parton theory for individual holes doped into the spin system \cite{Grusdt2018} with linear spin wave theory (LST) above the magnetically ordered spin background. The main achievement of this approach is to combine advantages of both methods: We keep the clear physical picture afforded by the parton theory while including a back-action of partons on their spin-environment. 

Concretely the idea of the method is to include small quantum fluctuations about a classical magnetic state of Ising spins in the lattice. In a key distinction from earlier approaches \cite{Kane1989,Sachdev1989,Martinez1991}, we allow the Ising configuration around which we expand to be displaced by the quantum motion of the doped hole. This method can be formalized by using a generalization of the $1/S$ expansion in the length $S$ of the considered spins \cite{Grusdt2018}. As usual, we send $S \to 1/2$ in the end to obtain predictions for the spin-$1/2$ model in Eq.~\eqref{model}.

\subsection{Generalized 1/S expansion}
We begin the discussion by using a Schwinger boson representation of the (fermionic) $t-J$ model. We introduce a spin-less fermionic chargon $\hd_j$ and a Schwinger boson $\hat{\beta}^\dagger_{j,\sigma}$ and write the original fermion operators as $\cd_{j,\sigma}=\h_j \hat{\beta}^\dagger_{j,\sigma}$ for $S=1/2$, see e.g. \cite{Auerbach1994}. For general values of spin $S$ of the underlying fermions, the physical Hilbert space we consider is realized by states satisfying
\begin{equation}\label{single-occ-constraint}
    \sum_{\sigma} \hat{\beta}^\dagger_{j,\sigma}\hat{\beta}_{j,\sigma}= 2S \l 1- \hd_j \h_j\r. 
\end{equation}
This constraint ensures that there is either one vacancy and no spin, or one spin and no vacancy localized at the specified site $j$. We emphasize that for $S\neq 1/2$ this constraint -- which is at the heart of the generalized $1/S$ expansion -- is different from the condition $\sum_\sigma \hat{\beta}^\dagger_{j,\sigma}\hat{\beta}_{j,\sigma} + \hd_j \h_j = 2S$ more commonly used in the conventional $1/S$ expansions \cite{Kane1989}. Using the new constraint in Eq.~\eqref{single-occ-constraint}  has the advantage to treat spin and charge as mutually exclusive degrees of freedom per lattice site, but leads to a highly non-linear chargon hopping term correlated with the surrounding spins when the $t-J$ model is expressed in terms of $\h_j$ and $\hat{\beta}_{j,\sigma}$.

Our strategy to deal with the complicated constraint in Eq.~\eqref{single-occ-constraint} in the following is two-fold. First, on the level of the Ising part of the Hamiltonian Eq.~\eqref{EqHIsing} we can keep track of the full constraint Eq.~\eqref{single-occ-constraint} due to the classical nature of the Ising spins. Second, we make a Holstein-Primakoff-approximation around the Ising configuration to take into account transverse spin fluctuations from Eq.~\eqref{eqHJperp}.

\subsubsection{Zero Doping}
At zero doping, the ground state of the model in Eq.~\eqref{model} has long-range AFM correlations along the $z$-direction for any non-zero $h$, which breaks the ${\rm SU}(2)$ symmetry of the bare $t-J$ model explicitly. The corresponding low-energy magnon excitations can be described using linear spin-wave theory, which is equivalent to a first order expansion of the model in powers of $1/S$ \cite{Auerbach1994}. To lowest order in $1/S$ one obtains a classical configuration $\Sz_j=\tau_j^z/2$ where the Ising variables $\tau_j^z= \l-1 \r^{j}$ describe a N\'eel configuration.

Up to first order in $1/S$ the linear spin wave theory corresponds to the Holstein-Primakoff (HP) approximation, where the spin operators are represented as
\begin{eqnarray}\label{HPapprox}
    \label{HPapproxSz}
    \Sz_j= \tau_j^z\l S - \ad_j\a_j \r,\\
    \label{HPapproxSpm}
    \hat{S}^{\tau_j^z}= \sqrt{2S}\a_j, \quad \hat{S}^{-\tau_j^z}= \sqrt{2S}\ad_j.
\end{eqnarray}
Note that in principle this expansion can be performed for arbitrary configurations of the classical Ising field $\tau_j^z$. The bosonic operators $\a_j$ are related to the Schwinger bosons by $\hat{\beta}_{j,-\tau^z_j}=\a_j$ and $\hat{\beta}_{j,\tau^z_j} = \sqrt{S}$, i.e. bosons $\hat{\beta}_{j,\sigma}$ with $\sigma=\tau^z_j$ condense and to leading order fluctuations of the condensate fraction are ignored.

Magnon excitations in the undoped AFM are obtained by setting $\tau^z_j=\l -1 \r^{j} $ and inserting Eqs.~\eqref{HPapproxSz} and \eqref{HPapproxSpm}  in the Heisenberg Hamiltonian. This results in the well-known free spin-wave Hamiltonian,
\begin{equation}
    \H_{\text{mag}}^{(0)}= \sum_q \omega_q \bd_q \b_q
\end{equation}
where the sum is taken over lattice momenta $q\in [-\pi,\pi]$.\\
The spin-wave dispersion is given by 
\begin{equation}
    \omega_q = J_z \sqrt{\l 1+ h/J_z \r² - \l J_\perp/J_z\r²\cos²{q} }
\end{equation}
where we allowed for anisotropic interactions $J_z$ ($J_\perp$) along $z$ ($xy$) direction in spin space; for our model in Eq.~\eqref{model}, $J_\perp=J_z=J$. The Bogoliubov operators $\b_q$ are related to the HP bosons $\a_j$ by a Fourier- and Bogoliubov transformation.

\subsubsection{Single Hole Doping - Hilbert Space}
To describe the properties of a single hole doped into a N\'eel state, we apply the generalized $1/S$ expansion outlined above and extrapolate our result to the case $S=1/2$ in the end. In order to include distortions of the N\'eel state by the chargon, we work with the constraint \eqref{single-occ-constraint} and promote the Ising field $\tau^z_{j}$, around which we perform the linear spin-wave expansion later on, to a dynamical field: i.e. $\tau^z_{j}$ depends explicitly on the instantaneous configuration of the spinon, string of displaced spins, and chargon. We will make this precise in the following, by constructing a complete set of low-energy basis states for the partons.

Before we proceed, we mention already that changes in the Ising fields will lead to corresponding changes in the magnon terms resulting from the HP approximation, see Eq.~\eqref{HPapproxSz}. The idea is to include HP bosons on all bonds of the lattice, as done in the zero doping case, and describe how each term in the Hamiltonian leads to parton and / or magnon processes in the effective model.

To leading order in $1/S$ we can ignore magnons completely and only take into account changes in the Ising fields $\tau^z_{j}$ induced by the chargon motion. This is equivalent to solving only the $t-J_z$ part, Eq.~\eqref{EqHIsing}, in our model, which we now do for one doped hole. To this end we construct a set of low-energy basis states with one hole: we start from a N\'eel state, i.e. $\tau^z_{j}=\left(-1\right)^{j}$, and remove a fermion from some lattice site $j^s$. In accordance with the spin representation \eqref{HPapproxSz} and the constraint \eqref{single-occ-constraint}, we enlarge the allowed values of $\tau^z$ and set $\tau^z_{j^s}=0$. Next, we construct all relevant basis states by applying the hopping part $\H_t$ in Eq.~\eqref{model}, still ignoring magnons. In 1D these states can be labeled by the position $j^s$, where the hole has been initially created, and the position $j^h$ reached by the chargon,
\begin{equation}\label{parton-Hilbert-space}
    \{\ket{j^h,j^s}\}, \quad j^h= j^s + \Sigma,\quad \Sigma\in \mathbb{Z}.
\end{equation}

When the chargon moves it displaces all spins along its path $\Sigma$ by one lattice site, which changes the Ising fields $\tau^z_j$ on the corresponding lattice sites. Thereby it distorts the N\'eel pattern. However, since $\Sz_j = S\tau^z_j$ in the absence of magnons, each displacement can be associated with a potential energy cost $\propto h$. Thus, to leading order in $1/S$, the problem is described by a single hole moving in a classical spin background $\tau^z_j$ with AFM Ising interactions in the staggered field $\pm h$. 

The site $j^s$ where the hole has been initially created carries a surplus of spin and corresponds to a domain wall of two nearest neighbor aligned spins in the $t-J_z$ model. The domain wall can be interpreted as the charge-neutral spinon which carries a spin $\sigma$ opposite to the spin of the removed fermion. In addition to the already introduced fermionic chargon operator $\h_j$, we define a new bosonic spinon operator $\s_{j,\sigma}$, which will become useful when mapping the $t-J$ Hamiltonian to the parton basis, Eq.~\eqref{parton-Hilbert-space}. To this end we make the following identification for the parton basis,
\begin{equation}\label{2nd-parton-Hilbert-space}
    \ket{j^h,j^s}\equiv \sd_{j^s,\sigma}\hd_{j^h}\ket{0}.
\end{equation}

As described above, every spinon-chargon configuration given by Eq.~\eqref{2nd-parton-Hilbert-space} is uniquely related to a configuration of the Ising fields $\tau^z_j$: at the position of the chargon, $\tau^z_{j^h}=0$, and along the string $\Sigma$ of displaced spins connecting the chargon and spinon, the Ising fields have a reversed sign as compared to the original N\'eel state, $\tau^z_j= -(-1)^{j}$ for $j=j^s+ \delta \Sigma $ where $\delta \Sigma= 1,...,\Sigma$. The spinon corresponds to the domain wall located at the beginning of the string, formed by two aligned spins.

Since we are working in a subspace with only one spinon and chargon, we can omit the spin index $\sigma$ at the spinon operator. Once the hole is created in the spin chain, the spin of the spinon is specified by the sub-lattice index of the removed fermion, and within our approximation this sub-lattice index of the spinon cannot change.

Before we proceed to construct the effective parton Hamiltonian, we note that our construction above is equivalent to assuming that $\Sigma$ describes a $\mathbb{Z}_2$ electric string connecting a pair of $\mathbb{Z}_2$-charged spinon and chargon in a $\mathbb{Z}_2$ LGT; see Refs.~\cite{Grusdt2020PRL,Kebric2021} for a detailed discussion of this general mapping. 

\subsubsection{Parton Hamiltonian}
Mapping our model \eqref{model} to the parton basis in Eq.~\eqref{parton-Hilbert-space} and introducing HP magnons yields an effective Hamiltonian of the form
\begin{equation}\label{partonHam}
    \H= -L \varepsilon_0 +\H_{\rm h}+\H_{\rm s}+\H_{\rm sh}+\H_{\text{mag}},
\end{equation}
where $\varepsilon_0= S^2 J_z+Sh $ denotes the classical Néel ground state energy per lattice site; the total number of sites in the system is denoted by $L$. 

The next two terms describe free chargon and spinon terms, namely
\begin{equation}\label{freechargonH}
    \H_{\rm h}= \varepsilon_0^{\rm h}\sum_{j}\hd_j\h_j+ t \sum_{j} \l\hd_{j+1}\h_j+\hc \r
\end{equation}
and
\begin{equation}\label{freespinonH}
    \H_{\rm s}= \varepsilon_0^{\rm s}\sum_{j}\sd_j\s_j+ \frac{J_\perp}{2} \sum_{j} \l \sd_{j+2}\s_j+\hc\r.
\end{equation}
Here, $\varepsilon_0^{\rm h}= 2S^2 J_z+Sh$ and $\varepsilon_0^{\rm s}= S^2 J_z$ are the rest energies of the chargon and spinon in the $t-J_z$ model with the staggered field, respectively. We emphasize that we use second-quantized operators for the spinon and the chargon for notational convenience, keeping in mind that our derivation is valid for a single spinon-chargon pair.

An important caveat is that the spinon tunneling term, second term in Eq.~\eqref{freespinonH}, is only valid for the case $S=1/2$. It appears to first order in $1/S$ and its derivation is given in Appendix \ref{ApdxSpnDyn}.  The spinon dynamics results similarly as in the case of the 1D $t-J$ model without an external magnetic field where genuine spin-charge separation occurs \cite{Giamarchi2004}; in the present case the flip-flop terms $\propto J_\perp \Sp_{j+1}\Sm_j$ acting on bonds adjacent to the domain wall let the spinon move by two lattice sites. In our formalism, the action of these terms on other bonds away from the spinon creates HP bosons, as will be shown below. After this subsection we will consider the limit $S=1/2$, as the spinon tunneling is only valid in this case. 

Next, the term $\H_{\rm sh}$ describes spinon-chargon interactions and consists of two terms specified below: a density-density interaction and a kinetic interaction,
\begin{equation}\label{spinon-chargon-int}
    \H_{\rm sh}=\hat{\mathcal{H}}_{\rm sh}^{\rm dd}+\hat{\mathcal{H}}_{ \rm sh}^{\text{kin}}.
\end{equation}
The first spinon-chargon interaction describes the linear confining potential stemming from the string of displaced spins in the $t-J_z$ Hamiltonian \eqref{EqHIsing},
\begin{equation}
    \hat{\mathcal{H}}_{\rm sh}^{dd}= \sum_{i,j} ~ \hd_i \h_i ~ \sd_j \s_j ~ V_{\rm sh}(|i-j|)
\end{equation}
with
\begin{equation}\label{linear-pot}
    V_{\rm sh}(\ell)= |h| \ell -S^2 J_z \delta_{\ell,0}, \quad \ell\ge0.
\end{equation}
The first term in \eqref{linear-pot} accounts for the spins residing on the energetically unfavorable sub-lattice along the string $\Sigma$ connecting the spinon and chargon. This leads to a string tension $\propto |h|$ which grows linearly with the length of the string $\ell=|\Sigma|$. The second term in Eq.~\eqref{linear-pot} describes a point-like attraction between the spinon and chargon. To understand its origin, note that the rest energies for the partons are different when they occupy the same lattice site compared to being on different lattice sites. The difference between these rest energies is accounted for by the point-like parton attraction.

The kinetic spinon-chargon term in Eq.~\eqref{spinon-chargon-int} describes how the spinon dynamics is constrained by the chargon. It only appears for the case $S=1/2$ as the spinon tunneling term,
\begin{equation}
    \hat{\mathcal{H}}_{\rm sh}^{\rm kin}= -\dfrac{J_\perp}{2} \sum_{j} \hd_{j+1}\h_{j+1}\left(\sd_{j+2}\s_j+\hc\right).
    \label{eqHshKin}
\end{equation}
This term describes how the presence of a chargon disables the flip-flop term $J_\perp\Sp\Sm$ across it, which was originally assumed to give rise to spinon dynamics in Eq.~\eqref{freespinonH}. The minus sign in Eq.~\eqref{eqHshKin} subtracts this contribution.

The last term in the parton Hamiltonian \eqref{partonHam} summarizes all magnon contributions,
\begin{equation}\label{mag-contributions}
    \hat{\mathcal{H}}_{\text{mag}}= \sum_q \omega_q\l \bd_q\b_q+1/2\r + \hat{\mathcal{H}}_{\text{mag}}^{\text{int}}
\end{equation}
where the first term results from diagonalization of our model \eqref{model} in the case of zero doping. $\H_{\text{mag}}^{\text{int}}$ describes magnon-parton interactions -- to be specified below --  which involve HP magnons to quadratic order and have to be included due to the effects of the partons on the underlying N\'eel state. The full expressions are derived in detail in Appendix \ref{ApdxMagnonCont}.

\subsection{Strong coupling approximation}
A full solution of the spinon-chargon problem in the presence of magnons is not possible. To simplify our analytical formalism further, we now introduce a strong coupling (SC) theory of the meson formed by the spinon and the chargon. The meson, in turn, interacts with the bath of low-energy magnon excitations. We will demonstrate that magnons lead to polaronic dressing of the meson, or even to the formation of a meson-magnon bound state. The strong-coupling approach is valid for $t\gg J,h$.
It is based on the separation of time scales between the chargon motion and the spin degrees of freedom, i.e. the spinon dynamics and the magnon creation and annihilation processes. Namely, the chargon dynamics takes place on a much shorter time scale in this limit.

\subsubsection{Born-Oppenheimer Approximation - Meson Operator}
At strong couplings, $t\gg J,h$, the fast chargon can adiabatically follow the slow spinon dynamics and magnon creation and annihilation processes. Thus, the chargon motion can be treated first while fixing the spinon position. Moreover, we assume that the meson state is weakly affected by magnons and neglect spin fluctuations at first. This allows us to solve the parton part of the Hamiltonian, \eqref{partonHam}, i.e. excluding the magnon contribution $\H_{\rm mag}$, by making a Born-Oppenheimer ansatz for the meson state,
\begin{equation}\label{Born-Oppenheimer-ansatz}
    \ket{\psi_{\rm sh}^{(n,\xi)}(k)} = \dfrac{1}{\sqrt{L}} \sum_{j^s} e^{-i k j^s} \sd_{j^s}\ket{0} \otimes \ket{\psi_h^{(n\xi)}(j^s)}
\end{equation}

This wavefunction corresponds to plane waves where $k$ denotes the total momentum of the meson, which is carried by the heavy spinon. For a fixed spinon position $j^s$, the chargon wavefunction is given by $\ket{\psi_h^{(n\xi)}(j^s)}$. It only depends on the distance from the chargon to the spinon, and can be characterized by two quantum numbers $n \in \mathbb{Z}_{>0}$ and $\xi=\pm 1$. They correspond to vibrational and rotational states of the chargon. The derivation of the meson energies and states can be found in Appendix \ref{ApdxMesonStatesSC}.

The ansatz in Eq.~\eqref{Born-Oppenheimer-ansatz} leads to a meson dispersion of the form 
\begin{equation}\label{meson-dispersion}
    \varepsilon_k^{(n\xi)} = E_h^{(n\xi)} + J_\perp^{(n\xi)} \cos(2k)
\end{equation}
which describes a tight-binding dispersion for the meson with a two-site tunneling term.
Here, the meson hopping amplitude $J_\perp^{(n\xi)}$ is related to the spin exchange coupling $J_\perp$ by a Franck-Condon factor,
\begin{equation}
    \label{FCoverlap}
    J_\perp^{(n\xi)}/J_\perp = \bra{\psi_h^{(n\xi)}(j^s+2)} 1- \hd_{j+1}\h_j \ket{\psi_h^{(n\xi)}(j^s)}.
\end{equation}
The energy offset $E_h^{(n\xi)}$ corresponds to the chargon eigenenergy characterized by the quantum numbers $n$ and $\xi$. As shown in Appendix \ref{ApdxMesonStatesSC}, for $t \gg h$ it scales as,
\begin{equation}
    E_h^{(n\xi)} \approx E_0 -2t + a_{\rm sh}^{(n\xi)} t^{1/3}h^{2/3} + \mathcal{O}(h),
\end{equation}
with numerical coefficients $a_{\rm sh}^{(n\xi)}$ related to the Airy-function \cite{Bulaevski1968}.

The center-of-mass momentum of the mesonic bound state is carried by the heavy spinon and the binding of the spinon to the fluctuating chargon leads to a renormalization of the hopping amplitude, $J^{(n\xi)}\le J$, see Fig.~\ref{Fig2}. Because the Franck-Condon factor $J^{(n\xi)}/J$ is independent of $k$ at strong couplings, the shape of the meson dispersion is identical to that of the spinon up to an overall rescaling.

Formally, we can define meson operators $\fd _{j^s,n\xi}$ to describe the bound state,
\begin{equation}
\label{mesonop}
    \fd _{j^s,n\xi} \ket{0} = \sd_{j^s}\ket{0}_s \otimes \ket{\psi_h^{(n\xi)}(j^s)}.
\end{equation}
This will help us to describe interactions between the meson and the bath of magnons in the next step. Note that in this description, the chargon quantum numbers $(n,\xi)$ take the role of band indices of the meson.

\begin{table*}[t]
\renewcommand{\arraystretch}{1.5}
    \centering
    \begin{tabular}{p{2.2cm}||p{5.5cm}|p{7.5cm}||p{1.2cm}}
        Theory & free variables & alternative representation & $\H$ \\
        \hline \hline
        $t-J$ model & $\c_{j,\sigma}$ (spin-$1/2$ fermions) & $\c_{j,\sigma} =  \hat{\beta}_{j,\sigma} \hd_{j}$ (Schwinger boson, slave fermion) & Eq.~\eqref{model}\\
        \hline
        parton theory &  $\hat{\tau}^z_j$ (Ising variables), $\a_j$ (HP magnon) & $\hat{s}_{j,\sigma}$ (spinon), $\h_j$ (chargon), $\b_q$ (Bogoliubov magnon)& Eq.~\eqref{partonHam} \\
        \hline
        polaron theory & $\f_{k,n\xi}$ meson, $\b_q$ (Bogoliubov magnon) & $\qquad \qquad \qquad$ -- & Eq.~\eqref{PolHam}
    \end{tabular}
    \caption{Overview of the different levels of approximation used in our semi-analytical theory, as described in the text.}
    \label{tabOverview}
\end{table*}

\subsubsection{Polaron Hamiltonian}
\label{SecPolHam}
Next we include magnon processes. As in the case of spinon dynamics, the time scales associated with magnon processes also correspond to longer times compared to the chargon motion. It is thus legitimate to extend the SC ansatz to the magnon contributions and treat magnon terms on a mean-field level. We assume that the meson is formed even in the presence of magnons, and neglect any back-action of the latter on the underlying SC meson wavefunction, such as a possible weak renormalization of the linear string tension.

Within the SC theory, the effective magnon Hamiltonian is obtained by averaging the parton Hamiltonian over the SC spinon-chargon wavefunction Eq.~\eqref{Born-Oppenheimer-ansatz}. Working in second quantized notation with the meson operators $\fd_{j^s,n\xi}$ we obtain:
\begin{equation}
    \label{PolHamProjection}
    \hat{\mathcal{H}}_{\rm pol}= \sum_{\substack{j^s,n\xi \\ j^{s\prime},n^\prime\xi^\prime}} \bra{\psi_{sh}^{(n^\prime\xi^\prime)}(j^{s\prime})}{\hat{\mathcal{H}}}\ket{\psi_{sh}^{(n\xi)}(j^s)} \fd_{j^{s\prime},n^\prime\xi^\prime}\f_{j^s,n\xi}.
\end{equation}
The resulting effective polaron Hamiltonian can be decomposed into different contributions,
\begin{equation}\label{PolHam}
    \hat{\mathcal{H}}_{\text{pol}}= \hat{\mathcal{H}}_{\text{mes}}^{(0)}+\hat{\mathcal{H}}_{\rm mag}^{(0)}+\hat{\mathcal{H}}_{\rm pol}^{\rm int}.
\end{equation}

The free meson and magnon Hamiltonians are
\begin{subequations}
\begin{eqnarray}
    \hat{\mathcal{H}}_{\text{mes}}^{(0)}&=& \sum_{k,n\xi}\varepsilon^{(n\xi)}_{k, \rm eff}\fd_{k,n\xi}\f_{k,n\xi}, \label{eqHmesFree}\\
    \hat{\mathcal{H}}_{\text{mag}}^{(0)}&=&\sum_q \omega_q \bd_q\b_q.
\end{eqnarray}
 \end{subequations}
Note that the free meson dispersion gets weakly renormalized by zero-point contributions of the magnons resulting from the Bogoliubov transformation. However, the analytic form of Eq.~\eqref{meson-dispersion} remains unchanged: $\epsilon^{(n\xi)}_{k} \to \epsilon^{(n\xi)}_{k, \rm eff} = E_{h,\rm eff}^{(n\xi)}+ J_{\perp, \rm eff}^{(n\xi)} \cos(2k)$. We do not include  full expressions for the small corrections here but refer the reader to Appendix \ref{ApdxMesMagInt} for details.

The meson-magnon interactions in Eq.~\eqref{PolHam} take the compact form
\begin{eqnarray}
    \label{PolInt}
    \nonumber\hat{\mathcal{H}}_{\rm pol}^{\rm int}=- \dfrac{1}{2L}\sum_{n\xi}\sum_{k,pq} \left[ \mathcal{V}^{(n\xi)}_{k,pq} ~ \fd_{k-p+q,n\xi}\f_{k,n\xi}~ \bd_{p}\b_q \right.\\
    \left. + \l \mathcal{W}^{(n\xi)}_{k,pq}~ \fd_{k-p-q,n\xi}\f_{k,n\xi}~ \bd_{p}\bd_q +\hc \r \right],
\end{eqnarray}
where we ignore band-changing collisions, i.e. we only consider terms diagonal in $(n,\xi)$. This is justified in the SC limit by the large separation of energy scales.
The Bogoliubov operators $\b_q$ are related to the HP bosons $\a_j$ by a Fourier- and Bogoliubov transformation where the last transformation diagonalizes the free magnon terms. This leads to explicit expressions for the two couplings $\mathcal{V}^{(n\xi)}_{k,pq}$ and $\mathcal{W}^{(n\xi)}_{k,pq}$ describing normal and anomalous magnon terms, as derived in Appendix \ref{ApdxMesMagInt}. 

This effective strong coupling Hamiltonian describes a polaron model of the mesonic impurity coupled to the bath of quantum spin fluctuations, represented by collective magnon excitations. 

\subsection{Overview of the approach}
Above we described how the original $t-J$ model with a staggered field, Eq.~\eqref{model}, first leads to an effective parton model which subsequently maps to the simplified polaron Hamiltonian in Eq.~\eqref{PolHam}. In Tab.~\ref{tabOverview} we provide a summary of the involved fields and Hamiltonians appearing in the various stages of approximations. Next we will solve the effective polaron theory.

\section{Solutions of the strong-\\coupling theory}
\label{secSolSCthy}
In this section we apply and compare different methods to solve the effective polaron Hamiltonian from Eq.~\eqref{PolHam}. These are based on known analytical polaron techniques which have been successfully used to study Bose polaron problems in the past\cite{Chevy2010,Devreese2020,Grusdt2015Varenna,Rath2013,Shi2018}.

\subsection{LLP + Gaussian approach: Meson-magnon binding}
The above effective meson-magnon theory does not couple states where the total momentum of the meson plus magnon excitations is changed. Thus the total momentum is a conserved quantity which reflects the underlying translational invariance of the system. Below we make this conservation of the total momentum explicit by applying a Lee-Low-Pines (LLP) transformation \cite{Lee1953} to the Hamiltonian \eqref{PolHam}. To solve the resulting Hamiltonian $\tilde{\mathcal{H}} = \Ud_{\text{LLP}}\hat{\mathcal{H}}\U_\text{LLP}$ in the LLP frame, we simplify it further by expanding to quadratic order in magnons. This allows us to solve it explicitly using multi-mode Gaussian states of magnons.

\begin{figure*}[t]
    \centering
    \includegraphics[width=19cm]{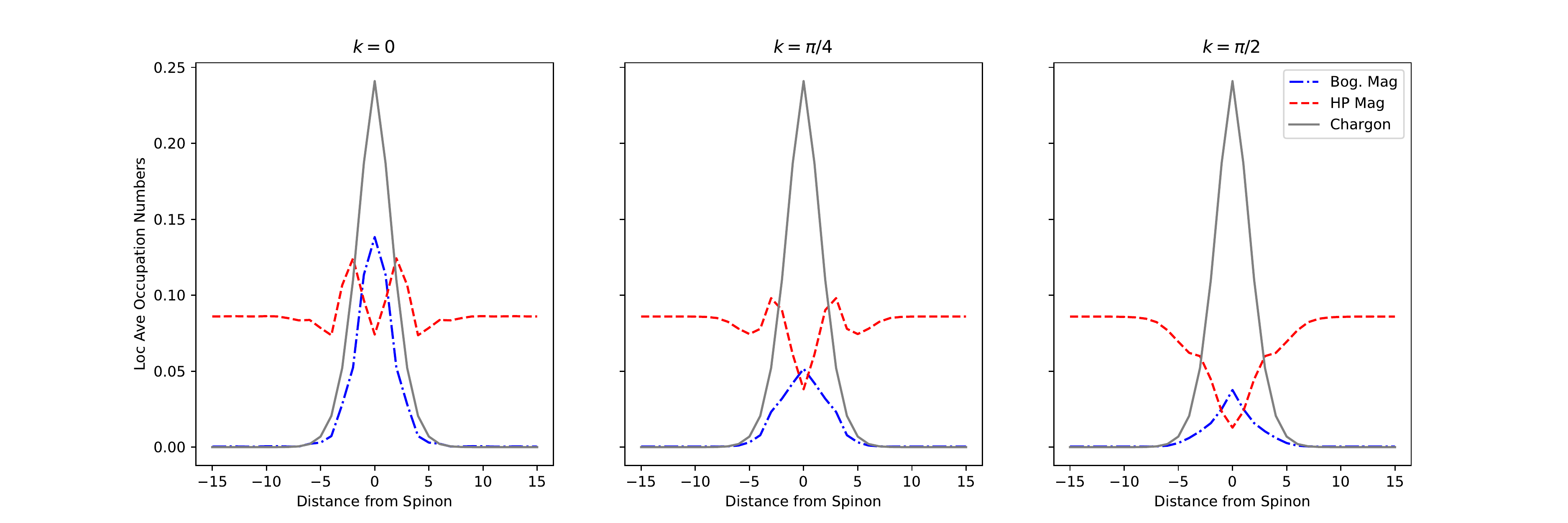}
    \caption{Average densities of bare HP bosons $\langle \ad_j \a_j \rangle$ (red), Bogoliubov bosons  $\langle \bd_j \b_j \rangle$ (magnons, blue) and the chargon $\langle \hd_j \h_j \rangle$ (gray) for total momenta $k=0$, $k=\pi/4$ and $k=\pi/2$. The staggered magnetic field is chosen as $h=0.45J$ and the chargon tunneling as $t=5J$. At this particular magnetic field value our theoretical treatment shows that at a total momentum $k=0$ (left panel) the tetra-parton bound state of the meson with a magnon comes close in energy to the dressed meson state, while for $k=\pi/2$ (right panel) they are further apart form each other, see Fig.~\ref{Fig2}. The formation of the bound state is also visible here in the quasi-particle densities: While in the right panel the HP boson density is strongly suppressed around the spinon, the left panel shows that local spin flips start to accumulate around the meson.}
    \label{Densities}
\end{figure*}

\subsubsection{LLP Transformation}
The LLP transformation shifts the entire magnon state into the frame co-moving with the center-of-mass of the impurity -- in our case the mesonic bound state. It is represented by the unitary transformation,
\begin{equation}
\U_\text{LLP}= e^{-i \hat{X}_\text{mes}\hat{Q}_{b}},
\end{equation}
where we introduced the meson position operator, $\hat{X}_{\text{mes}}=\sum_{j^s,n\xi}j^s \fd_{j^s,n\xi}\f_{j^s,n\xi}$ and the total magnon momentum operator, $\hat{Q}_{b}= \sum_q q\bd_q\b_q$.

Now we apply the unitary LLP transformation to the effective polaron Hamiltonian \eqref{PolHam}. This is established by determining how the meson and magnon operators transform, namely
\begin{eqnarray}
    \Ud_\text{LLP} \fd_{k,n\xi}\U_\text{LLP}= \fd_{k + \hat{Q}_b,n\xi} \\   \Ud_\text{LLP}\bd_q\U_\text{LLP}= e^{iq\hat{X}_{\text{mes}}}\bd_q.
\end{eqnarray}
Insertion of these relations into $\Ud_{\text{LLP}}\hat{\mathcal{H}}\U_\text{LLP}=\tilde{\mathcal{H}}$ yields a Hamiltonian which is block-diagonal in the total system momentum $K$. Thus we get a Hamiltonian of the form,
\begin{equation}
    \tilde{\mathcal{H}}= \sum_{K,n\xi}\fd_{K,n\xi}\f_{K,n\xi}\otimes \hat{\mathcal{H}}_a (K),
\end{equation}
where the term $\hat{\mathcal{H}}_a (K)$ depends only on magnon operators $\bd_j$. I.e. we eliminated the impurity degree of freedom from the problem. However, the transformed magnon Hamiltonian $\hat{\mathcal{H}}_a (K)$ now includes non-linearities in the magnon operators.

Specifically, the LLP transformation leads to a shift of the meson momenta by the total magnon momentum operator. By applying the LLP transformation, the following replacement occurs in the free meson term Eq.~\eqref{eqHmesFree}, 
\begin{equation}
    \Ud_\text{LLP} \H_{\rm mes}^{(0)} \U_\text{LLP}= \sum_{K,n\xi} J_{\perp,\rm eff}^{(n\xi)} \cos(2K-2\hat{Q}_b) \fd_{K,n\xi} \f_{K,n\xi},
\end{equation}
i.e. effectively $\cos(2k)\to \cos(2K-2\hat{Q}_b)$. Expressing the cosine in terms of exponentials, we observe strong magnon non-linearities corresponding to the factors of the form $e^{-i2\hat{Q}_b}$ in the Hamiltonian.

\subsubsection{Linearization and Gaussian states}
To deal with the non-linearities encountered during the LLP transformation, we use the approximation
\begin{equation}
    e^{-i2\hat{Q}_{b}}\approx \openone + \sum_{q}\left(e^{-i2q}-1\right)\bd_q\b_q.
    \label{eqLinearization}
\end{equation}
This equation holds exactly within a subspace of no more than one magnon excitation. Hence it is similar in spirit to the HP approximation we made earlier, which also restricted us to consider low magnon densities only. Both approximations rely on the assumption that meson-magnon interactions are sufficiently weak to work at low excitation densities.

The resulting LLP Hamiltonian $\tilde{\mathcal{H}}$ following this truncation yields a quadratic Hamiltonian in the bosonic magnon operators $\bd_q$, making it exactly solvable via a multi-mode Bogoliubov transformation \cite{Xiao2009}. Within the LLP frame, our approach is Gaussian, in the sense that all non-Gaussian contributions are ignored. Nevertheless, we emphasize that the overall meson-magnon wavefunction includes non-Gaussian correlations, since we applied the non-Gaussian LLP transformation first. 

Our ansatz is similar in spirit to the more general class of non-Gaussian variational states discussed in Ref.~\cite{Shi2020}. Indeed, a more sophisticated approach would be to avoid the linearization in Eq.~\eqref{eqLinearization} and fully solve the variational problem in the LLP frame, $\langle \tilde{\mathcal{H}}(K)\rangle \stackrel{!}{=} {\rm min}$, in the class of Gaussian states. Here we chose the simpler linearization method because it admits more direct analytical insights and immediately yields a full magnon excitation spectrum, but extensions beyond this simple limit constitute an interesting future direction.

In the following we solve the meson-magnon polaron problem as described, and discuss the properties of our solution in more detail.

\subsubsection{Magnon Distribution}
To gain better understanding of how magnons lead to polaronic dressing of the meson, we first plot the average number of bare HP bosons $n_a(j) = \langle \ad_j \a_j \rangle$ and Bogoliubov bosons $n_b(j) = \langle \bd_j \b_j \rangle$ in Fig. \ref{Densities}. Since we work in the LLP frame $j$ corresponds to the distance to the spinon, which is located at the core of the meson and thus at the origin of the LLP frame. Additionally, we plot the local distribution of the chargon in its ro-vibrational ground state, which indicates the extension of the mesonic bound state and the chargon cloud around the spinon. Far away from the meson, the number of HP bosons approaches a constant value (vacuum fluctuations) which is a consequence of the spin-flip terms $\Sp_{j+1} \Sm_j$. The asymptotic value of $n_a(|j|\gg 1)$ can be calculated straightforwardly from linear spin-wave theory.

In the ground state at $k=\pi/2$, Fig.~\ref{Densities} right, we observe a suppression of the number of HP bosons, i.e. fewer local spin flips, in the vicinity of the spinon. The suppression of $n_a(j)$ close to the spinon is a direct consequence of the formation of the geometric string. In the region where the chargon delocalizes, i.e. the spinon-chargon string fluctuates, the magnetization of the spin background is reduced and therefore we expect suppressed quantum fluctuations around the spinon. Note that this suppression of magnon fluctuations is dictated by the chargon distribution.

The substantial \emph{reduction} of quantum fluctuations around the spinon also provides an a-posteriori justification for the use of the lowest order Holstein-Primakoff approximation in Eqs.~\eqref{HPapproxSz}, \eqref{HPapproxSpm}. This is in contrast to the conventional $1/S$-expansion\cite{Kane1989,Sachdev1989,SchmittRink1988}, where non-linear terms in the magnons must be included to prevent excessive densities of excitations at strong couplings. Indeed, in the conventional $1/S$-expansion where magnons are defined relative to the undoped N\'eel state, a local \emph{enhancement} of bare HP magnon fluctuations is obtained around the mobile hole.

\subsubsection{Ground State Energy}
\begin{figure}
    \centering
    \includegraphics[width=0.42\textwidth]{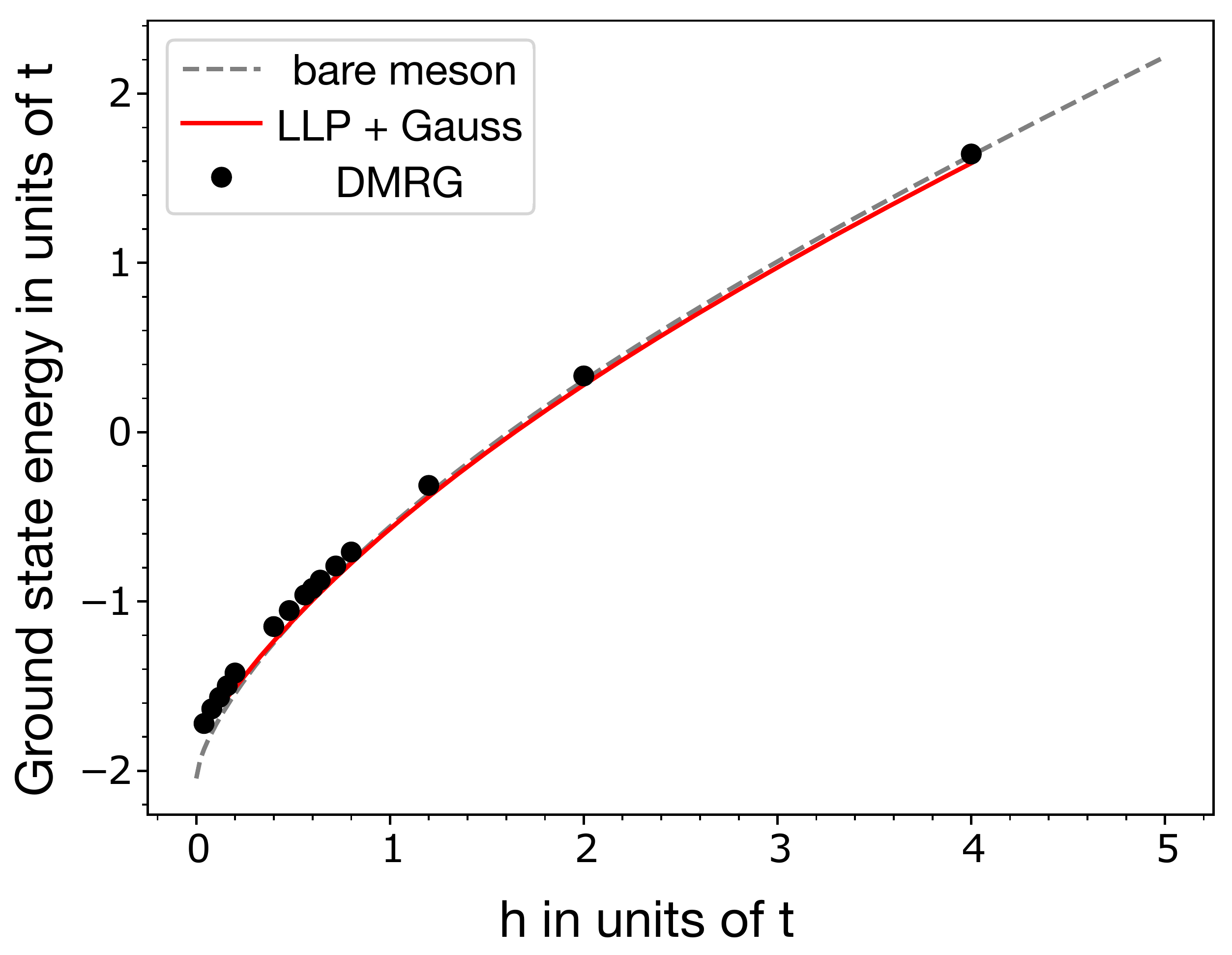}
    \caption{Ground state energy in our Gaussian LLP approach (solid red) compared to the numerical DMRG data (symbols). 
    We also provide the ground state (G.S., dashed line) energy for the bare meson without including magnon couplings, to show the influence of spin fluctuations on the energy. The theory shows good agreement to the numerical data. We used  parameters $J_\perp=J_z=J=1$ and assumed that the meson carries a total momentum of $k=\pi/2$ in our calculations.}
    \label{GS}
\end{figure}

To acquire a better knowledge of how magnon excitations influence the dressed meson we show in Fig. \ref{GS} the ground state energy of one hole for various values of the staggered magnetic field $h/t$. We compare the ground state energy computed from our SC theory to direct numerical DMRG calculations of the ground state energy. The results show good agreement between our SC theory and the numerical data for all chosen magnetic field values. Note that we also display the ground state energy for the bare meson - neglecting magnon excitations - which follows from a pure parton theory for the doped hole. It shows good agreement to the numerics, especially for large values of $h$ where magnon dressing only leads to a weak renormalization of the meson dispersion, see Fig.~\ref{Fig2}. Thus the bare parton theory for the hole already provides a good description of the ground state properties of the meson.

\subsubsection{Polaron Spectrum}
\label{SecPolSpec}
As already elaborated, we have simplified the effective meson-magnon Hamiltonian~\eqref{PolHam} to a quadratic Hamiltonian in the Bogoliubov operators $\bd_q, \b_p$,
\begin{eqnarray}
    \label{H_aK}
    \nonumber \H_a (K) = E_{h,\rm eff}^{(n\xi)} + J_{\perp, \eff}^{(n\xi)}\cos(2K)+ \sum_q \Omega^{(n\xi)}_{K,q} \bd_q \b_q\\
    - \dfrac{1}{2L} \sum_{pq} \left[\mathcal{V}^{(n\xi)}_{K,pq} \bd_p\b_q + \l \mathcal{W}^{(n\xi)}_{K,pq} \bd_p\bd_q + \hc\r  \right]
\end{eqnarray}
by transforming the system into the LLP-frame and linearization -- corresponding to the frame co-moving with the meson -- which eliminates the meson degree of freedom. The total system momentum, $K$, is conserved in this frame and thus labels different total momentum blocks of the Hamiltonian. 
The linearization~\eqref{eqLinearization} results in a renormalized free magnon dispersion at fixed total momentum $K$:
\begin{equation}
    \Omega_{K,q}^{(n\xi)} = \omega_q + J_{\perp,\eff}^{(n\xi)} \l \cos(2K-2q) - \cos(2K) \r. 
\end{equation}

To solve the quadratic Hamiltonian in Eq.~\eqref{H_aK}, we use a  multi-mode Bogoliubov transformation\cite{Xiao2009}. In the following we will sketch the idea and mention peculiarities arising with this multi-mode technique. It starts by introducing transformed Bogoliubov operators,
\begin{eqnarray}
    \label{MultiModeTrafo}
    \d_l = \sum_q \left[U_{l,q}\b_q -V_{l,q}\bd_q\right],\\
    \dd_l = \sum_q \left[U_{l,q}\bd_q - V_{l,q}\b_q \right],
\end{eqnarray}
where $U,V$ are real $L\times L$ matrices; $L$ is the number of lattice sites which equals the number of momentum modes. 

Next one searches for $U$ and $V$ such that the following equation, 
\begin{equation}
    \comm{\H_a (K)}{\dd_l} = w_l(K) \dd_l,
    \label{eqHaKDiagonalization}
\end{equation}
is fulfilled. Up to a constant energy shift, this ensures that the Hamiltonian takes the diagonal form $\H_d(K)= \sum_l w_l(K) \dd_l \d_l + {\rm const.}$  with dispersion $w_l(K)$. Eq.~\eqref{eqHaKDiagonalization} leads us to an eigenvalue equation of the form
\begin{equation}
    \label{MultiEig}
    \begin{pmatrix} A & B \\ -B & -A \end{pmatrix}
    \begin{pmatrix} U_l\\ V_l\end{pmatrix}
    = w_l(K) \begin{pmatrix} U_l\\ V_l \end{pmatrix}.
\end{equation}
$U_l$($V_l$) denotes the $l$-th column of the matrix $U$($V$) and $w_l(k)$ then correspond to the eigenenergies of the resulting matrix in the left hand site of Eq. \eqref{MultiEig}. Note that this matrix is not Hermitian and thus may lead to complex eigenvalues $w_l(K)$. If this is the case the Hamiltonian \eqref{H_aK} is not diagonalizable\cite{Xiao2009}.

The matrices $A$ and $B$ have the components 
\begin{eqnarray*}
A_{pq} &=& \Omega_{K,q}^{(n\xi)} \delta_{pq} -\dfrac{1}{2L} \mathcal{V}^{(n\xi)}_{K,pq}\\
B_{pq} &=& -\dfrac{1}{L} \mathcal{W}^{(n\xi)}_{K,pq}.\\
\end{eqnarray*}
The eigenvalue equation \eqref{MultiEig} is solved numerically by exact diagonalization. Finally, inserting the transformation \eqref{MultiModeTrafo} and rearranging terms such that we can use the eigenvalue equation \eqref{MultiEig}, brings us to the diagonalized form of the Hamiltonian\cite{Xiao2009} \eqref{H_aK}, 
\begin{multline}
    \H_a(K) = E_{h,\rm eff}^{(n\xi)} + J_{\perp, \eff}^{(n\xi)}\cos(2K) \\+ \sum_l w_l(K) \dd_l\d_l + \dfrac{1}{2} \l \sum_l w_l(K) - \tr(A)\r.
\end{multline}
The second term in the second line corresponds to the zero-point energy of the $\dd$ bosons. 

\begin{figure}
    \centering
    \includegraphics[width=0.47\textwidth]{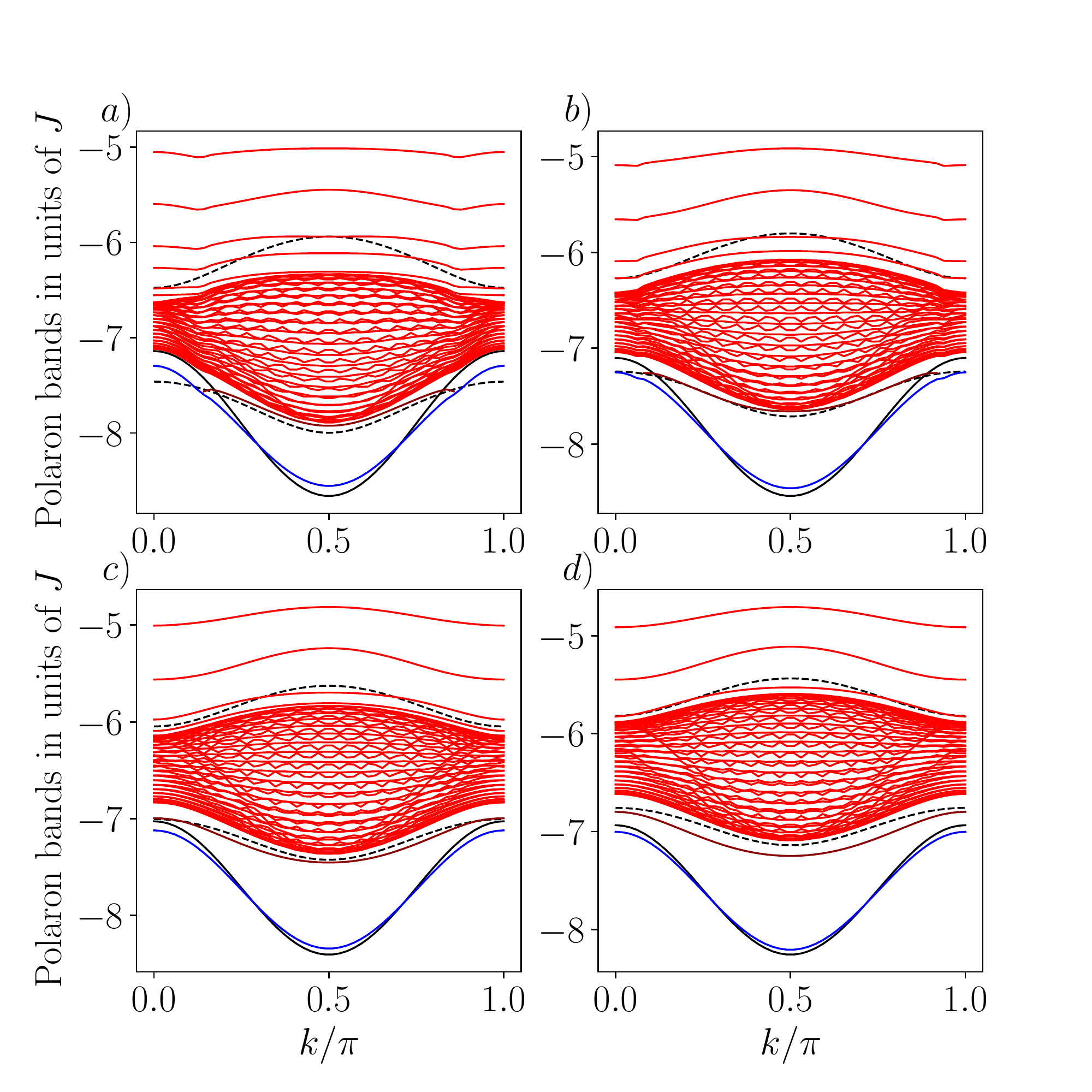}
    \caption{Magnetic polaron bands plotted for various magnetic field values with the meson in its ro-vibrational ground state, $n=1,\xi=+1$. The chosen field values are a) $h=0.2J$, b) $h=0.3J$, c) $h=0.4J$ and d) $h=0.5J$. The black solid curve shows the bare meson band without magnon influence and the black dashed lines show the lower and upper edges of the non-interacting meson-magnon scattering continuum; blue lines correspond to the ground state magnetic polaron band; the dark red line corresponds to a meson-magnon bound state. We used system sizes of $L=51$ in our exact diagonalization of the effective polaron Hamiltonian.}
    \label{4PolBands}
\end{figure}

We can further simplify the Hamiltonian and bring it to the form
\begin{multline}
    \label{PolHLLP}
    \H_a(K) =  E_{h}^{(n\xi)} + \Gamma^{(n\xi)} + J_\perp^{(n\xi)} \cos(2K) \\ + \Delta^{(n\xi)}(K) + \sum_l w_l(K) \dd_l\d_l.
\end{multline}
$\Gamma^{(n\xi)}$ is the sum of all contributions coming from the vacuum fluctuations, not depending on the total momentum $K$. In addition we introduced $\Delta^{(n\xi)}(K)$ which sums up all the remaining $K$-dependent terms. 
The first four terms of the polaron Hamiltonian~\eqref{PolHLLP} describe the polaron ground state band, i.e. the meson dressed by virtual magnon excitations, see the blue curves in Figs. \ref{Fig2} and \ref{4PolBands}. Each higher band (red) in the mentioned figures corresponds to the creation of one multi-mode boson $\dd_l$ with a specific $l$ value. In order to get a better understanding of the resulting polaron spectrum we included curves for the bare meson, $E^{(n\xi)}_h + J_\perp^{(n\xi)}\cos(2K)$ (black solid curve) and the lower and upper edges (black dashed curves) of the non-interacting meson-magnon continuum, defined by $E_h^{n\xi}+J_\perp^{(n\xi)}\cos(2P) + \omega_{K-P}$ with $P=\pi/2$ and $P=0$, respectively. 

From inspection of Fig.~\ref{4PolBands} we see that the bandwidth of the meson-magnon continuum (red region) is reduced compared the non-interacting case, due to the interaction with the mesonic impurity, for the shown magnetic field values $h$. We note that for large $h\gg J_z$ the interacting continuum agrees with the non-interacting one. We also notice that the meson bandwidth gets smaller when lowering the magnetic field $h$, while for large $h\gg J_z$ it is equal to the one of the bare meson. Thus for growing magnetic fields $h$ the renormalization of the meson band and the meson-magnon continuum becomes less pronounced. The reason is that the energy gap for creating magnon excitations is growing with $h$. From the polaron spectrum in Fig.~\ref{4PolBands} it is clear that the multi-mode boson operators $\dd_l$ at some $l$ resemble a magnon excitation $\bd_{K-q}$ with the meson carrying momentum $K$.

\subsubsection{Meson Magnon Bound State}
\label{SubSecBoundState}
Our theory contains further structure beyond a mere dressing of the meson. We predict a stable bound state of the meson with one local spin flip excitation, i.e. one HP boson. In Fig.~\ref{4PolBands} we see that one magnon band emerges below the meson-magnon scattering continuum but above the polaron ground state band for the chosen values of the staggered magnetic field $h$ (dark red line in Fig.~\ref{4PolBands}). We interpret this isolated band as a meson-magnon bound state as it lies energetically below the scattering continuum. Even for very large values of $h$ (not shown in Fig.~\ref{4PolBands}) we predict the existence of this stable bound state, which is further shown by an analytical calculation in appendix \ref{ApdxBoundStateLargeh}. 

Another hint supporting the claim that a HP magnon may bind to the meson was already found in the density plot in Fig.~\ref{Densities}: In the region where the bound state band crosses the polaron ground state band, we see that local spin flips start to accumulate around the meson. This can be treated more formally by making a simplified variational Chevy-type ansatz, as we show explicitly in Sec.~\ref{SubSecChevy1} below.

On several occasions in this work we referred to the bound state as a tetra-parton bound state. The reasoning is that the spin-$1$ magnon excitation can be interpreted as a confined state of two spinons, developing an internal structure itself in the weak-field limit $h \to 0$ \cite{Piazza2015}. Indeed, an interesting future extension for the small-$h$ limit would be to construct an effective spinon-magnon model describing how the magnon resonates in and out of the spinon continuum \cite{Shao2017}, and adding couplings to the chargons as done in the present work. We expect this could improve results for the magnon dispersion, which is poorly represented by our simple linear-spin wave theory when $h/J \to 0$: we checked numerically by calculating the zero-doping dynamical spin structure factor that while the magnon gap is captured well by linear spin-wave theory, the overall shape of the magnon dispersion resembles more closely the spinon dispersion.

\subsection{Chevy approach}
\label{SubSecAvdMgnDcy}
In this section we will discuss an alternative solution of the effective polaron theory, based on Chevy's variational polaron wavefunction \cite{Chevy}. We distinguish two scenarios, depending on the number of magnon excitations we allow in the expansion. For simplicity, we omit the band indices $n,\xi$ of the meson from now. 

\subsubsection{Two-magnon state and avoided magnon decay}
\label{SubSecChevy2}
Inspired by the so-called Chevy ansatz, originally introduced for an imbalanced Fermi gas \cite{Chevy}, we expand the ground state of the polaron Hamiltonian \eqref{PolHam} up to two magnon excitations above the bare meson. Taking into account the conservation of the total system momentum, we make the following variational ansatz,
\begin{equation}
    \ket{\psi_k}=  \sqrt{Z_k} \fd_{k} \ket{0} + \sum_{pq} \alpha_{k,pq} \fd_{k-p-q} \bd_p\bd_q\ket{0},
    \label{eq2MagChevy}
\end{equation}
where $\sqrt{Z}_k$ and $\alpha_{k,pq}$ are variational parameters satisfying the normalization condition $Z_k + 2 \sum_{pq} |\alpha_{k,pq}|^2=1$.

Minimizing the functional $\mathcal{L} := \bra{\psi} \H_{\text{pol}} - E \ket{\psi}$ with respect to the variational parameters $\sqrt{Z_k}$ and $\alpha_{k,pq}$, leads to the following coupled equations: 
\begin{eqnarray}
    \label{sqrtZ}
    \l \varepsilon_k - E\r \sqrt{Z_k} = \dfrac{1}{L} \sum_{pq} \mathcal{W}_{k,pq}^* \alpha_{k,pq},\\
    \label{alpha}
    \sum_{q^\prime}\l \uuline{G}^{-1}_{kp,E}\r_{qq^\prime}\alpha_{k,pq^\prime} = -\dfrac{\sqrt{Z_k}}{2L}\mathcal{W}_{k,pq} ,
\end{eqnarray}
where the $L \times L$ matrix $\uuline{G}^{-1}_{kp,E}$ has the following components: 
\begin{multline}
\label{GreensInv}
\l \uuline{G}^{-1}_{kp,E}\r_{qq^\prime} = \l E - \varepsilon_{k-p-q^\prime,\eff} - \omega_p -\omega_{q^\prime} \r \delta_{qq^\prime}\\+ \dfrac{1}{L} \mathcal{V}_{k-p-q^\prime,qq^\prime}.
\end{multline}
The matrix $\uuline{G}_{kp,E}$ can be understood as the free retarted Green's function for two excited magnon excitations propagating along with the meson.

Solving the two above relations for the variational parameters, $\sqrt{Z_k}$ and $\alpha_{k,pq}$, by using a matrix inversion in Eq.~\eqref{alpha}, we get a variational energy of the form 
\begin{equation}
    \label{VarEn}
    E_k = \varepsilon_{k,\eff} + \Sigma_k(E_k)
\end{equation}
where we defined the self-energy,
\begin{equation}
    \Sigma_k(E_k) =  \dfrac{1}{2L^2} \sum_{pqq^\prime}\mathcal{W}^*_{k,pq} \l \uuline{G}_{kp,E_k}\r_{qq^\prime}\mathcal{W}_{k,pq^\prime}.
\end{equation}
It is important to note that $\l\uuline{G}_{kp,E_k}\r_{qq^\prime}$ are the components of the inverse of the matrix formed by the elements in Eq.~\eqref{GreensInv}.
Eq.~\eqref{VarEn} has to be solved self-consistently to get the variational energy $E_k$. The result is non-perturbative: it corresponds to re-summation over all diagrams describing two-magnon excitations \cite{Grusdt2015Varenna}. We will present our numerical results below.

Now we use the Chevy approach to argue that the meson at $k=0$ is stable for small magnetic fields $h$ and magnon decay as described in Sec.~\ref{secAvdMagDcySmry} is prevented by meson-magnon interactions. We already elaborated in previous chapters that the meson-magnon scattering continuum and the dressed meson band are repelling each other in this parameter regime. Now we will follow closely the arguments by Verresen et al.~\cite{Verresen2018}: they determined a threshold value where quasi-particle decay of a particle coupled into a continuum of states is prevented by strong interactions. 

The condition for the existence of a stable state below the continuum is given by~\cite{Verresen2018}
\begin{equation}
    \dfrac{\varepsilon_{k,\eff} - \omega^{-}_{k}}{|\Sigma_k(\omega^-_{k}+0^-)|} < 1   .
\label{eqCondStbleQP}
\end{equation}
Here $\omega^{-}_{k}$ denotes the lower edge of the meson plus two-magnon scattering continuum which is defined by $\omega^{-}_{k} = \min_{pq} (\varepsilon_{k-p-q,\eff} + \omega_p +\omega_q )$. When the condition \eqref{eqCondStbleQP} is fulfilled, we expect an avoided band crossing between the two-magnon continuum and the meson band. 

There are two cases to distinguish in the analysis of the condition~\eqref{eqCondStbleQP}:
\begin{itemize}
    \item When $\varepsilon_{k,\eff} < \omega^{-}_{k}$ the condition~\eqref{eqCondStbleQP} is trivially fulfilled for any interaction strength. In this case the bare meson dispersion $\varepsilon_{k,\eff}$ does not cross the two-magnon scattering continuum.
    \item For $\varepsilon_{k,\eff} > \omega^{-}_{k}$, the integrated meson-magnon interactions must be strong enough, such that the self-energy in the denominator, $|\Sigma_k(\omega^-_{k}+0^-)|$, dominates. At a given interaction strength this can also be achieved by a sufficiently high density of states at low energies. This case corresponds to a non-trivial avoided quasi-particle decay.
\end{itemize}

In the following we analyze the self-energy obtained from the two-magnon Chevy ansatz introduced above. It turns out that the condition~\eqref{eqCondStbleQP} is trivially fulfilled for all values of the staggered magnetic field $h>0$. The reason is that the renormalization of the meson tunneling amplitude, $J_{\perp}^{(n\xi)} \to J_{\perp,\eff}^{(n\xi)}$ by magnon zero-point contributions is strong enough to prevent the meson band from reaching the meson-magnon continuum. We emphasize that this effect results from meson-magnon interactions, which lead to the described renormalization, see Sec.~\ref{SecPolHam}.

\begin{figure}
    \centering
    \includegraphics[width=0.5 \textwidth]{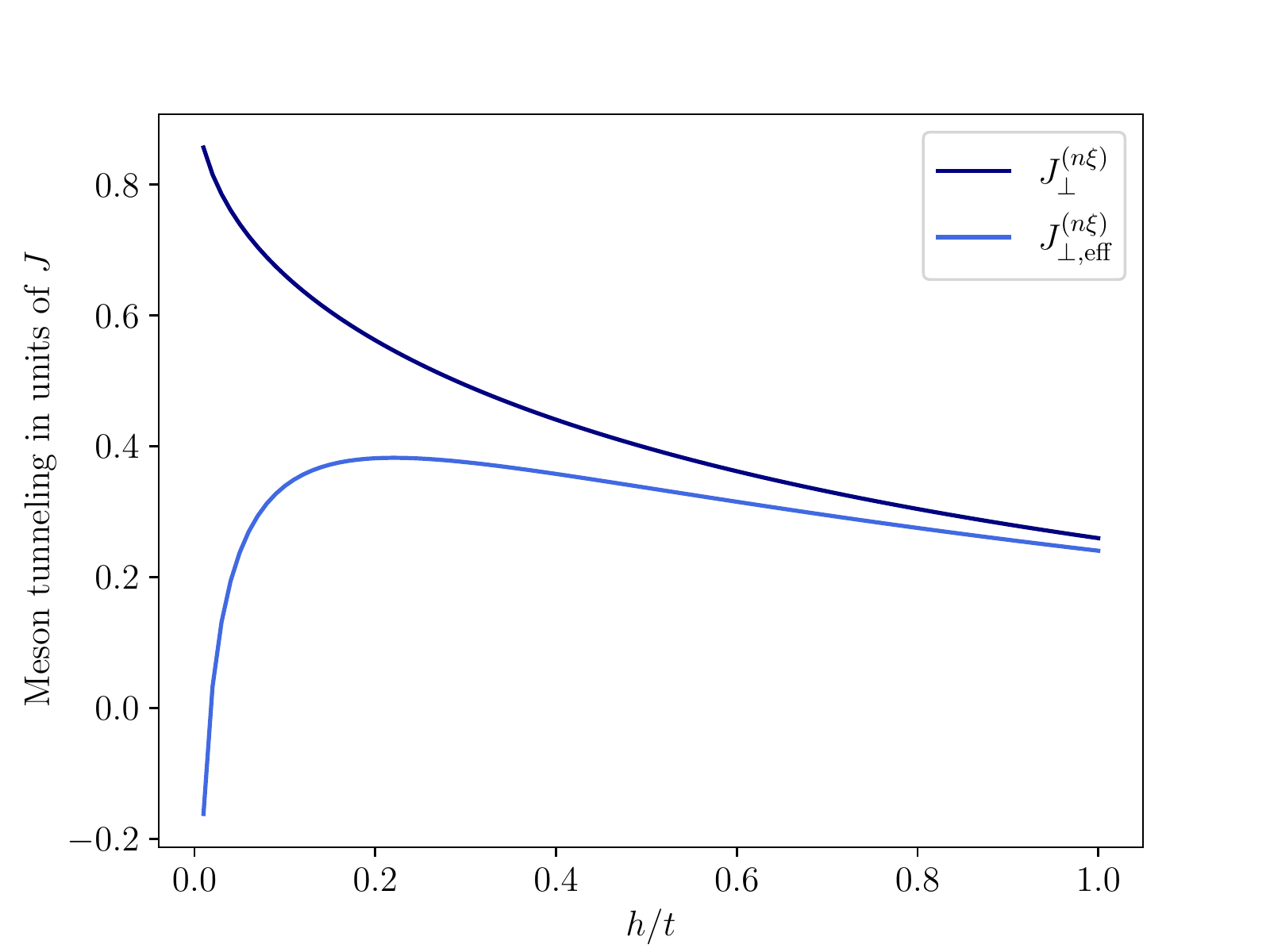}
    \caption{Effective meson tunneling amplitude $J_{\perp, \eff}^{(n\xi)}$ for $n=1,\xi=1$, which includes a reduction of the spinon tunneling due to magnon vacuum fluctuations compared to the meson tunneling amplitude $J_{\perp}^{(n\xi)}$ for $n=1,\xi=1$. The magnon zero-point contributions stabilize the meson band and prevent a decay into the two-magnon scattering continuum.}
    \label{figFCeff}
\end{figure}

In Fig.~\ref{figFCeff} we compare the bare meson tunneling amplitude $J_{\perp}^{(n\xi)}$, neglecting magnon fluctuations, with $J_{\perp,{\rm eff}}^{(n\xi)}$ renormalized by magnon zero-point contributions, assuming the  meson is in its ro-vibrational ground state $n=1,\xi=1$. For small values of the staggered magnetic field $h\lesssim 0.1$, the renormalized tunneling amplitude drops drastically. This is explained by the fact that the zero-point contributions of the magnons start to diverge for $h\to 0$ as we get a large magnon accumulation on all lattice sites of the chain at such low magnetic field values. In this regime, the linear spin-wave theory loses its applicability.

Finally, it is interesting to ask whether the condition~\eqref{eqCondStbleQP} would be fulfilled even if the bare meson band manages to leak into the meson-magnon scattering continuum. To this end, we repeat our above analysis and ignore the renormalization of the meson tunneling amplitude; i.e. we work with $\varepsilon_{k}$ instead of $\varepsilon_{k,\eff}$ in Eq.~\eqref{eqCondStbleQP}. Moreover, we re-scale the magnon dispersion obtained from linear spin-wave theory by a numerical factor $\lambda$, $\omega_q \to \lambda \omega_q$, such that the analytically known spinon bandwidth $J \pi/2$ is correctly captured in the limit $h\to0$. 

\begin{figure}
    \centering
    \includegraphics[width=0.5 \textwidth]{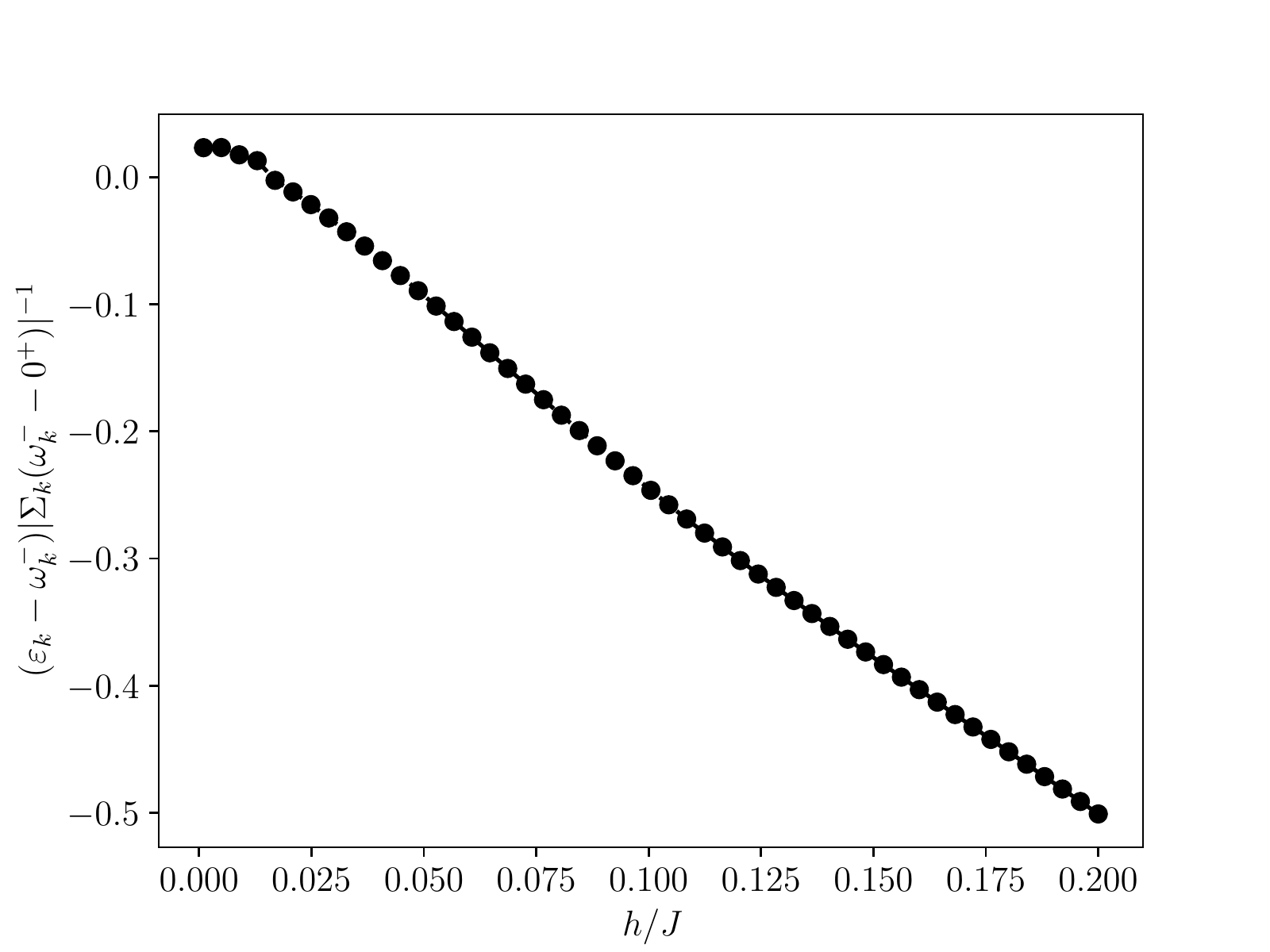}
    \caption{Verification of the stability of the meson. We show the left hand side (l.h.s.) of Eq.~\eqref{eqCondStbleQP}, ignoring the renormalization of magnon tunneling due to vacuum magnon fluctuations as described in the text. For values smaller than zero the condition is trivially fulfilled and if the l.h.s. of the condition is between zero and one it is non-trivially fulfilled. We assume the meson to be in its ro-vibrational ground state $n=1,\xi=1$ and the chosen parameters are $t=5, J = 1$.}
    \label{figCond}
\end{figure}

As shown in Fig.~\ref{figCond} we find that the resulting meson is non-trivially stable at $k=0$ due to meson-magnon interactions. There we plot the left hand side of Eq.~\eqref{eqCondStbleQP} under our simplifying assumptions, for different values of the staggered magnetic field $h$. For $h \lesssim 0.02 J$ non-trivial stabilization is found.

\begin{figure}
    \centering
    \includegraphics[width=0.45\textwidth]{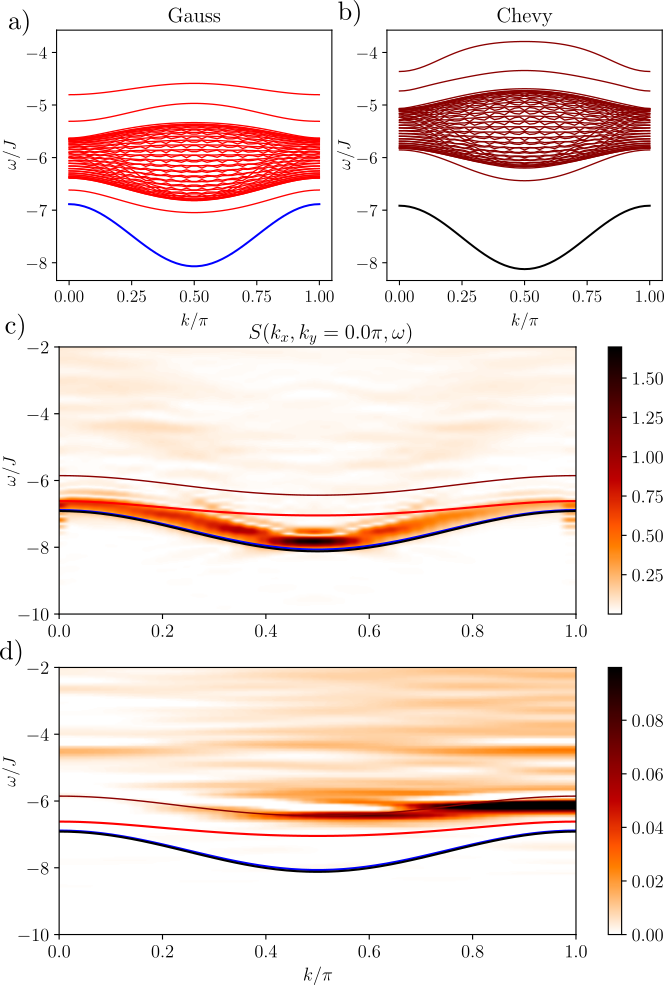}
    \caption{Comparison of the results of all of our methods at low energies for $h=0.6J$ and $t=5J$: The upper row shows the semi-analytical results, the polaron bands from our LLP treatment with a Gaussian linearization (a) and the curves for two-magnon and one-magnon Chevy ansatz (b). The middle panel c) is the standard one-hole ARPES spectrum, with our theoretical curves plotted on top using the same color scheme as in a) and b). The lower figure d) shows the spin-flip ARPES spectrum with total spin-$3/2$, see \ref{ARPESBound}. In c) and d), for the Gaussian LLP approach we only included the curves corresponding to the polaron ground state and the meson-magnon bound state. The shapes of the theoretical curves and the ARPES dispersion are in good agreement while there is a small shift in total energy. This has already been seen in the ground state energy, see Fig. \ref{GS}.}
    \label{FigComparison}
\end{figure}

\subsubsection{One-magnon state and meson-magnon bound state}
\label{SubSecChevy1}
Instead of the two-magnon ansatz Eq.~\eqref{eq2MagChevy}, mixing even-magnon number states, we can also make an odd-magnon number Chevy ansatz. The lowest order, all one-magnon states should be considered. Since the next order is rather involved, including three magnons, we restrict our discussion to the simplest one-magnon states in the following. 

In order to get a variational energy for one-magnon states, we project the full polaron Hamiltonian Eq.~\eqref{PolHam} onto the basis
\begin{equation}
    \ket{q}_k = \fd_{k-q}\bd_q \ket{0}
\end{equation}
which includes exactly one magnon excitation with variable momentum $q$. The state is constructed such that the total conserved momentum is $k$. We then diagonalize the projected Hamiltonian, defined by its matrix elements 
\begin{eqnarray}
    \label{OneMagChevyHam}
    \nonumber \hat{H}_{pq} &=& {}_k\bra{p} \H_{\rm pol} \ket{q}_k\\
    &=& \l \varepsilon_{k-q,\eff} + \omega_q \r \delta_{pq} - \dfrac{1}{2L} \mathcal{V}_{k-q,pq}.
\end{eqnarray}
Since our model Hamiltonian Eq.~\eqref{PolHam} only couples states of equal magnon-number parity, the one- and two-magnon Chevy wavefunctions must be considered independently and cannot couple. 

In Fig.~\ref{FigComparison} b) we show the variational one-magnon eigenenergies (dark red curves) obtained by diagonalizing the Hamiltonian matrix in Eq.~\eqref{OneMagChevyHam}. In the same figure we include the result for the variational ground state energy of the two-magnon Chevy ansatz, \eqref{eq2MagChevy} obtained by solving Eq.~\eqref{VarEn} self-consistently (black curve). The obtained spectrum compares qualitatively well to those from the LLP + Gaussian approach, see Fig.~\ref{FigComparison} a). In particular the Chevy approach correctly predicts the broad meson-magnon continuum at high energies, as well as a stable tetra-parton bound state between the meson and the meson-magnon continuum.

\subsection{Comparison of Methods}
In Fig. \ref{FigComparison} we compare the numerical ARPES spectra with our two theoretical methods. We already discussed the polaron spectrum obtained by a Gaussian LLP approach in Sec. \ref{SecPolSpec} which is again shown here in panel a). The variational energies obtained from the two-magnon and one-magnon Chevy ansatz are shown in panel b). We put these semi-analytical curves on top of the numerically obtained standard one-hole ARPES spectrum [panel c)] where the ground state can be compared, and the spin-flip ARPES spectrum probing the sector with $S^z_{\rm tot}=3/2$ [panel d)] where the meson-magnon bound state can be compared. 

In our discussion of the ground state energy we observed that there is a small deviation between the numerical and theoretical result, see Fig. \ref{GS}. This small deviation is also seen here in the comparison in Fig. \ref{FigComparison} c). The predicted shapes of the meson dispersion agree very well with each other, and with the full numerical result. We note that this agreement holds for all magnetic field values $h$ that we considered. 

For the spin-$3/2$ spin-flip ARPES spectrum the situation is less clear. On the one hand, we find good qualitative agreement in that our numerical td-MPS simulations predict a pronounced quasiparticle peak at low energies, for all considered values of $h$. For the value of $h=0.6 J$ shown in Fig.~\ref{FigComparison} we find remarkable quantitative agreement with our 1-magnon Chevy prediction, which is also relatively close to the LLP prediction. However, for other values of the staggered field $h/J$, we observed larger quantitative differences. 

\section{Summary and Outlook}
We have investigated the problem of a single hole doped into a spin chain with an external staggered Zeeman field and studied the interplay of the doped hole with quantum fluctuations of the spin background. In this simplistic setting, we found a remarkably rich zoo of quasiparticle excitations arising from the interplay of charge fluctuations and quantum magnetism. Our predictions can be tested in state of the art quantum gas microscopy experiments.

Based on a parton construction for the hole, capturing the Ising-limit of our model, we developed a simple semi-analytically solvable theory describing the hole as having an internal structure composed of two constituents: a spin-less chargon and a charge-neutral spinon. The hole is a confined mesonic bound state of these constituents, similar to mesons in high-energy physics \cite{Greensite2003}, and their binding potential is linear in nature. Formally, this setting is directly related to a $\mathbb{Z}_2$ lattice gauge theory \cite{Grusdt2020PRL,Kebric2021,Borla2020}. Similar to previous proposals for an analogous problem in 2D~\cite{Bohrdt2020PRB,Bohrdt2021} we find stable long-lived ro-vibrational excitations of the doped hole. Their spectroscopic measurement would constitute compelling evidence that mobile holes have a rich internal structure and can be understood as spinon-chargon bound states. 

The main theoretical advancement of our work was the systematic inclusion of quantum fluctuations in the parton description. We used a generalized $1/S$ expansion technique\cite{Grusdt2018} to treat quantum spin fluctuations -- resulting from transverse spin couplings $\propto J_\perp$ in the $t-J$ model -- around the N\'eel state distorted by dominant charge fluctuations. This allowed us to derive an effective polaronic model, describing how a pre-formed meson interacts weakly with additional magnon excitations. As a main result of this approach, we were able to predict an additional stable meson-magnon bound state, i.e. a tetra-parton state of a chargon bound to three spinons. We confirmed this prediction by numerical DMRG simulations and an exact perturbative analysis in the strong-field limit, $h \gg t,J$. Another significant insight obtained by our method concerns the stability of the mesonic quasiparticle when its energy approaches the meson-magnon continuum around zero momentum. Namely, we found evidence that meson-magnon interactions are sufficiently strong to stabilize the quasiparticle peak for arbitrarily weak confining fields $h$.

Our theoretical formalism paves the way for many future extensions. For example, the simplification of the problem to a well-known and weakly coupled Bose-polaron model allows to study far-from equilibrium dynamics \cite{Zvonarev2007,Bruderer2008,Grusdt2014BO,Lampo2017bosepolaronas} of a mobile hole \cite{Ji2021}, going to non-zero temperatures and much longer times than accessible by the more accurate tensor-network methods \cite{Bohrdt2020Dyn,Hubig2020}. Moreover, our approach can be generalized to higher dimensions, where exact numerical methods become significantly more challenging. The dressing of strongly paired states of holes \cite{bohrdt2021strong} can also be investigated. Another promising direction would be the study of magnon excitations in mixed-dimensional systems \cite{Grusdt2018SciPost,Grusdt2020PRL} at finite doping. Finally, the microscopic connection we establish to an underlying $\mathbb{Z}_2$ lattice gauge theory may be more general, suggesting a new route how emergent gauge structures can arise in strongly correlated quantum matter. 

The model Hamiltonian we considered in one dimension, namely a $t-J$ model in a staggered Zeeman field, also constitutes an interesting platform for future studies. Its close connections to other interesting models on one hand, such as the 1D $t-J_z$ or $t-J$ models or the 2D $t-J$ model which also has long-range magnetic correlations at zero doping, and its direct experimental realizability in ultracold atoms on the other hand make it an appealing system to study. In this article we limited our discussion to a single doped hole, but extensions to finite doping are straightforward. For example, it will be interesting to search for pairing or charge order at finite doping and investigate the role played by couplings to magnon excitations in the spin background. Exploring the connection to an underlying lattice gauge structure, and a possible breakdown of the meson picture with doping will also be worthwhile endeavours.

\section{Acknowledgements}
The authors benefited particularly from discussions with E. Demler and L. Pollet, and gratefully acknowledge discussions with U. Schollwöck, I. Bloch, F. Pollmann, M. Knap, R. Verresen, T. Shi, P. Bermees, Z. Zhu, P. Wrzosek, K. Wohlfeld and G. Uhrig. This research was funded by the Deutsche Forschungsgemeinschaft (DFG, German Research Foundation) under Germany's Excellence Strategy -- EXC-2111 -- 390814868, from the European Research Council (ERC) under the European Union’s Horizon 2020 research and innovation programm (Grant Agreement no 948141) — ERC Starting Grant SimUcQuam, and by the NSF through a grant for the Institute for Theoretical Atomic, Molecular, and Optical Physics at Harvard University and the Smithsonian Astrophysical Observatory.



\begin{thebibliography}{72}%
\makeatletter
\providecommand \@ifxundefined [1]{%
 \@ifx{#1\undefined}
}%
\providecommand \@ifnum [1]{%
 \ifnum #1\expandafter \@firstoftwo
 \else \expandafter \@secondoftwo
 \fi
}%
\providecommand \@ifx [1]{%
 \ifx #1\expandafter \@firstoftwo
 \else \expandafter \@secondoftwo
 \fi
}%
\providecommand \natexlab [1]{#1}%
\providecommand \enquote  [1]{``#1''}%
\providecommand \bibnamefont  [1]{#1}%
\providecommand \bibfnamefont [1]{#1}%
\providecommand \citenamefont [1]{#1}%
\providecommand \href@noop [0]{\@secondoftwo}%
\providecommand \href [0]{\begingroup \@sanitize@url \@href}%
\providecommand \@href[1]{\@@startlink{#1}\@@href}%
\providecommand \@@href[1]{\endgroup#1\@@endlink}%
\providecommand \@sanitize@url [0]{\catcode `\\12\catcode `\$12\catcode
  `\&12\catcode `\#12\catcode `\^12\catcode `\_12\catcode `\%12\relax}%
\providecommand \@@startlink[1]{}%
\providecommand \@@endlink[0]{}%
\providecommand \url  [0]{\begingroup\@sanitize@url \@url }%
\providecommand \@url [1]{\endgroup\@href {#1}{\urlprefix }}%
\providecommand \urlprefix  [0]{URL }%
\providecommand \Eprint [0]{\href }%
\providecommand \doibase [0]{https://doi.org/}%
\providecommand \selectlanguage [0]{\@gobble}%
\providecommand \bibinfo  [0]{\@secondoftwo}%
\providecommand \bibfield  [0]{\@secondoftwo}%
\providecommand \translation [1]{[#1]}%
\providecommand \BibitemOpen [0]{}%
\providecommand \bibitemStop [0]{}%
\providecommand \bibitemNoStop [0]{.\EOS\space}%
\providecommand \EOS [0]{\spacefactor3000\relax}%
\providecommand \BibitemShut  [1]{\csname bibitem#1\endcsname}%
\let\auto@bib@innerbib\@empty
\bibitem [{\citenamefont {Hauschild}\ \emph {et~al.}(2018)\citenamefont
  {Hauschild}, \citenamefont {Mong}, \citenamefont {Pollmann}, \citenamefont
  {Schulz}, \citenamefont {Schoonderwoert}, \citenamefont {Unfried},
  \citenamefont {Tzeng},\ and\ \citenamefont {Zaletel}}]{Hauschild2018}%
  \BibitemOpen
  \bibfield  {author} {\bibinfo {author} {\bibfnamefont {J.}~\bibnamefont
  {Hauschild}}, \bibinfo {author} {\bibfnamefont {R.}~\bibnamefont {Mong}},
  \bibinfo {author} {\bibfnamefont {F.}~\bibnamefont {Pollmann}}, \bibinfo
  {author} {\bibfnamefont {M.}~\bibnamefont {Schulz}}, \bibinfo {author}
  {\bibfnamefont {L.}~\bibnamefont {Schoonderwoert}}, \bibinfo {author}
  {\bibfnamefont {J.}~\bibnamefont {Unfried}}, \bibinfo {author} {\bibfnamefont
  {Y.}~\bibnamefont {Tzeng}},\ and\ \bibinfo {author} {\bibfnamefont
  {M.}~\bibnamefont {Zaletel}},\ }\bibfield  {title} {\bibinfo {title} {Tensor
  network python},\ }\href@noop {} {\bibfield  {journal} {\bibinfo  {journal}
  {\normalfont The code is available online at https://github.com/tenpy/tenpy/,
  the documentation can be found at https://tenpy.github.com/.}\ } (\bibinfo
  {year} {2018})}\BibitemShut {NoStop}%
\bibitem [{\citenamefont {Hauschild}\ and\ \citenamefont
  {Pollmann}(2018)}]{Hauschild2019}%
  \BibitemOpen
  \bibfield  {author} {\bibinfo {author} {\bibfnamefont {J.}~\bibnamefont
  {Hauschild}}\ and\ \bibinfo {author} {\bibfnamefont {F.}~\bibnamefont
  {Pollmann}},\ }\bibfield  {title} {\bibinfo {title} {{Efficient numerical
  simulations with Tensor Networks: Tensor Network Python (TeNPy)}},\ }\href
  {https://doi.org/10.21468/SciPostPhysLectNotes.5} {\bibfield  {journal}
  {\bibinfo  {journal} {SciPost Phys. Lect. Notes}\ ,\ \bibinfo {pages} {5}}
  (\bibinfo {year} {2018})}\BibitemShut {NoStop}%
\bibitem [{\citenamefont {Haldane}(1983)}]{Haldane1983a}%
  \BibitemOpen
  \bibfield  {author} {\bibinfo {author} {\bibfnamefont {F.}~\bibnamefont
  {Haldane}},\ }\bibfield  {title} {\bibinfo {title} {Continuum dynamics of the
  1-d heisenberg antiferromagnet: Identification with the o(3) nonlinear sigma
  model},\ }\href
  {https://doi.org/https://doi.org/10.1016/0375-9601(83)90631-X} {\bibfield
  {journal} {\bibinfo  {journal} {Physics Letters A}\ }\textbf {\bibinfo
  {volume} {93}},\ \bibinfo {pages} {464} (\bibinfo {year} {1983})}\BibitemShut
  {NoStop}%
\bibitem [{\citenamefont {Sachdev}(2011)}]{Sachdev2011}%
  \BibitemOpen
  \bibfield  {author} {\bibinfo {author} {\bibfnamefont {S.}~\bibnamefont
  {Sachdev}},\ }\href {https://books.google.de/books?id=F3IkpxwpqSgC} {\emph
  {\bibinfo {title} {Quantum Phase Transitions}}}\ (\bibinfo  {publisher}
  {Cambridge University Press},\ \bibinfo {year} {2011})\BibitemShut {NoStop}%
\bibitem [{\citenamefont {Chevy}\ and\ \citenamefont {Mora}(2010)}]{Chevy2010}%
  \BibitemOpen
  \bibfield  {author} {\bibinfo {author} {\bibfnamefont {F.}~\bibnamefont
  {Chevy}}\ and\ \bibinfo {author} {\bibfnamefont {C.}~\bibnamefont {Mora}},\
  }\bibfield  {title} {\bibinfo {title} {Ultra-cold polarized fermi gases},\
  }\href {https://doi.org/10.1088/0034-4885/73/11/112401} {\bibfield  {journal}
  {\bibinfo  {journal} {Reports on Progress in Physics}\ }\textbf {\bibinfo
  {volume} {73}},\ \bibinfo {pages} {112401} (\bibinfo {year}
  {2010})}\BibitemShut {NoStop}%
\bibitem [{\citenamefont {Devreese}(2020)}]{Devreese2020}%
  \BibitemOpen
  \bibfield  {author} {\bibinfo {author} {\bibfnamefont {J.~T.}\ \bibnamefont
  {Devreese}},\ }\href@noop {} {\bibinfo {title} {Fr\"ohlich polarons. lecture
  course including detailed theoretical derivations -- 10th edition}} (\bibinfo
  {year} {2020}),\ \Eprint {https://arxiv.org/abs/1611.06122} {arXiv:1611.06122
  [cond-mat.other]} \BibitemShut {NoStop}%
\bibitem [{\citenamefont {Grusdt}\ and\ \citenamefont
  {Demler}(2015)}]{Grusdt2015Varenna}%
  \BibitemOpen
  \bibfield  {author} {\bibinfo {author} {\bibfnamefont {F.}~\bibnamefont
  {Grusdt}}\ and\ \bibinfo {author} {\bibfnamefont {E.~A.}\ \bibnamefont
  {Demler}},\ }\bibfield  {title} {\bibinfo {title} {New theoretical approaches
  to bose polarons},\ }\href {https://arxiv.org/abs/1510.04934} {\bibfield
  {journal} {\bibinfo  {journal} {Proceedings of the International School of
  Physics Enrico Fermi, arXiv:1510.04934}\ } (\bibinfo {year}
  {2015})}\BibitemShut {NoStop}%
\bibitem [{\citenamefont {Rath}\ and\ \citenamefont
  {Schmidt}(2013)}]{Rath2013}%
  \BibitemOpen
  \bibfield  {author} {\bibinfo {author} {\bibfnamefont {S.~P.}\ \bibnamefont
  {Rath}}\ and\ \bibinfo {author} {\bibfnamefont {R.}~\bibnamefont {Schmidt}},\
  }\bibfield  {title} {\bibinfo {title} {Field-theoretical study of the bose
  polaron},\ }\href {https://doi.org/10.1103/PhysRevA.88.053632} {\bibfield
  {journal} {\bibinfo  {journal} {Physical Review A}\ }\textbf {\bibinfo
  {volume} {88}},\ \bibinfo {pages} {053632} (\bibinfo {year}
  {2013})}\BibitemShut {NoStop}%
\bibitem [{\citenamefont {Shi}\ \emph {et~al.}(2018)\citenamefont {Shi},
  \citenamefont {Demler},\ and\ \citenamefont {Ignacio~Cirac}}]{Shi2018}%
  \BibitemOpen
  \bibfield  {author} {\bibinfo {author} {\bibfnamefont {T.}~\bibnamefont
  {Shi}}, \bibinfo {author} {\bibfnamefont {E.}~\bibnamefont {Demler}},\ and\
  \bibinfo {author} {\bibfnamefont {J.}~\bibnamefont {Ignacio~Cirac}},\
  }\bibfield  {title} {\bibinfo {title} {Variational study of fermionic and
  bosonic systems with non-gaussian states: Theory and applications},\ }\href
  {https://doi.org/10.1016/j.aop.2017.11.014} {\bibfield  {journal} {\bibinfo
  {journal} {Annals of Physics}\ }\textbf {\bibinfo {volume} {390}},\ \bibinfo
  {pages} {245–302} (\bibinfo {year} {2018})}\BibitemShut {NoStop}%
\bibitem [{\citenamefont {Kim}\ \emph {et~al.}(1996)\citenamefont {Kim},
  \citenamefont {Matsuura}, \citenamefont {Shen}, \citenamefont {Motoyama},
  \citenamefont {Eisaki}, \citenamefont {Uchida}, \citenamefont {Tohyama},\
  and\ \citenamefont {Maekawa}}]{Kim1996}%
  \BibitemOpen
  \bibfield  {author} {\bibinfo {author} {\bibfnamefont {C.}~\bibnamefont
  {Kim}}, \bibinfo {author} {\bibfnamefont {A.~Y.}\ \bibnamefont {Matsuura}},
  \bibinfo {author} {\bibfnamefont {Z.-X.}\ \bibnamefont {Shen}}, \bibinfo
  {author} {\bibfnamefont {N.}~\bibnamefont {Motoyama}}, \bibinfo {author}
  {\bibfnamefont {H.}~\bibnamefont {Eisaki}}, \bibinfo {author} {\bibfnamefont
  {S.}~\bibnamefont {Uchida}}, \bibinfo {author} {\bibfnamefont
  {T.}~\bibnamefont {Tohyama}},\ and\ \bibinfo {author} {\bibfnamefont
  {S.}~\bibnamefont {Maekawa}},\ }\bibfield  {title} {\bibinfo {title}
  {Observation of spin-charge separation in one-dimensional
  srcu${\mathrm{o}}_{2}$},\ }\href
  {https://doi.org/10.1103/PhysRevLett.77.4054} {\bibfield  {journal} {\bibinfo
   {journal} {Phys. Rev. Lett.}\ }\textbf {\bibinfo {volume} {77}},\ \bibinfo
  {pages} {4054} (\bibinfo {year} {1996})}\BibitemShut {NoStop}%
\bibitem [{\citenamefont {Sing}\ \emph {et~al.}(2003)\citenamefont {Sing},
  \citenamefont {Schwingenschl\"ogl}, \citenamefont {Claessen}, \citenamefont
  {Blaha}, \citenamefont {Carmelo}, \citenamefont {Martelo}, \citenamefont
  {Sacramento}, \citenamefont {Dressel},\ and\ \citenamefont
  {Jacobsen}}]{Sing2003}%
  \BibitemOpen
  \bibfield  {author} {\bibinfo {author} {\bibfnamefont {M.}~\bibnamefont
  {Sing}}, \bibinfo {author} {\bibfnamefont {U.}~\bibnamefont
  {Schwingenschl\"ogl}}, \bibinfo {author} {\bibfnamefont {R.}~\bibnamefont
  {Claessen}}, \bibinfo {author} {\bibfnamefont {P.}~\bibnamefont {Blaha}},
  \bibinfo {author} {\bibfnamefont {J.~M.~P.}\ \bibnamefont {Carmelo}},
  \bibinfo {author} {\bibfnamefont {L.~M.}\ \bibnamefont {Martelo}}, \bibinfo
  {author} {\bibfnamefont {P.~D.}\ \bibnamefont {Sacramento}}, \bibinfo
  {author} {\bibfnamefont {M.}~\bibnamefont {Dressel}},\ and\ \bibinfo {author}
  {\bibfnamefont {C.~S.}\ \bibnamefont {Jacobsen}},\ }\bibfield  {title}
  {\bibinfo {title} {Electronic structure of the quasi-one-dimensional organic
  conductor ttf-tcnq},\ }\href {https://doi.org/10.1103/PhysRevB.68.125111}
  {\bibfield  {journal} {\bibinfo  {journal} {Phys. Rev. B}\ }\textbf {\bibinfo
  {volume} {68}},\ \bibinfo {pages} {125111} (\bibinfo {year}
  {2003})}\BibitemShut {NoStop}%
\bibitem [{\citenamefont {Ogata}\ and\ \citenamefont
  {Shiba}(1990)}]{Ogata1990}%
  \BibitemOpen
  \bibfield  {author} {\bibinfo {author} {\bibfnamefont {M.}~\bibnamefont
  {Ogata}}\ and\ \bibinfo {author} {\bibfnamefont {H.}~\bibnamefont {Shiba}},\
  }\bibfield  {title} {\bibinfo {title} {Bethe-ansatz wave function, momentum
  distribution, and spin correlation in the one-dimensional strongly correlated
  hubbard model},\ }\href {https://doi.org/10.1103/PhysRevB.41.2326} {\bibfield
   {journal} {\bibinfo  {journal} {Phys. Rev. B}\ }\textbf {\bibinfo {volume}
  {41}},\ \bibinfo {pages} {2326} (\bibinfo {year} {1990})}\BibitemShut
  {NoStop}%
\bibitem [{\citenamefont {Hilker}\ \emph {et~al.}(2017)\citenamefont {Hilker},
  \citenamefont {Salomon}, \citenamefont {Grusdt}, \citenamefont {Omran},
  \citenamefont {Boll}, \citenamefont {Demler}, \citenamefont {Bloch},\ and\
  \citenamefont {Gross}}]{Hilker2017}%
  \BibitemOpen
  \bibfield  {author} {\bibinfo {author} {\bibfnamefont {T.~A.}\ \bibnamefont
  {Hilker}}, \bibinfo {author} {\bibfnamefont {G.}~\bibnamefont {Salomon}},
  \bibinfo {author} {\bibfnamefont {F.}~\bibnamefont {Grusdt}}, \bibinfo
  {author} {\bibfnamefont {A.}~\bibnamefont {Omran}}, \bibinfo {author}
  {\bibfnamefont {M.}~\bibnamefont {Boll}}, \bibinfo {author} {\bibfnamefont
  {E.}~\bibnamefont {Demler}}, \bibinfo {author} {\bibfnamefont
  {I.}~\bibnamefont {Bloch}},\ and\ \bibinfo {author} {\bibfnamefont
  {C.}~\bibnamefont {Gross}},\ }\bibfield  {title} {\bibinfo {title} {Revealing
  hidden antiferromagnetic correlations in doped hubbard chains via string
  correlators},\ }\href {https://doi.org/10.1126/science.aam8990} {\bibfield
  {journal} {\bibinfo  {journal} {Science}\ }\textbf {\bibinfo {volume}
  {357}},\ \bibinfo {pages} {484–487} (\bibinfo {year} {2017})}\BibitemShut
  {NoStop}%
\bibitem [{\citenamefont {Vijayan}\ \emph {et~al.}(2020)\citenamefont
  {Vijayan}, \citenamefont {Sompet}, \citenamefont {Salomon}, \citenamefont
  {Koepsell}, \citenamefont {Hirthe}, \citenamefont {Bohrdt}, \citenamefont
  {Grusdt}, \citenamefont {Bloch},\ and\ \citenamefont {Gross}}]{Vijayan2020}%
  \BibitemOpen
  \bibfield  {author} {\bibinfo {author} {\bibfnamefont {J.}~\bibnamefont
  {Vijayan}}, \bibinfo {author} {\bibfnamefont {P.}~\bibnamefont {Sompet}},
  \bibinfo {author} {\bibfnamefont {G.}~\bibnamefont {Salomon}}, \bibinfo
  {author} {\bibfnamefont {J.}~\bibnamefont {Koepsell}}, \bibinfo {author}
  {\bibfnamefont {S.}~\bibnamefont {Hirthe}}, \bibinfo {author} {\bibfnamefont
  {A.}~\bibnamefont {Bohrdt}}, \bibinfo {author} {\bibfnamefont
  {F.}~\bibnamefont {Grusdt}}, \bibinfo {author} {\bibfnamefont
  {I.}~\bibnamefont {Bloch}},\ and\ \bibinfo {author} {\bibfnamefont
  {C.}~\bibnamefont {Gross}},\ }\bibfield  {title} {\bibinfo {title}
  {Time-resolved observation of spin-charge deconfinement in fermionic hubbard
  chains},\ }\href {https://doi.org/10.1126/science.aay2354} {\bibfield
  {journal} {\bibinfo  {journal} {Science}\ }\textbf {\bibinfo {volume}
  {367}},\ \bibinfo {pages} {186–189} (\bibinfo {year} {2020})}\BibitemShut
  {NoStop}%
\bibitem [{\citenamefont {Giamarchi}\ and\ \citenamefont
  {Press}(2004)}]{Giamarchi2004}%
  \BibitemOpen
  \bibfield  {author} {\bibinfo {author} {\bibfnamefont {T.}~\bibnamefont
  {Giamarchi}}\ and\ \bibinfo {author} {\bibfnamefont {O.~U.}\ \bibnamefont
  {Press}},\ }\href {https://books.google.de/books?id=1MwTDAAAQBAJ} {\emph
  {\bibinfo {title} {Quantum Physics in One Dimension}}},\ International Series
  of Monogr\ (\bibinfo  {publisher} {Clarendon Press},\ \bibinfo {year}
  {2004})\BibitemShut {NoStop}%
\bibitem [{\citenamefont {Borla}\ \emph {et~al.}(2020)\citenamefont {Borla},
  \citenamefont {Verresen}, \citenamefont {Grusdt},\ and\ \citenamefont
  {Moroz}}]{Borla2020}%
  \BibitemOpen
  \bibfield  {author} {\bibinfo {author} {\bibfnamefont {U.}~\bibnamefont
  {Borla}}, \bibinfo {author} {\bibfnamefont {R.}~\bibnamefont {Verresen}},
  \bibinfo {author} {\bibfnamefont {F.}~\bibnamefont {Grusdt}},\ and\ \bibinfo
  {author} {\bibfnamefont {S.}~\bibnamefont {Moroz}},\ }\bibfield  {title}
  {\bibinfo {title} {Confined phases of one-dimensional spinless fermions
  coupled to ${Z}_ {2}$ gauge theory},\ }\href
  {https://doi.org/10.1103/PhysRevLett.124.120503} {\bibfield  {journal}
  {\bibinfo  {journal} {Phys. Rev. Lett.}\ }\textbf {\bibinfo {volume} {124}},\
  \bibinfo {pages} {120503} (\bibinfo {year} {2020})}\BibitemShut {NoStop}%
\bibitem [{\citenamefont {Kebric}\ \emph {et~al.}(2021)\citenamefont {Kebric},
  \citenamefont {Barbiero}, \citenamefont {Reinmoser}, \citenamefont
  {Schollw\"ock},\ and\ \citenamefont {Grusdt}}]{Kebric2021}%
  \BibitemOpen
  \bibfield  {author} {\bibinfo {author} {\bibfnamefont {M.}~\bibnamefont
  {Kebric}}, \bibinfo {author} {\bibfnamefont {L.}~\bibnamefont {Barbiero}},
  \bibinfo {author} {\bibfnamefont {C.}~\bibnamefont {Reinmoser}}, \bibinfo
  {author} {\bibfnamefont {U.}~\bibnamefont {Schollw\"ock}},\ and\ \bibinfo
  {author} {\bibfnamefont {F.}~\bibnamefont {Grusdt}},\ }\bibfield  {title}
  {\bibinfo {title} {Confinement and mott transitions of dynamical charges in
  one-dimensional lattice gauge theories},\ }\href
  {https://doi.org/10.1103/PhysRevLett.127.167203} {\bibfield  {journal}
  {\bibinfo  {journal} {Phys. Rev. Lett.}\ }\textbf {\bibinfo {volume} {127}},\
  \bibinfo {pages} {167203} (\bibinfo {year} {2021})}\BibitemShut {NoStop}%
\bibitem [{\citenamefont {B\'eran}\ \emph {et~al.}(1996)\citenamefont
  {B\'eran}, \citenamefont {Poilblanc},\ and\ \citenamefont
  {Laughlin}}]{Beran1996}%
  \BibitemOpen
  \bibfield  {author} {\bibinfo {author} {\bibfnamefont {P.}~\bibnamefont
  {B\'eran}}, \bibinfo {author} {\bibfnamefont {D.}~\bibnamefont {Poilblanc}},\
  and\ \bibinfo {author} {\bibfnamefont {R.}~\bibnamefont {Laughlin}},\
  }\bibfield  {title} {\bibinfo {title} {Evidence for composite nature of
  quasiparticles in the 2d t-j model},\ }\href
  {https://doi.org/10.1016/0550-3213(96)00196-4} {\bibfield  {journal}
  {\bibinfo  {journal} {Nuclear Physics B}\ }\textbf {\bibinfo {volume}
  {473}},\ \bibinfo {pages} {707–720} (\bibinfo {year} {1996})}\BibitemShut
  {NoStop}%
\bibitem [{\citenamefont {Brunner}\ \emph {et~al.}(2000)\citenamefont
  {Brunner}, \citenamefont {Assaad},\ and\ \citenamefont
  {Muramatsu}}]{Brunner2000}%
  \BibitemOpen
  \bibfield  {author} {\bibinfo {author} {\bibfnamefont {M.}~\bibnamefont
  {Brunner}}, \bibinfo {author} {\bibfnamefont {F.~F.}\ \bibnamefont
  {Assaad}},\ and\ \bibinfo {author} {\bibfnamefont {A.}~\bibnamefont
  {Muramatsu}},\ }\bibfield  {title} {\bibinfo {title} {Single-hole dynamics in
  the $t\ensuremath{-}j$ model on a square lattice},\ }\href
  {https://doi.org/10.1103/PhysRevB.62.15480} {\bibfield  {journal} {\bibinfo
  {journal} {Phys. Rev. B}\ }\textbf {\bibinfo {volume} {62}},\ \bibinfo
  {pages} {15480} (\bibinfo {year} {2000})}\BibitemShut {NoStop}%
\bibitem [{\citenamefont {Mishchenko}\ \emph {et~al.}(2001)\citenamefont
  {Mishchenko}, \citenamefont {Prokof’ev},\ and\ \citenamefont
  {Svistunov}}]{Mishchenko2001}%
  \BibitemOpen
  \bibfield  {author} {\bibinfo {author} {\bibfnamefont {A.~S.}\ \bibnamefont
  {Mishchenko}}, \bibinfo {author} {\bibfnamefont {N.~V.}\ \bibnamefont
  {Prokof’ev}},\ and\ \bibinfo {author} {\bibfnamefont {B.~V.}\ \bibnamefont
  {Svistunov}},\ }\bibfield  {title} {\bibinfo {title} {Single-hole spectral
  function and spin-charge separation in thet−jmodel},\ }\bibfield  {journal}
  {\bibinfo  {journal} {Physical Review B}\ }\textbf {\bibinfo {volume} {64}},\
  \href {https://doi.org/10.1103/physrevb.64.033101}
  {10.1103/physrevb.64.033101} (\bibinfo {year} {2001})\BibitemShut {NoStop}%
\bibitem [{\citenamefont {Bohrdt}\ \emph
  {et~al.}(2020{\natexlab{a}})\citenamefont {Bohrdt}, \citenamefont {Demler},
  \citenamefont {Pollmann}, \citenamefont {Knap},\ and\ \citenamefont
  {Grusdt}}]{Bohrdt2020PRB}%
  \BibitemOpen
  \bibfield  {author} {\bibinfo {author} {\bibfnamefont {A.}~\bibnamefont
  {Bohrdt}}, \bibinfo {author} {\bibfnamefont {E.}~\bibnamefont {Demler}},
  \bibinfo {author} {\bibfnamefont {F.}~\bibnamefont {Pollmann}}, \bibinfo
  {author} {\bibfnamefont {M.}~\bibnamefont {Knap}},\ and\ \bibinfo {author}
  {\bibfnamefont {F.}~\bibnamefont {Grusdt}},\ }\bibfield  {title} {\bibinfo
  {title} {Parton theory of angle-resolved photoemission spectroscopy spectra
  in antiferromagnetic mott insulators},\ }\href
  {https://doi.org/10.1103/PhysRevB.102.035139} {\bibfield  {journal} {\bibinfo
   {journal} {Phys. Rev. B}\ }\textbf {\bibinfo {volume} {102}},\ \bibinfo
  {pages} {035139} (\bibinfo {year} {2020}{\natexlab{a}})}\BibitemShut
  {NoStop}%
\bibitem [{\citenamefont {Grusdt}\ \emph
  {et~al.}(2018{\natexlab{a}})\citenamefont {Grusdt}, \citenamefont
  {K\'anasz-Nagy}, \citenamefont {Bohrdt}, \citenamefont {Chiu}, \citenamefont
  {Ji}, \citenamefont {Greiner}, \citenamefont {Greif},\ and\ \citenamefont
  {Demler}}]{Grusdt2018}%
  \BibitemOpen
  \bibfield  {author} {\bibinfo {author} {\bibfnamefont {F.}~\bibnamefont
  {Grusdt}}, \bibinfo {author} {\bibfnamefont {M.}~\bibnamefont
  {K\'anasz-Nagy}}, \bibinfo {author} {\bibfnamefont {A.}~\bibnamefont
  {Bohrdt}}, \bibinfo {author} {\bibfnamefont {C.~S.}\ \bibnamefont {Chiu}},
  \bibinfo {author} {\bibfnamefont {G.}~\bibnamefont {Ji}}, \bibinfo {author}
  {\bibfnamefont {M.}~\bibnamefont {Greiner}}, \bibinfo {author} {\bibfnamefont
  {D.}~\bibnamefont {Greif}},\ and\ \bibinfo {author} {\bibfnamefont
  {E.}~\bibnamefont {Demler}},\ }\bibfield  {title} {\bibinfo {title} {Parton
  theory of magnetic polarons: Mesonic resonances and signatures in dynamics},\
  }\href {https://doi.org/10.1103/PhysRevX.8.011046} {\bibfield  {journal}
  {\bibinfo  {journal} {Phys. Rev. X}\ }\textbf {\bibinfo {volume} {8}},\
  \bibinfo {pages} {011046} (\bibinfo {year} {2018}{\natexlab{a}})}\BibitemShut
  {NoStop}%
\bibitem [{\citenamefont {Bohrdt}\ \emph
  {et~al.}(2021{\natexlab{a}})\citenamefont {Bohrdt}, \citenamefont {Demler},\
  and\ \citenamefont {Grusdt}}]{Bohrdt2021}%
  \BibitemOpen
  \bibfield  {author} {\bibinfo {author} {\bibfnamefont {A.}~\bibnamefont
  {Bohrdt}}, \bibinfo {author} {\bibfnamefont {E.}~\bibnamefont {Demler}},\
  and\ \bibinfo {author} {\bibfnamefont {F.}~\bibnamefont {Grusdt}},\
  }\bibfield  {title} {\bibinfo {title} {Rotational resonances and regge-like
  trajectories in lightly doped antiferromagnets},\ }\href
  {https://doi.org/10.1103/PhysRevLett.127.197004} {\bibfield  {journal}
  {\bibinfo  {journal} {Phys. Rev. Lett.}\ }\textbf {\bibinfo {volume} {127}},\
  \bibinfo {pages} {197004} (\bibinfo {year} {2021}{\natexlab{a}})}\BibitemShut
  {NoStop}%
\bibitem [{\citenamefont {Hubig}\ \emph {et~al.}(2020)\citenamefont {Hubig},
  \citenamefont {Bohrdt}, \citenamefont {Knap}, \citenamefont {Grusdt},\ and\
  \citenamefont {Cirac}}]{Hubig2020}%
  \BibitemOpen
  \bibfield  {author} {\bibinfo {author} {\bibfnamefont {C.}~\bibnamefont
  {Hubig}}, \bibinfo {author} {\bibfnamefont {A.}~\bibnamefont {Bohrdt}},
  \bibinfo {author} {\bibfnamefont {M.}~\bibnamefont {Knap}}, \bibinfo {author}
  {\bibfnamefont {F.}~\bibnamefont {Grusdt}},\ and\ \bibinfo {author}
  {\bibfnamefont {J.~I.}\ \bibnamefont {Cirac}},\ }\bibfield  {title} {\bibinfo
  {title} {{Evaluation of time-dependent correlators after a local quench in
  iPEPS: hole motion in the t-J model}},\ }\href
  {https://doi.org/10.21468/SciPostPhys.8.2.021} {\bibfield  {journal}
  {\bibinfo  {journal} {SciPost Phys.}\ }\textbf {\bibinfo {volume} {8}},\
  \bibinfo {pages} {21} (\bibinfo {year} {2020})}\BibitemShut {NoStop}%
\bibitem [{\citenamefont {Bohrdt}\ \emph
  {et~al.}(2020{\natexlab{b}})\citenamefont {Bohrdt}, \citenamefont {Grusdt},\
  and\ \citenamefont {Knap}}]{Bohrdt2020Dyn}%
  \BibitemOpen
  \bibfield  {author} {\bibinfo {author} {\bibfnamefont {A.}~\bibnamefont
  {Bohrdt}}, \bibinfo {author} {\bibfnamefont {F.}~\bibnamefont {Grusdt}},\
  and\ \bibinfo {author} {\bibfnamefont {M.}~\bibnamefont {Knap}},\ }\bibfield
  {title} {\bibinfo {title} {Dynamical formation of a magnetic polaron in a
  two-dimensional quantum antiferromagnet},\ }\href
  {https://doi.org/10.1088/1367-2630/abcfee} {\bibfield  {journal} {\bibinfo
  {journal} {New Journal of Physics}\ }\textbf {\bibinfo {volume} {22}},\
  \bibinfo {pages} {123023} (\bibinfo {year} {2020}{\natexlab{b}})}\BibitemShut
  {NoStop}%
\bibitem [{\citenamefont {Ji}\ \emph {et~al.}(2021)\citenamefont {Ji},
  \citenamefont {Xu}, \citenamefont {Kendrick}, \citenamefont {Chiu},
  \citenamefont {Br\"uggenj\"urgen}, \citenamefont {Greif}, \citenamefont
  {Bohrdt}, \citenamefont {Grusdt}, \citenamefont {Demler}, \citenamefont
  {Lebrat},\ and\ \citenamefont {Greiner}}]{Ji2021}%
  \BibitemOpen
  \bibfield  {author} {\bibinfo {author} {\bibfnamefont {G.}~\bibnamefont
  {Ji}}, \bibinfo {author} {\bibfnamefont {M.}~\bibnamefont {Xu}}, \bibinfo
  {author} {\bibfnamefont {L.~H.}\ \bibnamefont {Kendrick}}, \bibinfo {author}
  {\bibfnamefont {C.~S.}\ \bibnamefont {Chiu}}, \bibinfo {author}
  {\bibfnamefont {J.~C.}\ \bibnamefont {Br\"uggenj\"urgen}}, \bibinfo {author}
  {\bibfnamefont {D.}~\bibnamefont {Greif}}, \bibinfo {author} {\bibfnamefont
  {A.}~\bibnamefont {Bohrdt}}, \bibinfo {author} {\bibfnamefont
  {F.}~\bibnamefont {Grusdt}}, \bibinfo {author} {\bibfnamefont
  {E.}~\bibnamefont {Demler}}, \bibinfo {author} {\bibfnamefont
  {M.}~\bibnamefont {Lebrat}},\ and\ \bibinfo {author} {\bibfnamefont
  {M.}~\bibnamefont {Greiner}},\ }\bibfield  {title} {\bibinfo {title}
  {Coupling a mobile hole to an antiferromagnetic spin background: Transient
  dynamics of a magnetic polaron},\ }\href
  {https://doi.org/10.1103/PhysRevX.11.021022} {\bibfield  {journal} {\bibinfo
  {journal} {Phys. Rev. X}\ }\textbf {\bibinfo {volume} {11}},\ \bibinfo
  {pages} {021022} (\bibinfo {year} {2021})}\BibitemShut {NoStop}%
\bibitem [{\citenamefont {Nielsen}\ \emph {et~al.}(2022)\citenamefont
  {Nielsen}, \citenamefont {Pohl},\ and\ \citenamefont
  {Bruun}}]{Nielsen2022arXiv}%
  \BibitemOpen
  \bibfield  {author} {\bibinfo {author} {\bibfnamefont {K.~K.}\ \bibnamefont
  {Nielsen}}, \bibinfo {author} {\bibfnamefont {T.}~\bibnamefont {Pohl}},\ and\
  \bibinfo {author} {\bibfnamefont {G.~M.}\ \bibnamefont {Bruun}},\ }\bibfield
  {title} {\bibinfo {title} {Non-equilibrium hole dynamics in antiferromagnets:
  damped strings and polarons},\ }\href {https://arxiv.org/abs/2203.04789}
  {\bibfield  {journal} {\bibinfo  {journal} {arXiv:2203.04789}\ } (\bibinfo
  {year} {2022})}\BibitemShut {NoStop}%
\bibitem [{\citenamefont {Bohrdt}\ \emph
  {et~al.}(2021{\natexlab{b}})\citenamefont {Bohrdt}, \citenamefont {Homeier},
  \citenamefont {Reinmoser}, \citenamefont {Demler},\ and\ \citenamefont
  {Grusdt}}]{Bohrdt2021review}%
  \BibitemOpen
  \bibfield  {author} {\bibinfo {author} {\bibfnamefont {A.}~\bibnamefont
  {Bohrdt}}, \bibinfo {author} {\bibfnamefont {L.}~\bibnamefont {Homeier}},
  \bibinfo {author} {\bibfnamefont {C.}~\bibnamefont {Reinmoser}}, \bibinfo
  {author} {\bibfnamefont {E.}~\bibnamefont {Demler}},\ and\ \bibinfo {author}
  {\bibfnamefont {F.}~\bibnamefont {Grusdt}},\ }\bibfield  {title} {\bibinfo
  {title} {Exploration of doped quantum magnets with ultracold atoms},\ }\href
  {https://doi.org/https://doi.org/10.1016/j.aop.2021.168651} {\bibfield
  {journal} {\bibinfo  {journal} {Annals of Physics}\ }\textbf {\bibinfo
  {volume} {435}},\ \bibinfo {pages} {168651} (\bibinfo {year}
  {2021}{\natexlab{b}})},\ \bibinfo {note} {special issue on Philip W.
  Anderson}\BibitemShut {NoStop}%
\bibitem [{\citenamefont {Grusdt}\ \emph
  {et~al.}(2018{\natexlab{b}})\citenamefont {Grusdt}, \citenamefont {Zhu},
  \citenamefont {Shi},\ and\ \citenamefont {Demler}}]{Grusdt2018SciPost}%
  \BibitemOpen
  \bibfield  {author} {\bibinfo {author} {\bibfnamefont {F.}~\bibnamefont
  {Grusdt}}, \bibinfo {author} {\bibfnamefont {Z.}~\bibnamefont {Zhu}},
  \bibinfo {author} {\bibfnamefont {T.}~\bibnamefont {Shi}},\ and\ \bibinfo
  {author} {\bibfnamefont {E.}~\bibnamefont {Demler}},\ }\bibfield  {title}
  {\bibinfo {title} {{Meson formation in mixed-dimensional t-J models}},\
  }\href {https://doi.org/10.21468/SciPostPhys.5.6.057} {\bibfield  {journal}
  {\bibinfo  {journal} {SciPost Phys.}\ }\textbf {\bibinfo {volume} {5}},\
  \bibinfo {pages} {57} (\bibinfo {year} {2018}{\natexlab{b}})}\BibitemShut
  {NoStop}%
\bibitem [{\citenamefont {Hirthe}\ \emph {et~al.}(2022)\citenamefont {Hirthe},
  \citenamefont {Chalopin}, \citenamefont {Bourgund}, \citenamefont {Bojovic},
  \citenamefont {Bohrdt}, \citenamefont {Demler}, \citenamefont {Grusdt},
  \citenamefont {Bloch},\ and\ \citenamefont {Hilker}}]{Hirthe2022arXiv}%
  \BibitemOpen
  \bibfield  {author} {\bibinfo {author} {\bibfnamefont {S.}~\bibnamefont
  {Hirthe}}, \bibinfo {author} {\bibfnamefont {T.}~\bibnamefont {Chalopin}},
  \bibinfo {author} {\bibfnamefont {D.}~\bibnamefont {Bourgund}}, \bibinfo
  {author} {\bibfnamefont {P.}~\bibnamefont {Bojovic}}, \bibinfo {author}
  {\bibfnamefont {A.}~\bibnamefont {Bohrdt}}, \bibinfo {author} {\bibfnamefont
  {E.}~\bibnamefont {Demler}}, \bibinfo {author} {\bibfnamefont
  {F.}~\bibnamefont {Grusdt}}, \bibinfo {author} {\bibfnamefont
  {I.}~\bibnamefont {Bloch}},\ and\ \bibinfo {author} {\bibfnamefont {T.~A.}\
  \bibnamefont {Hilker}},\ }\bibfield  {title} {\bibinfo {title} {Magnetically
  mediated hole pairing in fermionic ladders of ultracold atoms},\ }\href
  {https://arxiv.org/abs/2203.10027} {\bibfield  {journal} {\bibinfo  {journal}
  {arXiv:2203.10027}\ } (\bibinfo {year} {2022})}\BibitemShut {NoStop}%
\bibitem [{\citenamefont {Stewart}\ \emph {et~al.}(2008)\citenamefont
  {Stewart}, \citenamefont {Gaebler},\ and\ \citenamefont {Jin}}]{Stewart2008}%
  \BibitemOpen
  \bibfield  {author} {\bibinfo {author} {\bibfnamefont {J.~T.}\ \bibnamefont
  {Stewart}}, \bibinfo {author} {\bibfnamefont {J.~P.}\ \bibnamefont
  {Gaebler}},\ and\ \bibinfo {author} {\bibfnamefont {D.~S.}\ \bibnamefont
  {Jin}},\ }\bibfield  {title} {\bibinfo {title} {Using photoemission
  spectroscopy to probe a strongly interacting fermi gas},\ }\href
  {http://dx.doi.org/10.1038/nature07172} {\bibfield  {journal} {\bibinfo
  {journal} {Nature}\ }\textbf {\bibinfo {volume} {454}},\ \bibinfo {pages}
  {744} (\bibinfo {year} {2008})}\BibitemShut {NoStop}%
\bibitem [{\citenamefont {Feld}\ \emph {et~al.}(2011)\citenamefont {Feld},
  \citenamefont {Fr\"ohlich}, \citenamefont {Vogt}, \citenamefont
  {Koschorreck},\ and\ \citenamefont {K\"ohl}}]{Feld2011}%
  \BibitemOpen
  \bibfield  {author} {\bibinfo {author} {\bibfnamefont {M.}~\bibnamefont
  {Feld}}, \bibinfo {author} {\bibfnamefont {B.}~\bibnamefont {Fr\"ohlich}},
  \bibinfo {author} {\bibfnamefont {E.}~\bibnamefont {Vogt}}, \bibinfo {author}
  {\bibfnamefont {M.}~\bibnamefont {Koschorreck}},\ and\ \bibinfo {author}
  {\bibfnamefont {M.}~\bibnamefont {K\"ohl}},\ }\bibfield  {title} {\bibinfo
  {title} {Observation of a pairing pseudogap in a two-dimensional fermi gas},\
  }\href {https://doi.org/10.1038/nature10627} {\bibfield  {journal} {\bibinfo
  {journal} {Nature}\ }\textbf {\bibinfo {volume} {480}},\ \bibinfo {pages}
  {75} (\bibinfo {year} {2011})}\BibitemShut {NoStop}%
\bibitem [{\citenamefont {Bohrdt}\ \emph {et~al.}(2018)\citenamefont {Bohrdt},
  \citenamefont {Greif}, \citenamefont {Demler}, \citenamefont {Knap},\ and\
  \citenamefont {Grusdt}}]{Bohrdt2018}%
  \BibitemOpen
  \bibfield  {author} {\bibinfo {author} {\bibfnamefont {A.}~\bibnamefont
  {Bohrdt}}, \bibinfo {author} {\bibfnamefont {D.}~\bibnamefont {Greif}},
  \bibinfo {author} {\bibfnamefont {E.}~\bibnamefont {Demler}}, \bibinfo
  {author} {\bibfnamefont {M.}~\bibnamefont {Knap}},\ and\ \bibinfo {author}
  {\bibfnamefont {F.}~\bibnamefont {Grusdt}},\ }\bibfield  {title} {\bibinfo
  {title} {Angle-resolved photoemission spectroscopy with quantum gas
  microscopes},\ }\href {https://doi.org/10.1103/PhysRevB.97.125117} {\bibfield
   {journal} {\bibinfo  {journal} {Phys. Rev. B}\ }\textbf {\bibinfo {volume}
  {97}},\ \bibinfo {pages} {125117} (\bibinfo {year} {2018})}\BibitemShut
  {NoStop}%
\bibitem [{\citenamefont {Brown}\ \emph {et~al.}(2019)\citenamefont {Brown},
  \citenamefont {Guardado-Sanchez}, \citenamefont {Spar}, \citenamefont
  {Huang}, \citenamefont {Devereaux},\ and\ \citenamefont {Bakr}}]{Brown2019}%
  \BibitemOpen
  \bibfield  {author} {\bibinfo {author} {\bibfnamefont {P.~T.}\ \bibnamefont
  {Brown}}, \bibinfo {author} {\bibfnamefont {E.}~\bibnamefont
  {Guardado-Sanchez}}, \bibinfo {author} {\bibfnamefont {B.~M.}\ \bibnamefont
  {Spar}}, \bibinfo {author} {\bibfnamefont {E.~W.}\ \bibnamefont {Huang}},
  \bibinfo {author} {\bibfnamefont {T.~P.}\ \bibnamefont {Devereaux}},\ and\
  \bibinfo {author} {\bibfnamefont {W.~S.}\ \bibnamefont {Bakr}},\ }\bibfield
  {title} {\bibinfo {title} {Angle-resolved photoemission spectroscopy of a
  fermi–hubbard system},\ }\href {https://doi.org/10.1038/s41567-019-0696-0}
  {\bibfield  {journal} {\bibinfo  {journal} {Nature Physics}\ }\textbf
  {\bibinfo {volume} {16}},\ \bibinfo {pages} {26–31} (\bibinfo {year}
  {2019})}\BibitemShut {NoStop}%
\bibitem [{\citenamefont {Endres}\ \emph {et~al.}(2011)\citenamefont {Endres},
  \citenamefont {Cheneau}, \citenamefont {Fukuhara}, \citenamefont
  {Weitenberg}, \citenamefont {Schauss}, \citenamefont {Gross}, \citenamefont
  {Mazza}, \citenamefont {Banuls}, \citenamefont {Pollet}, \citenamefont
  {Bloch},\ and\ \citenamefont {Kuhr}}]{Endres2011}%
  \BibitemOpen
  \bibfield  {author} {\bibinfo {author} {\bibfnamefont {M.}~\bibnamefont
  {Endres}}, \bibinfo {author} {\bibfnamefont {M.}~\bibnamefont {Cheneau}},
  \bibinfo {author} {\bibfnamefont {T.}~\bibnamefont {Fukuhara}}, \bibinfo
  {author} {\bibfnamefont {C.}~\bibnamefont {Weitenberg}}, \bibinfo {author}
  {\bibfnamefont {P.}~\bibnamefont {Schauss}}, \bibinfo {author} {\bibfnamefont
  {C.}~\bibnamefont {Gross}}, \bibinfo {author} {\bibfnamefont
  {L.}~\bibnamefont {Mazza}}, \bibinfo {author} {\bibfnamefont {M.~C.}\
  \bibnamefont {Banuls}}, \bibinfo {author} {\bibfnamefont {L.}~\bibnamefont
  {Pollet}}, \bibinfo {author} {\bibfnamefont {I.}~\bibnamefont {Bloch}},\ and\
  \bibinfo {author} {\bibfnamefont {S.}~\bibnamefont {Kuhr}},\ }\bibfield
  {title} {\bibinfo {title} {Observation of correlated particle-hole pairs and
  string order in low-dimensional mott insulators},\ }\href
  {https://doi.org/10.1126/science.1209284} {\bibfield  {journal} {\bibinfo
  {journal} {Science}\ }\textbf {\bibinfo {volume} {334}},\ \bibinfo {pages}
  {200} (\bibinfo {year} {2011})}\BibitemShut {NoStop}%
\bibitem [{\citenamefont {Chiu}\ \emph {et~al.}(2019)\citenamefont {Chiu},
  \citenamefont {Ji}, \citenamefont {Bohrdt}, \citenamefont {Xu}, \citenamefont
  {Knap}, \citenamefont {Demler}, \citenamefont {Grusdt}, \citenamefont
  {Greiner},\ and\ \citenamefont {Greif}}]{Chiu2019}%
  \BibitemOpen
  \bibfield  {author} {\bibinfo {author} {\bibfnamefont {C.~S.}\ \bibnamefont
  {Chiu}}, \bibinfo {author} {\bibfnamefont {G.}~\bibnamefont {Ji}}, \bibinfo
  {author} {\bibfnamefont {A.}~\bibnamefont {Bohrdt}}, \bibinfo {author}
  {\bibfnamefont {M.}~\bibnamefont {Xu}}, \bibinfo {author} {\bibfnamefont
  {M.}~\bibnamefont {Knap}}, \bibinfo {author} {\bibfnamefont {E.}~\bibnamefont
  {Demler}}, \bibinfo {author} {\bibfnamefont {F.}~\bibnamefont {Grusdt}},
  \bibinfo {author} {\bibfnamefont {M.}~\bibnamefont {Greiner}},\ and\ \bibinfo
  {author} {\bibfnamefont {D.}~\bibnamefont {Greif}},\ }\bibfield  {title}
  {\bibinfo {title} {String patterns in the doped hubbard model},\ }\href
  {https://doi.org/10.1126/science.aav3587} {\bibfield  {journal} {\bibinfo
  {journal} {Science}\ }\textbf {\bibinfo {volume} {365}},\ \bibinfo {pages}
  {251} (\bibinfo {year} {2019})},\ \Eprint
  {https://arxiv.org/abs/https://www.science.org/doi/pdf/10.1126/science.aav3587}
  {https://www.science.org/doi/pdf/10.1126/science.aav3587} \BibitemShut
  {NoStop}%
\bibitem [{\citenamefont {Koepsell}\ \emph {et~al.}(2019)\citenamefont
  {Koepsell}, \citenamefont {Vijayan}, \citenamefont {Sompet}, \citenamefont
  {Grusdt}, \citenamefont {Hilker}, \citenamefont {Demler}, \citenamefont
  {Salomon}, \citenamefont {Bloch},\ and\ \citenamefont
  {Gross}}]{Koepsell2019}%
  \BibitemOpen
  \bibfield  {author} {\bibinfo {author} {\bibfnamefont {J.}~\bibnamefont
  {Koepsell}}, \bibinfo {author} {\bibfnamefont {J.}~\bibnamefont {Vijayan}},
  \bibinfo {author} {\bibfnamefont {P.}~\bibnamefont {Sompet}}, \bibinfo
  {author} {\bibfnamefont {F.}~\bibnamefont {Grusdt}}, \bibinfo {author}
  {\bibfnamefont {T.~A.}\ \bibnamefont {Hilker}}, \bibinfo {author}
  {\bibfnamefont {E.}~\bibnamefont {Demler}}, \bibinfo {author} {\bibfnamefont
  {G.}~\bibnamefont {Salomon}}, \bibinfo {author} {\bibfnamefont
  {I.}~\bibnamefont {Bloch}},\ and\ \bibinfo {author} {\bibfnamefont
  {C.}~\bibnamefont {Gross}},\ }\bibfield  {title} {\bibinfo {title} {Imaging
  magnetic polarons in the doped fermi–hubbard model},\ }\href
  {https://doi.org/10.1038/s41586-019-1463-1} {\bibfield  {journal} {\bibinfo
  {journal} {Nature}\ }\textbf {\bibinfo {volume} {572}},\ \bibinfo {pages}
  {358–362} (\bibinfo {year} {2019})}\BibitemShut {NoStop}%
\bibitem [{\citenamefont {Auerbach}(1994)}]{Auerbach1994}%
  \BibitemOpen
  \bibfield  {author} {\bibinfo {author} {\bibfnamefont {A.}~\bibnamefont
  {Auerbach}},\ }\href {https://books.google.de/books?id=jAmbQgAACAAJ} {\emph
  {\bibinfo {title} {Interacting Electrons and Quantum Magnetism}}},\ Graduate
  texts in contemporary physics\ (\bibinfo  {publisher} {Springer-Verlag},\
  \bibinfo {year} {1994})\BibitemShut {NoStop}%
\bibitem [{\citenamefont {Kogoj}\ \emph {et~al.}(2014)\citenamefont {Kogoj},
  \citenamefont {Lenarcic}, \citenamefont {Golez}, \citenamefont
  {Mierzejewski}, \citenamefont {Prelovcek},\ and\ \citenamefont
  {Bonca}}]{Kogoj2014}%
  \BibitemOpen
  \bibfield  {author} {\bibinfo {author} {\bibfnamefont {J.}~\bibnamefont
  {Kogoj}}, \bibinfo {author} {\bibfnamefont {Z.}~\bibnamefont {Lenarcic}},
  \bibinfo {author} {\bibfnamefont {D.}~\bibnamefont {Golez}}, \bibinfo
  {author} {\bibfnamefont {M.}~\bibnamefont {Mierzejewski}}, \bibinfo {author}
  {\bibfnamefont {P.}~\bibnamefont {Prelovcek}},\ and\ \bibinfo {author}
  {\bibfnamefont {J.}~\bibnamefont {Bonca}},\ }\bibfield  {title} {\bibinfo
  {title} {Multistage dynamics of the spin-lattice polaron formation},\ }\href
  {https://doi.org/10.1103/PhysRevB.90.125104} {\bibfield  {journal} {\bibinfo
  {journal} {Phys. Rev. B}\ }\textbf {\bibinfo {volume} {90}},\ \bibinfo
  {pages} {125104} (\bibinfo {year} {2014})}\BibitemShut {NoStop}%
\bibitem [{\citenamefont {Kruis}\ \emph {et~al.}(2004)\citenamefont {Kruis},
  \citenamefont {McCulloch}, \citenamefont {Nussinov},\ and\ \citenamefont
  {Zaanen}}]{Kruis2004a}%
  \BibitemOpen
  \bibfield  {author} {\bibinfo {author} {\bibfnamefont {H.~V.}\ \bibnamefont
  {Kruis}}, \bibinfo {author} {\bibfnamefont {I.~P.}\ \bibnamefont
  {McCulloch}}, \bibinfo {author} {\bibfnamefont {Z.}~\bibnamefont
  {Nussinov}},\ and\ \bibinfo {author} {\bibfnamefont {J.}~\bibnamefont
  {Zaanen}},\ }\bibfield  {title} {\bibinfo {title} {Geometry and the hidden
  order of luttinger liquids: The universality of squeezed space},\ }\href
  {https://doi.org/10.1103/PhysRevB.70.075109} {\bibfield  {journal} {\bibinfo
  {journal} {Phys. Rev. B}\ }\textbf {\bibinfo {volume} {70}},\ \bibinfo
  {pages} {075109} (\bibinfo {year} {2004})}\BibitemShut {NoStop}%
\bibitem [{\citenamefont {Grusdt}\ and\ \citenamefont
  {Pollet}(2020)}]{Grusdt2020PRL}%
  \BibitemOpen
  \bibfield  {author} {\bibinfo {author} {\bibfnamefont {F.}~\bibnamefont
  {Grusdt}}\ and\ \bibinfo {author} {\bibfnamefont {L.}~\bibnamefont
  {Pollet}},\ }\bibfield  {title} {\bibinfo {title} {Z2 parton phases in the
  mixed-dimensional t−jz model},\ }\bibfield  {journal} {\bibinfo  {journal}
  {Physical Review Letters}\ }\textbf {\bibinfo {volume} {125}},\ \href
  {https://doi.org/10.1103/physrevlett.125.256401}
  {10.1103/physrevlett.125.256401} (\bibinfo {year} {2020})\BibitemShut
  {NoStop}%
\bibitem [{\citenamefont {{Bulaevski}}\ \emph {et~al.}(1968)\citenamefont
  {{Bulaevski}}, \citenamefont {{Nagaev}},\ and\ \citenamefont
  {{Khomski{\v{i}}}}}]{Bulaevski1968}%
  \BibitemOpen
  \bibfield  {author} {\bibinfo {author} {\bibfnamefont {L.~N.}\ \bibnamefont
  {{Bulaevski}}}, \bibinfo {author} {\bibfnamefont {{\'E}.~L.}\ \bibnamefont
  {{Nagaev}}},\ and\ \bibinfo {author} {\bibfnamefont {D.~I.}\ \bibnamefont
  {{Khomski{\v{i}}}}},\ }\bibfield  {title} {\bibinfo {title} {{A New Type of
  Auto-localized State of a Conduction Electron in an Antiferromagnetic
  Semiconductor}},\ }\href@noop {} {\bibfield  {journal} {\bibinfo  {journal}
  {Soviet Journal of Experimental and Theoretical Physics}\ }\textbf {\bibinfo
  {volume} {27}},\ \bibinfo {pages} {836} (\bibinfo {year} {1968})}\BibitemShut
  {NoStop}%
\bibitem [{\citenamefont {Trugman}(1988)}]{Trugman1988}%
  \BibitemOpen
  \bibfield  {author} {\bibinfo {author} {\bibfnamefont {S.~A.}\ \bibnamefont
  {Trugman}},\ }\bibfield  {title} {\bibinfo {title} {Interaction of holes in a
  hubbard antiferromagnet and high-temperature superconductivity},\ }\href
  {https://doi.org/10.1103/PhysRevB.37.1597} {\bibfield  {journal} {\bibinfo
  {journal} {Phys. Rev. B}\ }\textbf {\bibinfo {volume} {37}},\ \bibinfo
  {pages} {1597} (\bibinfo {year} {1988})}\BibitemShut {NoStop}%
\bibitem [{\citenamefont {Verresen}\ \emph {et~al.}(2019)\citenamefont
  {Verresen}, \citenamefont {Moessner},\ and\ \citenamefont
  {Pollmann}}]{Verresen2018}%
  \BibitemOpen
  \bibfield  {author} {\bibinfo {author} {\bibfnamefont {R.}~\bibnamefont
  {Verresen}}, \bibinfo {author} {\bibfnamefont {R.}~\bibnamefont {Moessner}},\
  and\ \bibinfo {author} {\bibfnamefont {F.}~\bibnamefont {Pollmann}},\
  }\bibfield  {title} {\bibinfo {title} {Avoided quasiparticle decay from
  strong quantum interactions},\ }\href
  {https://doi.org/10.1038/s41567-019-0535-3} {\bibfield  {journal} {\bibinfo
  {journal} {Nature Physics}\ }\textbf {\bibinfo {volume} {15}} (\bibinfo
  {year} {2019})}\BibitemShut {NoStop}%
\bibitem [{\citenamefont {Kane}\ \emph {et~al.}(1989)\citenamefont {Kane},
  \citenamefont {Lee},\ and\ \citenamefont {Read}}]{Kane1989}%
  \BibitemOpen
  \bibfield  {author} {\bibinfo {author} {\bibfnamefont {C.~L.}\ \bibnamefont
  {Kane}}, \bibinfo {author} {\bibfnamefont {P.~A.}\ \bibnamefont {Lee}},\ and\
  \bibinfo {author} {\bibfnamefont {N.}~\bibnamefont {Read}},\ }\bibfield
  {title} {\bibinfo {title} {Motion of a single hole in a quantum
  antiferromagnet},\ }\href {https://doi.org/10.1103/PhysRevB.39.6880}
  {\bibfield  {journal} {\bibinfo  {journal} {Phys. Rev. B}\ }\textbf {\bibinfo
  {volume} {39}},\ \bibinfo {pages} {6880} (\bibinfo {year}
  {1989})}\BibitemShut {NoStop}%
\bibitem [{\citenamefont {Sachdev}(1989)}]{Sachdev1989}%
  \BibitemOpen
  \bibfield  {author} {\bibinfo {author} {\bibfnamefont {S.}~\bibnamefont
  {Sachdev}},\ }\bibfield  {title} {\bibinfo {title} {Hole motion in a quantum
  n\'eel state},\ }\href {https://doi.org/10.1103/PhysRevB.39.12232} {\bibfield
   {journal} {\bibinfo  {journal} {Phys. Rev. B}\ }\textbf {\bibinfo {volume}
  {39}},\ \bibinfo {pages} {12232} (\bibinfo {year} {1989})}\BibitemShut
  {NoStop}%
\bibitem [{\citenamefont {Martinez}\ and\ \citenamefont
  {Horsch}(1991)}]{Martinez1991}%
  \BibitemOpen
  \bibfield  {author} {\bibinfo {author} {\bibfnamefont {G.}~\bibnamefont
  {Martinez}}\ and\ \bibinfo {author} {\bibfnamefont {P.}~\bibnamefont
  {Horsch}},\ }\bibfield  {title} {\bibinfo {title} {Spin polarons in the t-j
  model},\ }\href {https://doi.org/10.1103/PhysRevB.44.317} {\bibfield
  {journal} {\bibinfo  {journal} {Phys. Rev. B}\ }\textbf {\bibinfo {volume}
  {44}},\ \bibinfo {pages} {317} (\bibinfo {year} {1991})}\BibitemShut
  {NoStop}%
\bibitem [{\citenamefont {Weitenberg}\ \emph {et~al.}(2011)\citenamefont
  {Weitenberg}, \citenamefont {Endres}, \citenamefont {Sherson}, \citenamefont
  {Cheneau}, \citenamefont {Schauss}, \citenamefont {Fukuhara}, \citenamefont
  {Bloch},\ and\ \citenamefont {Kuhr}}]{Weitenberg2011}%
  \BibitemOpen
  \bibfield  {author} {\bibinfo {author} {\bibfnamefont {C.}~\bibnamefont
  {Weitenberg}}, \bibinfo {author} {\bibfnamefont {M.}~\bibnamefont {Endres}},
  \bibinfo {author} {\bibfnamefont {J.~F.}\ \bibnamefont {Sherson}}, \bibinfo
  {author} {\bibfnamefont {M.}~\bibnamefont {Cheneau}}, \bibinfo {author}
  {\bibfnamefont {P.}~\bibnamefont {Schauss}}, \bibinfo {author} {\bibfnamefont
  {T.}~\bibnamefont {Fukuhara}}, \bibinfo {author} {\bibfnamefont
  {I.}~\bibnamefont {Bloch}},\ and\ \bibinfo {author} {\bibfnamefont
  {S.}~\bibnamefont {Kuhr}},\ }\bibfield  {title} {\bibinfo {title}
  {Single-spin addressing in an atomic mott insulator},\ }\href
  {https://doi.org/10.1038/nature09827} {\bibfield  {journal} {\bibinfo
  {journal} {Nature}\ }\textbf {\bibinfo {volume} {471}},\ \bibinfo {pages}
  {319} (\bibinfo {year} {2011})}\BibitemShut {NoStop}%
\bibitem [{\citenamefont {Cheuk}\ \emph {et~al.}(2015)\citenamefont {Cheuk},
  \citenamefont {Nichols}, \citenamefont {Okan}, \citenamefont {Gersdorf},
  \citenamefont {Ramasesh}, \citenamefont {Bakr}, \citenamefont {Lompe},\ and\
  \citenamefont {Zwierlein}}]{Cheuk2015}%
  \BibitemOpen
  \bibfield  {author} {\bibinfo {author} {\bibfnamefont {L.~W.}\ \bibnamefont
  {Cheuk}}, \bibinfo {author} {\bibfnamefont {M.~A.}\ \bibnamefont {Nichols}},
  \bibinfo {author} {\bibfnamefont {M.}~\bibnamefont {Okan}}, \bibinfo {author}
  {\bibfnamefont {T.}~\bibnamefont {Gersdorf}}, \bibinfo {author}
  {\bibfnamefont {V.~V.}\ \bibnamefont {Ramasesh}}, \bibinfo {author}
  {\bibfnamefont {W.~S.}\ \bibnamefont {Bakr}}, \bibinfo {author}
  {\bibfnamefont {T.}~\bibnamefont {Lompe}},\ and\ \bibinfo {author}
  {\bibfnamefont {M.~W.}\ \bibnamefont {Zwierlein}},\ }\bibfield  {title}
  {\bibinfo {title} {Quantum-gas microscope for fermionic atoms},\ }\href
  {https://doi.org/10.1103/PhysRevLett.114.193001} {\bibfield  {journal}
  {\bibinfo  {journal} {Phys. Rev. Lett.}\ }\textbf {\bibinfo {volume} {114}},\
  \bibinfo {pages} {193001} (\bibinfo {year} {2015})}\BibitemShut {NoStop}%
\bibitem [{\citenamefont {Nichols}\ \emph {et~al.}(2019)\citenamefont
  {Nichols}, \citenamefont {Cheuk}, \citenamefont {Okan}, \citenamefont
  {Hartke}, \citenamefont {Mendez}, \citenamefont {Senthil}, \citenamefont
  {Khatami}, \citenamefont {Zhang},\ and\ \citenamefont
  {Zwierlein}}]{Nichols2018}%
  \BibitemOpen
  \bibfield  {author} {\bibinfo {author} {\bibfnamefont {M.~A.}\ \bibnamefont
  {Nichols}}, \bibinfo {author} {\bibfnamefont {L.~W.}\ \bibnamefont {Cheuk}},
  \bibinfo {author} {\bibfnamefont {M.}~\bibnamefont {Okan}}, \bibinfo {author}
  {\bibfnamefont {T.~R.}\ \bibnamefont {Hartke}}, \bibinfo {author}
  {\bibfnamefont {E.}~\bibnamefont {Mendez}}, \bibinfo {author} {\bibfnamefont
  {T.}~\bibnamefont {Senthil}}, \bibinfo {author} {\bibfnamefont
  {E.}~\bibnamefont {Khatami}}, \bibinfo {author} {\bibfnamefont
  {H.}~\bibnamefont {Zhang}},\ and\ \bibinfo {author} {\bibfnamefont {M.~W.}\
  \bibnamefont {Zwierlein}},\ }\bibfield  {title} {\bibinfo {title} {Spin
  transport in a mott insulator of ultracold fermions},\ }\bibfield  {journal}
  {\bibinfo  {journal} {Science 363, 383}\ }\href
  {https://doi.org/10.1126/science.aat4387} {10.1126/science.aat4387} (\bibinfo
  {year} {2019}),\ \Eprint {https://arxiv.org/abs/1802.10018v2} {1802.10018v2}
  \BibitemShut {NoStop}%
\bibitem [{\citenamefont {Guardado-Sanchez}\ \emph {et~al.}(2018)\citenamefont
  {Guardado-Sanchez}, \citenamefont {Brown}, \citenamefont {Mitra},
  \citenamefont {Devakul}, \citenamefont {Huse}, \citenamefont {Schauss},\ and\
  \citenamefont {Bakr}}]{GuardadoSanchez2017}%
  \BibitemOpen
  \bibfield  {author} {\bibinfo {author} {\bibfnamefont {E.}~\bibnamefont
  {Guardado-Sanchez}}, \bibinfo {author} {\bibfnamefont {P.~T.}\ \bibnamefont
  {Brown}}, \bibinfo {author} {\bibfnamefont {D.}~\bibnamefont {Mitra}},
  \bibinfo {author} {\bibfnamefont {T.}~\bibnamefont {Devakul}}, \bibinfo
  {author} {\bibfnamefont {D.~A.}\ \bibnamefont {Huse}}, \bibinfo {author}
  {\bibfnamefont {P.}~\bibnamefont {Schauss}},\ and\ \bibinfo {author}
  {\bibfnamefont {W.~S.}\ \bibnamefont {Bakr}},\ }\bibfield  {title} {\bibinfo
  {title} {Probing quench dynamics across a quantum phase transition into a 2d
  ising antiferromagnet},\ }\bibfield  {journal} {\bibinfo  {journal} {Phys.
  Rev. X 8, 021069}\ }\href {https://doi.org/10.1103/PhysRevX.8.021069}
  {10.1103/PhysRevX.8.021069} (\bibinfo {year} {2018}),\ \Eprint
  {https://arxiv.org/abs/1711.00887v1} {1711.00887v1} \BibitemShut {NoStop}%
\bibitem [{\citenamefont {Trotzky}\ \emph {et~al.}(2008)\citenamefont
  {Trotzky}, \citenamefont {Cheinet}, \citenamefont {F\"olling}, \citenamefont
  {Feld}, \citenamefont {Schnorrberger}, \citenamefont {Rey}, \citenamefont
  {Polkovnikov}, \citenamefont {Demler}, \citenamefont {Lukin},\ and\
  \citenamefont {Bloch}}]{Trotzky2008}%
  \BibitemOpen
  \bibfield  {author} {\bibinfo {author} {\bibfnamefont {S.}~\bibnamefont
  {Trotzky}}, \bibinfo {author} {\bibfnamefont {P.}~\bibnamefont {Cheinet}},
  \bibinfo {author} {\bibfnamefont {S.}~\bibnamefont {F\"olling}}, \bibinfo
  {author} {\bibfnamefont {M.}~\bibnamefont {Feld}}, \bibinfo {author}
  {\bibfnamefont {U.}~\bibnamefont {Schnorrberger}}, \bibinfo {author}
  {\bibfnamefont {A.~M.}\ \bibnamefont {Rey}}, \bibinfo {author} {\bibfnamefont
  {A.}~\bibnamefont {Polkovnikov}}, \bibinfo {author} {\bibfnamefont {E.~A.}\
  \bibnamefont {Demler}}, \bibinfo {author} {\bibfnamefont {M.~D.}\
  \bibnamefont {Lukin}},\ and\ \bibinfo {author} {\bibfnamefont
  {I.}~\bibnamefont {Bloch}},\ }\bibfield  {title} {\bibinfo {title}
  {Time-resolved observation and control of superexchange interactions with
  ultracold atoms in optical lattices},\ }\href
  {http://science.sciencemag.org/content/319/5861/295.abstract} {\bibfield
  {journal} {\bibinfo  {journal} {Science}\ }\textbf {\bibinfo {volume}
  {319}},\ \bibinfo {pages} {295} (\bibinfo {year} {2008})}\BibitemShut
  {NoStop}%
\bibitem [{\citenamefont {Dimitrova}\ \emph {et~al.}(2020)\citenamefont
  {Dimitrova}, \citenamefont {Jepsen}, \citenamefont {Buyskikh}, \citenamefont
  {Venegas-Gomez}, \citenamefont {Amato-Grill}, \citenamefont {Daley},\ and\
  \citenamefont {Ketterle}}]{Dimitrova2019}%
  \BibitemOpen
  \bibfield  {author} {\bibinfo {author} {\bibfnamefont {I.}~\bibnamefont
  {Dimitrova}}, \bibinfo {author} {\bibfnamefont {N.}~\bibnamefont {Jepsen}},
  \bibinfo {author} {\bibfnamefont {A.}~\bibnamefont {Buyskikh}}, \bibinfo
  {author} {\bibfnamefont {A.}~\bibnamefont {Venegas-Gomez}}, \bibinfo {author}
  {\bibfnamefont {J.}~\bibnamefont {Amato-Grill}}, \bibinfo {author}
  {\bibfnamefont {A.}~\bibnamefont {Daley}},\ and\ \bibinfo {author}
  {\bibfnamefont {W.}~\bibnamefont {Ketterle}},\ }\bibfield  {title} {\bibinfo
  {title} {Enhanced superexchange in a tilted mott insulator},\ }\bibfield
  {journal} {\bibinfo  {journal} {Phys. Rev. Lett. 124, 043204}\ }\href
  {https://doi.org/10.1103/PhysRevLett.124.043204}
  {10.1103/PhysRevLett.124.043204} (\bibinfo {year} {2020}),\ \Eprint
  {https://arxiv.org/abs/1908.09870v2} {1908.09870v2} \BibitemShut {NoStop}%
\bibitem [{\citenamefont {\text{Arute}~et al.}(2020)}]{Arute2020}%
  \BibitemOpen
  \bibfield  {author} {\bibinfo {author} {\bibnamefont {\text{Arute}~et al.}},\
  }\href@noop {} {\bibinfo {title} {Observation of separated dynamics of charge
  and spin in the fermi-hubbard model}} (\bibinfo {year} {2020}),\ \Eprint
  {https://arxiv.org/abs/2010.07965} {arXiv:2010.07965 [quant-ph]} \BibitemShut
  {NoStop}%
\bibitem [{\citenamefont {Paeckel}\ \emph {et~al.}(2019)\citenamefont
  {Paeckel}, \citenamefont {K\"ohler}, \citenamefont {Swoboda}, \citenamefont
  {Manmana}, \citenamefont {Schollw\"ock},\ and\ \citenamefont
  {Hubig}}]{Paeckel2019}%
  \BibitemOpen
  \bibfield  {author} {\bibinfo {author} {\bibfnamefont {S.}~\bibnamefont
  {Paeckel}}, \bibinfo {author} {\bibfnamefont {T.}~\bibnamefont {K\"ohler}},
  \bibinfo {author} {\bibfnamefont {A.}~\bibnamefont {Swoboda}}, \bibinfo
  {author} {\bibfnamefont {S.~R.}\ \bibnamefont {Manmana}}, \bibinfo {author}
  {\bibfnamefont {U.}~\bibnamefont {Schollw\"ock}},\ and\ \bibinfo {author}
  {\bibfnamefont {C.}~\bibnamefont {Hubig}},\ }\bibfield  {title} {\bibinfo
  {title} {Time-evolution methods for matrix-product states},\ }\href
  {https://doi.org/10.1016/j.aop.2019.167998} {\bibfield  {journal} {\bibinfo
  {journal} {Annals of Physics}\ }\textbf {\bibinfo {volume} {411}},\ \bibinfo
  {pages} {167998} (\bibinfo {year} {2019})}\BibitemShut {NoStop}%
\bibitem [{\citenamefont {Kj\"all}\ \emph {et~al.}(2013)\citenamefont
  {Kj\"all}, \citenamefont {Zaletel}, \citenamefont {Mong}, \citenamefont
  {Bardarson},\ and\ \citenamefont {Pollmann}}]{Kjaell2013}%
  \BibitemOpen
  \bibfield  {author} {\bibinfo {author} {\bibfnamefont {J.}~\bibnamefont
  {Kj\"all}}, \bibinfo {author} {\bibfnamefont {M.}~\bibnamefont {Zaletel}},
  \bibinfo {author} {\bibfnamefont {R.}~\bibnamefont {Mong}}, \bibinfo {author}
  {\bibfnamefont {J.}~\bibnamefont {Bardarson}},\ and\ \bibinfo {author}
  {\bibfnamefont {F.}~\bibnamefont {Pollmann}},\ }\bibfield  {title} {\bibinfo
  {title} {Phase diagram of the anisotropic spin-2 xxz model: Infinite-system
  density matrix renormalization group study},\ }\href
  {https://doi.org/10.1103/PhysRevB.87.235106} {\bibfield  {journal} {\bibinfo
  {journal} {Physical Review B}\ }\textbf {\bibinfo {volume} {87}} (\bibinfo
  {year} {2013})}\BibitemShut {NoStop}%
\bibitem [{\citenamefont {Zaletel}\ \emph {et~al.}(2015)\citenamefont
  {Zaletel}, \citenamefont {Mong}, \citenamefont {Karrasch}, \citenamefont
  {Moore},\ and\ \citenamefont {Pollmann}}]{Zaletel2015}%
  \BibitemOpen
  \bibfield  {author} {\bibinfo {author} {\bibfnamefont {M.~P.}\ \bibnamefont
  {Zaletel}}, \bibinfo {author} {\bibfnamefont {R.~S.~K.}\ \bibnamefont
  {Mong}}, \bibinfo {author} {\bibfnamefont {C.}~\bibnamefont {Karrasch}},
  \bibinfo {author} {\bibfnamefont {J.~E.}\ \bibnamefont {Moore}},\ and\
  \bibinfo {author} {\bibfnamefont {F.}~\bibnamefont {Pollmann}},\ }\bibfield
  {title} {\bibinfo {title} {Time-evolving a matrix product state with
  long-ranged interactions},\ }\href
  {https://doi.org/10.1103/PhysRevB.91.165112} {\bibfield  {journal} {\bibinfo
  {journal} {Phys. Rev. B}\ }\textbf {\bibinfo {volume} {91}},\ \bibinfo
  {pages} {165112} (\bibinfo {year} {2015})}\BibitemShut {NoStop}%
\bibitem [{\citenamefont {Verresen}\ \emph {et~al.}(2018)\citenamefont
  {Verresen}, \citenamefont {Pollmann},\ and\ \citenamefont
  {Moessner}}]{Verresen2018spec}%
  \BibitemOpen
  \bibfield  {author} {\bibinfo {author} {\bibfnamefont {R.}~\bibnamefont
  {Verresen}}, \bibinfo {author} {\bibfnamefont {F.}~\bibnamefont {Pollmann}},\
  and\ \bibinfo {author} {\bibfnamefont {R.}~\bibnamefont {Moessner}},\
  }\bibfield  {title} {\bibinfo {title} {Quantum dynamics of the square-lattice
  heisenberg model},\ }\href {https://doi.org/10.1103/PhysRevB.98.155102}
  {\bibfield  {journal} {\bibinfo  {journal} {Phys. Rev. B}\ }\textbf {\bibinfo
  {volume} {98}},\ \bibinfo {pages} {155102} (\bibinfo {year}
  {2018})}\BibitemShut {NoStop}%
\bibitem [{\citenamefont {Seetharam}\ \emph {et~al.}(2021)\citenamefont
  {Seetharam}, \citenamefont {Shchadilova}, \citenamefont {Grusdt},
  \citenamefont {Zvonarev},\ and\ \citenamefont {Demler}}]{Seetharam2021}%
  \BibitemOpen
  \bibfield  {author} {\bibinfo {author} {\bibfnamefont {K.}~\bibnamefont
  {Seetharam}}, \bibinfo {author} {\bibfnamefont {Y.}~\bibnamefont
  {Shchadilova}}, \bibinfo {author} {\bibfnamefont {F.}~\bibnamefont {Grusdt}},
  \bibinfo {author} {\bibfnamefont {M.~B.}\ \bibnamefont {Zvonarev}},\ and\
  \bibinfo {author} {\bibfnamefont {E.}~\bibnamefont {Demler}},\ }\bibfield
  {title} {\bibinfo {title} {Dynamical quantum cherenkov transition of fast
  impurities in quantum liquids},\ }\href
  {https://doi.org/10.1103/PhysRevLett.127.185302} {\bibfield  {journal}
  {\bibinfo  {journal} {Phys. Rev. Lett.}\ }\textbf {\bibinfo {volume} {127}},\
  \bibinfo {pages} {185302} (\bibinfo {year} {2021})}\BibitemShut {NoStop}%
\bibitem [{\citenamefont {Lee}\ \emph {et~al.}(1953)\citenamefont {Lee},
  \citenamefont {Low},\ and\ \citenamefont {Pines}}]{Lee1953}%
  \BibitemOpen
  \bibfield  {author} {\bibinfo {author} {\bibfnamefont {T.~D.}\ \bibnamefont
  {Lee}}, \bibinfo {author} {\bibfnamefont {F.~E.}\ \bibnamefont {Low}},\ and\
  \bibinfo {author} {\bibfnamefont {D.}~\bibnamefont {Pines}},\ }\bibfield
  {title} {\bibinfo {title} {The motion of slow electrons in a polar crystal},\
  }\href {https://doi.org/10.1103/PhysRev.90.297} {\bibfield  {journal}
  {\bibinfo  {journal} {Phys. Rev.}\ }\textbf {\bibinfo {volume} {90}},\
  \bibinfo {pages} {297} (\bibinfo {year} {1953})}\BibitemShut {NoStop}%
\bibitem [{\citenamefont {wen Xiao}(2009)}]{Xiao2009}%
  \BibitemOpen
  \bibfield  {author} {\bibinfo {author} {\bibfnamefont {M.}~\bibnamefont {wen
  Xiao}},\ }\href@noop {} {\bibinfo {title} {Theory of transformation for the
  diagonalization of quadratic hamiltonians}} (\bibinfo {year} {2009}),\
  \Eprint {https://arxiv.org/abs/0908.0787} {arXiv:0908.0787 [math-ph]}
  \BibitemShut {NoStop}%
\bibitem [{\citenamefont {Shi}\ \emph {et~al.}(2020)\citenamefont {Shi},
  \citenamefont {Demler},\ and\ \citenamefont {Cirac}}]{Shi2020}%
  \BibitemOpen
  \bibfield  {author} {\bibinfo {author} {\bibfnamefont {T.}~\bibnamefont
  {Shi}}, \bibinfo {author} {\bibfnamefont {E.}~\bibnamefont {Demler}},\ and\
  \bibinfo {author} {\bibfnamefont {J.~I.}\ \bibnamefont {Cirac}},\ }\bibfield
  {title} {\bibinfo {title} {Variational approach for many-body systems at
  finite temperature},\ }\href {https://doi.org/10.1103/PhysRevLett.125.180602}
  {\bibfield  {journal} {\bibinfo  {journal} {Phys. Rev. Lett.}\ }\textbf
  {\bibinfo {volume} {125}},\ \bibinfo {pages} {180602} (\bibinfo {year}
  {2020})}\BibitemShut {NoStop}%
\bibitem [{\citenamefont {Schmitt-Rink}\ \emph {et~al.}(1988)\citenamefont
  {Schmitt-Rink}, \citenamefont {Varma},\ and\ \citenamefont
  {Ruckenstein}}]{SchmittRink1988}%
  \BibitemOpen
  \bibfield  {author} {\bibinfo {author} {\bibfnamefont {S.}~\bibnamefont
  {Schmitt-Rink}}, \bibinfo {author} {\bibfnamefont {C.~M.}\ \bibnamefont
  {Varma}},\ and\ \bibinfo {author} {\bibfnamefont {A.~E.}\ \bibnamefont
  {Ruckenstein}},\ }\bibfield  {title} {\bibinfo {title} {Spectral function of
  holes in a quantum antiferromagnet},\ }\href
  {https://doi.org/10.1103/PhysRevLett.60.2793} {\bibfield  {journal} {\bibinfo
   {journal} {Phys. Rev. Lett.}\ }\textbf {\bibinfo {volume} {60}},\ \bibinfo
  {pages} {2793} (\bibinfo {year} {1988})}\BibitemShut {NoStop}%
\bibitem [{\citenamefont {Dalla~Piazza}\ \emph {et~al.}(2015)\citenamefont
  {Dalla~Piazza}, \citenamefont {Mourigal}, \citenamefont {Christensen},
  \citenamefont {Nilsen}, \citenamefont {Tregenna-Piggott}, \citenamefont
  {Perring}, \citenamefont {Enderle}, \citenamefont {McMorrow}, \citenamefont
  {Ivanov},\ and\ \citenamefont {Ronnow}}]{Piazza2015}%
  \BibitemOpen
  \bibfield  {author} {\bibinfo {author} {\bibfnamefont {B.}~\bibnamefont
  {Dalla~Piazza}}, \bibinfo {author} {\bibfnamefont {M.}~\bibnamefont
  {Mourigal}}, \bibinfo {author} {\bibfnamefont {N.~B.}\ \bibnamefont
  {Christensen}}, \bibinfo {author} {\bibfnamefont {G.~J.}\ \bibnamefont
  {Nilsen}}, \bibinfo {author} {\bibfnamefont {P.}~\bibnamefont
  {Tregenna-Piggott}}, \bibinfo {author} {\bibfnamefont {T.~G.}\ \bibnamefont
  {Perring}}, \bibinfo {author} {\bibfnamefont {M.}~\bibnamefont {Enderle}},
  \bibinfo {author} {\bibfnamefont {D.~F.}\ \bibnamefont {McMorrow}}, \bibinfo
  {author} {\bibfnamefont {D.~A.}\ \bibnamefont {Ivanov}},\ and\ \bibinfo
  {author} {\bibfnamefont {H.~M.}\ \bibnamefont {Ronnow}},\ }\bibfield  {title}
  {\bibinfo {title} {Fractional excitations in the square-lattice quantum
  antiferromagnet},\ }\href {https://doi.org/10.1038/nphys3172} {\bibfield
  {journal} {\bibinfo  {journal} {Nat Phys}\ }\textbf {\bibinfo {volume}
  {11}},\ \bibinfo {pages} {62} (\bibinfo {year} {2015})}\BibitemShut {NoStop}%
\bibitem [{\citenamefont {Shao}\ \emph {et~al.}(2017)\citenamefont {Shao},
  \citenamefont {Qin}, \citenamefont {Capponi}, \citenamefont {Chesi},
  \citenamefont {Meng},\ and\ \citenamefont {Sandvik}}]{Shao2017}%
  \BibitemOpen
  \bibfield  {author} {\bibinfo {author} {\bibfnamefont {H.}~\bibnamefont
  {Shao}}, \bibinfo {author} {\bibfnamefont {Y.~Q.}\ \bibnamefont {Qin}},
  \bibinfo {author} {\bibfnamefont {S.}~\bibnamefont {Capponi}}, \bibinfo
  {author} {\bibfnamefont {S.}~\bibnamefont {Chesi}}, \bibinfo {author}
  {\bibfnamefont {Z.~Y.}\ \bibnamefont {Meng}},\ and\ \bibinfo {author}
  {\bibfnamefont {A.~W.}\ \bibnamefont {Sandvik}},\ }\bibfield  {title}
  {\bibinfo {title} {Nearly deconfined spinon excitations in the square-lattice
  spin-$1/2$ heisenberg antiferromagnet},\ }\href
  {https://doi.org/10.1103/PhysRevX.7.041072} {\bibfield  {journal} {\bibinfo
  {journal} {Phys. Rev. X}\ }\textbf {\bibinfo {volume} {7}},\ \bibinfo {pages}
  {041072} (\bibinfo {year} {2017})}\BibitemShut {NoStop}%
\bibitem [{\citenamefont {Chevy}(2006)}]{Chevy}%
  \BibitemOpen
  \bibfield  {author} {\bibinfo {author} {\bibfnamefont {F.}~\bibnamefont
  {Chevy}},\ }\bibfield  {title} {\bibinfo {title} {Universal phase diagram of
  a strongly interacting fermi gas with unbalanced spin populations},\ }\href
  {https://doi.org/10.1103/PhysRevA.74.063628} {\bibfield  {journal} {\bibinfo
  {journal} {Phys. Rev. A}\ }\textbf {\bibinfo {volume} {74}},\ \bibinfo
  {pages} {063628} (\bibinfo {year} {2006})}\BibitemShut {NoStop}%
\bibitem [{\citenamefont {Greensite}(2003)}]{Greensite2003}%
  \BibitemOpen
  \bibfield  {author} {\bibinfo {author} {\bibfnamefont {J.}~\bibnamefont
  {Greensite}},\ }\bibfield  {title} {\bibinfo {title} {The confinement problem
  in lattice gauge theory},\ }\href
  {http://www.sciencedirect.com/science/article/pii/S0146641003900123}
  {\bibfield  {journal} {\bibinfo  {journal} {Progress in Particle and Nuclear
  Physics}\ }\textbf {\bibinfo {volume} {51}},\ \bibinfo {pages} {1} (\bibinfo
  {year} {2003})}\BibitemShut {NoStop}%
\bibitem [{\citenamefont {Zvonarev}\ \emph {et~al.}(2007)\citenamefont
  {Zvonarev}, \citenamefont {Cheianov},\ and\ \citenamefont
  {Giamarchi}}]{Zvonarev2007}%
  \BibitemOpen
  \bibfield  {author} {\bibinfo {author} {\bibfnamefont {M.~B.}\ \bibnamefont
  {Zvonarev}}, \bibinfo {author} {\bibfnamefont {V.~V.}\ \bibnamefont
  {Cheianov}},\ and\ \bibinfo {author} {\bibfnamefont {T.}~\bibnamefont
  {Giamarchi}},\ }\bibfield  {title} {\bibinfo {title} {Spin dynamics in a
  one-dimensional ferromagnetic bose gas},\ }\href
  {https://doi.org/10.1103/PhysRevLett.99.240404} {\bibfield  {journal}
  {\bibinfo  {journal} {Physical Review Letters}\ }\textbf {\bibinfo {volume}
  {99}},\ \bibinfo {pages} {240404} (\bibinfo {year} {2007})}\BibitemShut
  {NoStop}%
\bibitem [{\citenamefont {Bruderer}\ \emph {et~al.}(2008)\citenamefont
  {Bruderer}, \citenamefont {Klein}, \citenamefont {Clark},\ and\ \citenamefont
  {Jaksch}}]{Bruderer2008}%
  \BibitemOpen
  \bibfield  {author} {\bibinfo {author} {\bibfnamefont {M.}~\bibnamefont
  {Bruderer}}, \bibinfo {author} {\bibfnamefont {A.}~\bibnamefont {Klein}},
  \bibinfo {author} {\bibfnamefont {S.~R.}\ \bibnamefont {Clark}},\ and\
  \bibinfo {author} {\bibfnamefont {D.}~\bibnamefont {Jaksch}},\ }\bibfield
  {title} {\bibinfo {title} {Transport of strong-coupling polarons in optical
  lattices},\ }\href {https://doi.org/10.1088/1367-2630/10/3/033015} {\bibfield
   {journal} {\bibinfo  {journal} {New Journal of Physics}\ }\textbf {\bibinfo
  {volume} {10}},\ \bibinfo {pages} {033015} (\bibinfo {year}
  {2008})}\BibitemShut {NoStop}%
\bibitem [{\citenamefont {Grusdt}\ \emph {et~al.}(2014)\citenamefont {Grusdt},
  \citenamefont {Shashi}, \citenamefont {Abanin},\ and\ \citenamefont
  {Demler}}]{Grusdt2014BO}%
  \BibitemOpen
  \bibfield  {author} {\bibinfo {author} {\bibfnamefont {F.}~\bibnamefont
  {Grusdt}}, \bibinfo {author} {\bibfnamefont {A.}~\bibnamefont {Shashi}},
  \bibinfo {author} {\bibfnamefont {D.}~\bibnamefont {Abanin}},\ and\ \bibinfo
  {author} {\bibfnamefont {E.}~\bibnamefont {Demler}},\ }\bibfield  {title}
  {\bibinfo {title} {Bloch oscillations of bosonic lattice polarons},\ }\href
  {https://doi.org/10.1103/PhysRevA.90.063610} {\bibfield  {journal} {\bibinfo
  {journal} {Phys. Rev. A}\ }\textbf {\bibinfo {volume} {90}},\ \bibinfo
  {pages} {063610} (\bibinfo {year} {2014})}\BibitemShut {NoStop}%
\bibitem [{\citenamefont {Lampo}\ \emph {et~al.}(2017)\citenamefont {Lampo},
  \citenamefont {Lim}, \citenamefont {Garc{\'{i}}a-March},\ and\ \citenamefont
  {Lewenstein}}]{Lampo2017bosepolaronas}%
  \BibitemOpen
  \bibfield  {author} {\bibinfo {author} {\bibfnamefont {A.}~\bibnamefont
  {Lampo}}, \bibinfo {author} {\bibfnamefont {S.~H.}\ \bibnamefont {Lim}},
  \bibinfo {author} {\bibfnamefont {M.~{\'{A}}.}\ \bibnamefont
  {Garc{\'{i}}a-March}},\ and\ \bibinfo {author} {\bibfnamefont
  {M.}~\bibnamefont {Lewenstein}},\ }\bibfield  {title} {\bibinfo {title} {Bose
  polaron as an instance of quantum {B}rownian motion},\ }\href
  {https://doi.org/10.22331/q-2017-09-27-30} {\bibfield  {journal} {\bibinfo
  {journal} {{Quantum}}\ }\textbf {\bibinfo {volume} {1}},\ \bibinfo {pages}
  {30} (\bibinfo {year} {2017})}\BibitemShut {NoStop}%
\bibitem [{\citenamefont {Bohrdt}\ \emph {et~al.}(2022)\citenamefont {Bohrdt},
  \citenamefont {Homeier}, \citenamefont {Bloch}, \citenamefont {Demler},\ and\
  \citenamefont {Grusdt}}]{bohrdt2021strong}%
  \BibitemOpen
  \bibfield  {author} {\bibinfo {author} {\bibfnamefont {A.}~\bibnamefont
  {Bohrdt}}, \bibinfo {author} {\bibfnamefont {L.}~\bibnamefont {Homeier}},
  \bibinfo {author} {\bibfnamefont {I.}~\bibnamefont {Bloch}}, \bibinfo
  {author} {\bibfnamefont {E.}~\bibnamefont {Demler}},\ and\ \bibinfo {author}
  {\bibfnamefont {F.}~\bibnamefont {Grusdt}},\ }\bibfield  {title} {\bibinfo
  {title} {Strong pairing in mixed-dimensional bilayer antiferromagnetic mott
  insulators},\ }\href {https://doi.org/10.1038/s41567-022-01561-8} {\bibfield
  {journal} {\bibinfo  {journal} {Nature Physics}\ }\textbf {\bibinfo {volume}
  {18}},\ \bibinfo {pages} {651} (\bibinfo {year} {2022})}\BibitemShut
  {NoStop}%
\end{thebibliography}

%


\appendix

\newpage

\section{Spinon Dynamics}
\label{ApdxSpnDyn}
To leading order in the $1/S$ expansion the spinon has no dynamics in 1D. Because we defined the spinon as a domain wall configuration of the Ising field $\tau_j^z$, the inclusion of magnon corrections does not introduce spinon dynamics either: Although the resulting Hamiltonian depends explicitly on $\tau_j^{z}$, see Eqs.~\eqref{HPapproxSz}, \eqref{HPapproxSpm}, it does not contain terms $\propto \hat{\tau}^{x,y}_j$ which are necessary to change the values of the Ising variables $\tau_j^{z}$.

In the large-$S$ limit our result that the spinon cannot move makes sense: a spinon corresponds to a domain wall of two aligned spins of length $|S_z|=S$ on neighboring sites, see Fig.~\ref{Fig1} (a). When $J_\perp = 0$ this excitation cannot move. Even when $J_\perp \neq 0$ the spin-exchange interaction $J_\perp \Sp \Sm$ can only reduce the length of the spin gradually from $|S_z|=S$ to $S-1$ to $S-2$, etc. in the large $S$ limit. Thus when $S\gg1$ and the number of magnons is small, $\langle \ad_j\a_j \rangle \ll 2S$, the direction of the N\'eel order, represented by the Ising variable $\tau_j^{z}$, cannot change.

\begin{figure}
	\centering 
	\includegraphics[width=4cm]{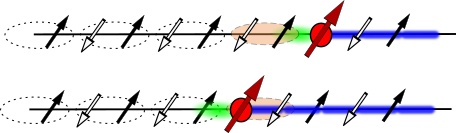}
	\quad
	\includegraphics[width=4cm]{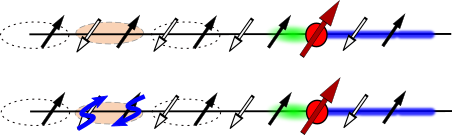}
	\caption{Possible processes resulting from the exchange term $J_\perp \Sp_{j+1}\Sm_j $. Left: spinon (domain-wall) motion accounted for by the free spinon Hamiltonian. Right: magnon vacuum fluctuations accounted for by the free magnon Hamiltonian.
	\label{spinontunneling}}
\end{figure}

From now on we will consider the situation $S=1/2$. In this case a single exchange process is sufficient to move the domain wall consisting of two aligned spins. Namely, by applying $J_\perp (\Sp_{j+1} \Sm_j + \hc)$ the N\'eel order parameter $\l -1\r^{j} S^z_j$ can change on two adjacent lattice sites next to the spinon, see Fig \ref{spinontunneling} (left). Indeed, in the 1D Heisenberg model without an external field, this process is well-known to lead to dynamics of deconfined domain wall excitations corresponding to spinons \cite{Giamarchi2004}.

In the free magnon part of the Hamiltonian \eqref{mag-contributions} we describe all exchange processes $\sim J_\perp (\Sp_{j+1} \Sm_j + \hc)$ using the bare HP operators $\ad_j$ introduced in Eq.~\eqref{HPapproxSpm}. Such terms will lead to the creation of HP bosons around the spinon, and from \eqref{HPapproxSz} we see that the corresponding physical eigenstates of $\Sz_j$ will correctly reflect the motion of the domain wall, even though the position of the spinon, defined by $\sum_j j \sd_j \s_j$, does not change. However, as we will show next, this is an artefact of using the overcomplete parton basis.

For a fixed configuration $\tau_j^z$ all physical eigenstates in a system of spin $S=1/2$ particles are correctly represented by Fock states $\ket{\{n^a_j\}}$ of HP occupation numbers $n_j^a = 0, 1$. 
By allowing different spinon positions $j^s$ associated with a different Ising configuration $\tau^z_j$ we get an enlarged effective Hilbert space $\{\ket{j^s}\otimes \ket{n^a_i}\}.$ When constructing the effective parton Hamiltonian each matrix element of the Hamiltonian \eqref{model} between two physical states has to correspond to one overlap in the over-complete basis $\{\ket{j^s}\otimes \ket{n^a_i}\}.$ In order to decide which matrix elements to associate to which terms in the Hamiltonian we compare the energy costs for flipping bonds being part of the spinon and those which are not. In particular, we will distinguish resonant from off-resonant processes, costing no or a finite amount of energy in a pure Ising configuration. In the end, we want to find a representation of spin-exchange terms which treats resonant terms as an effective spinon hopping process. 

Consider first bonds which are not part of the spinon. We introduced them into our formalism by writing the spin-exchange terms as $ J_\perp/2 \sum_j \left(\ad_{j+1} \ad_j+h.c.\right)$. These lead to creation and annihilation of HP bosons with an energy cost of $\left(J+h\right)$.
Using our $1/S$ expansion we can estimate the number of HP bosons per lattice to scale as $\propto J_z^2/\left(J_z+h\right)^2$. Thus for non-vanishing magnetic fields $h$ the number of bare HP bosons should be small, which justifies to use them on such bonds as it means that the local magnetization $\langle \Sz_j \rangle$ cannot change significantly. 

On the other hand, bonds adjacent to the spinon can be used to let it tunnel by two lattice sites as illustrated in Fig.~\ref{spinontunneling} (left). Such a process would increase the length of the string by two units leading to a energy cost of $\propto 2h$. But due to the strong coupling limit $t\gg J_z,J_\perp,h$ the chargon can instantly adjust to the new configuration and restore the average string length. Thus the spinon motion would not lead to any energy cost. If we would use such bonds for the HP bosons a huge number of them could be created at essentially no energy cost. However, this would violate the HP condition $\langle \ad_j \a_j \rangle\ll 1$.

To avoid this issue, we describe these resonant bonds by allowing for spinon tunneling. Thus, we choose to compute the matrix elements of the spin-exchange term on such bonds by letting only the spinon position change while excluding changes of the magnon occupation number states $\ket{\{n_i^a\}}$.

Using the overlap $J_\perp/2 \bra{j^s_1}\Sp_{j+1}\Sm_j\ket{j^s_2}$ we evaluate the described matrix elements of $\H_{J_\perp}$, which leads us to the spinon tunneling term 
\begin{align}
 \H^s_0=	\dfrac{J_\perp}{2} \sum_{j} \left(\sd_{j+2}\s_j+h.c.\right).
\end{align}
We emphasize that this process is only possible in the spin-$1/2$ case. Further, we note that we already used these bonds to include magnon excitations in the free magnon Hamiltonian in Eq.~\eqref{mag-contributions}. To avoid double-counting  we have to exclude these bonds again from the magnon Hamiltonian, which leads to spinon-magnon interactions as discussed in the next appendix.

Finally we also have to consider the situation when a chargon sits on a site next to the spinon. In this case the spinon tunneling process would not be possible because the chargon blocks a spin-exchange term. Again, to avoid double-counting processes, we must introduce an additional kinetic spinon-chargon interaction 
\begin{align}
	\hat{\mathcal{H}}^{\rm kin}_{\rm sh}= -\dfrac{J_\perp}{2} \sum_{j}\hd_{j+1}\h_{j+1}\left(\sd_{j+2}\s_j+\hc\right).
\end{align}

\section{Magnon Contributions}
\label{ApdxMagnonCont}
In this subsection we turn our attention to the magnon contributions in the effective Hamiltonian. They are of higher order in $1/S$ and here we derive the form of these parton-magnon interactions up to quadratic order in the HP boson operators $\ad_j$.

The parton-magnon interaction consists of two contributions,
\begin{equation}
    \hat{\mathcal{H}}_{\rm mag}^{\rm int}= \hat{\mathcal{H}}_{\rm mag}^{J}+\hat{\mathcal{H}}_{\rm mag}^{\rm kin}.
\end{equation}
The first describes how spin-exchange terms are modified by distortions of the Néel background due to the presence of the partons. The other includes effects on the motion of the partons due to the presence of magnons.

\subsection{Interactions due to \\ Distortions of the Spin Environment}
In the non-interacting Hamiltonian we already introduced magnons on all bonds of the lattice assuming $\tau^z_j=(-1)^j$. However, the Ising fields $\tau^z_j$ are not static but can change due to displacements of the spins by the chargon motion. The configuration of Ising fields is determined by the spinon-chargon configuration; i.e. $\tau^z_j$ explicitly depends on the quantum state of the partons: $\hat{\tau}^z_j\equiv \tau^z_j(j^h,j^s)$. 

Now we use the $\tau^z_j$-dependent HP representation of spin operators, see Eq.~\eqref{HPapprox}, to express all couplings involving spins in our $t-{\rm XXZ}$ model; the presence of the partons leads to the following changes for the magnon terms in the effective Hamiltonian:
\begin{itemize}
    \item[(i)] Bonds occupied by the chargon lead to no couplings to spins, since $\tau^z_{j^h}=0$. Therefore, contributions from such bonds have to be subtracted from the already included bonds in the free magnon Hamiltonian $\sum_q \omega_q \bd_q\b_q$. This yields a term
    \begin{multline}
        \dfrac{1}{2} \left(J_z+h\right)\left(\ad_j\a_j+\ad_{j+1} \a_{j+1}\right)+ \\ +\dfrac{J_\perp}{2} \left(\ad_{j+1}\ad_j+\a_j \a_{j+1}\right)
    \end{multline}
   on such bonds to subtract.
    \item[(ii)] Additionally, for $S=1/2$ we used flip-flop terms $~J_\perp \Sp_{j+1}\Sm_j$ to describe spinon tunnelings on bonds involving the spinon. To avoid double counting we do not include additional magnon couplings on these bonds. Again, magnon couplings already included in the free magnon Hamiltonian involving processes describing the spinon dynamics, have to be subtracted, similar to the procedure in (i).
    \item[(iii)] Along the string, the displaced spins occupy the wrong sub-lattice site relative to the Néel order. Thus, at those sites the Ising filed $\hat{\tau}^z_j$ has a reversed sign and leads to a potential energy cost $\propto h$, which we included in the string tension \eqref{linear-pot}. But this also leads to a separate energy cost for HP bosons along the string which we have not included so far. To account for this effect, we add the term
    \begin{equation}
    \label{MagStringInt}
        -h \sum_j \left| \l-1\r^j - \hat{\tau}_j^z \right| \ad_j \a_j
    \end{equation}
    to the effective Hamiltonian.
\end{itemize}
As a result, we obtain the following coupling resulting due to distortions of the spin environment,
\begin{multline}
    \hat{\mathcal{H}}_{\rm mag}^{J}= -\dfrac{1}{2} \sum_{\hat{X}_s, \hat{X}_h\in\langle j,j+1\rangle} \left[\left(J_z+h\right)\left(\ad_j\a_j+\ad_{j+1}\a_{j+1}\right)\right.\\
    \left.+J_\perp \left(\ad_{j}\ad_{j+1}+\a_{j}\a_{j+1}\right) \right] -\\
    -h \sum_j \left| \l-1\r^j - \hat{\tau}_j^z \right| \ad_j \a_j.
\end{multline}
Here, we introduced the position operator $\hat{X}_h= \sum_j j \hd_j\h_j$ for the chargon and analogously for the spinon.\\
If a bond involves both, spinon and chargon, it is only counted once in the sum.

\subsection{Magnon Influence on the Parton Dynamics}
\label{ApdxMagPartonDyn} 
\begin{figure}
    \centering
    \includegraphics[width=4cm]{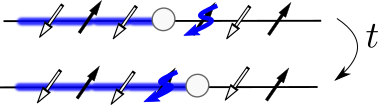}\quad
    \includegraphics[width=4cm]{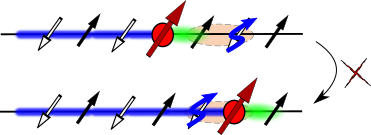}
    \caption{Illustration of kinetic chargon-magnon (left) and spinon-magnon (right) couplings. The left figure illustrates the translation of a HP magnon by the motion of the chargon. In the right sketch a configuration is shown where spinon tunneling is blocked by the presence of a HP magnon.}
    \label{KinPartonMagFig}
\end{figure}
When deriving the parton-magnon interaction $\hat{\mathcal{H}}_{\rm mag}^{J}$ accounting for distortions of the Néel background, we assumed that the parton configuration is static. Further, in the strong coupling theory introduced in subsection \ref{secStrongCplngThy} of the main text we ignored magnon contributions affecting parton dynamics.
Here, we introduce additional terms $\hat{\mathcal{H}}_{\rm mag}^{\rm kin}$ describing how the parton dynamics couples to magnon excitations.

We assume that the number of HP bosons per lattice site is small, $\langle \ad_j \a_j \rangle \ll 1$, which gives us the possibility to include only processes involving not more than one HP boson $\ad_j$ per lattice site. Within this approximation we obtain an effective Hamiltonian quadratic in the HP bosons.

We start by discussing processes involving the chargon, where it tunnels from some lattice site $i$ to $j$. Due to our chosen constraint, Eq.~\eqref{single-occ-constraint}, the whole spin state from site $j$ will be translated to the neighboring site $i$. This is illustrated in Fig.~\ref{KinPartonMagFig} (left).

From the $\tau^z_j$-dependent HP approximation, it is clear that the chargon tunneling has not just an effect on $\tau^z_j$ but also on the HP bosons $\ad_j$. In our discussion about the parton dynamics we already accounted for changes in the Ising fields, by using the parton basis, see Eq.~\eqref{2nd-parton-Hilbert-space}. Additionally, we have to include terms to our parton Hamiltonian which ensure that a HP boson $\ad_j$, residing on site $j$, is also translated to the site $i$, when the chargon tunnels, see Fig.~\ref{KinPartonMagFig} (left).

In our subspace of no more than one HP boson per lattice site the relevant kinetic chargon-magnon coupling is given by
\begin{eqnarray}
    \hat{\mathcal{H}}_{\rm mag}^{\rm kin,h}= t \sum_j \left(\hd_{j+1}\h_{j}+h.c.\right)\left(\ad_{j+1}\a_{j}+\ad_{j}\a_{j+1}\notag \right. \\
    \left. -\ad_j \a_j-\ad_{j+1}\a_{j+1}\right).
\end{eqnarray}
The terms in the first term describe the correlated hopping of the chargon and a HP magnon; they vanish when no HP magnon is present, in which case the bare chargon tunneling in the free chargon Hamiltonian correctly describes the hopping process. The terms in the second line subtract the bare chargon tunneling if a magnon is present. In summary, the so constructed effective Hamiltonian describes (free chargon hopping) purely correlated magnon-chargon hopping in the (absence) presence of a magnon next to the chargon.

A similar analysis has to be performed for the spinon tunneling. We consider a situation where the spinon moves from site $j^s$ to $j^s\pm 2$. As explained in Sec. \ref{ApdxSpnDyn} the spinon dynamics originates from spin-exchange interactions on bonds involving the spinon. One of the involved lattice sites, say $i$, is part of the domain wall of the Ising field $\tau^z$ defining the spinon, while the other, which we label $r$ is not. As in the case of the chargon, we assume that the HP boson density is low, $\langle \ad_j \a_j\rangle \ll 1$, and derive the kinetic spinon-magnon Hamiltonian by considering only states $\sd_{j^s}\ket{0}$ and $\sd_{j^s}\ad_r \ket{0}$: I.e. higher-order effects in the HP boson operators $\a_j$ are neglected.

A HP boson excitation  on site $r$ leads to a ferromagnetic configuration for the original spins $\Sz_i \Sz_r =1$ on the bond $\langle i,r \rangle$ where spin-exchange interactions introduce spinon dynamics. The action of the spin-exchange terms $\propto J_\perp\Sp_i \Sm_r$ on this state vanishes. Therefore the HP boson on site $r$ suppresses spinon dynamics. In order to cancel the dynamics already included in the free spinon Hamiltonian, Eq.~\eqref{freespinonH}, we add a counter term
\begin{equation}
    \label{KinSpMagInt}
    \H_{\rm mag}^{\rm kin,s} = - \dfrac{J_\perp}{2} \sum_{j^s,\mu=\pm}\sd_{j^s+2\mu} \s_{j^s} \ad_{r_\mu}\a_{r_\mu} + \hc,
\end{equation}
which corresponds to the kinetic spinon-magnon coupling. Note that the position $r_\pm$ introduced above explicitly depends on the parton configuration. When the spinon tunnels from $j^s$ to site $j^s\pm 2$, it is given by
\begin{equation}
    r_\pm = j^s \pm \left[ 3 \mp \sgn(\Sigma) \right]/2
\end{equation}
where $\sgn(\Sigma)$ denotes the direction of the string $\Sigma$ from the spinon to the chargon.

\section{Meson Spectrum in SC without Magnons}
\label{ApdxMesonStatesSC}
Here we look at the parton Hamiltonian \eqref{partonHam}, neglecting the magnon contributions, and show that the ansatz \eqref{Born-Oppenheimer-ansatz} is an appropriate eigenstate for the meson; We also derive its eigen-spectrum. Ignoring the constant energy shifts, the Hamiltonian is given by
\begin{eqnarray}
    \label{PartonHnoMag}
     \nonumber\H_{\text{mes}}^{(0)} &=& t \sum_j \l \hd_{j+1}\h_j + \hc \r + \sum_{ji} V_{\rm sh}(|i-j|)\sd_j\s_j \hd_i\h_i \\
    &+& \dfrac{J_\perp}{2} \sum_j \l 1- \hd_{j+1}\h_{j+1}\r \l \sd_{j+2}\s_j + \hc \r 
\end{eqnarray}
with the confinement potential \[V_{\rm sh}(\ell) = h\ell - \dfrac{J_z}{4} \delta_{\ell,0}.\]\\
Due to the strong coupling limit $t\gg J_z, J_\perp, h$, we can solve this Hamiltonian via a Born-Oppenheimer approximation.

The chargon follows instantly the slow spinon motion. Thus, we first fix the spinon motion at some site $j^s$ and solve the chargon problem independently,
\begin{eqnarray}
    \label{HchEff}
    \nonumber\H_{h}^\eff = \bra{0} \s_{j^s} \H_{\text{mes}}^{(0)}\sd_{j^s}\ket{0}= -L \varepsilon_0 + \dfrac{3}{4}J_z + \dfrac{h}{2}\\
    +t \sum_{\langle\Sigma,\Sigma^\prime \rangle} \l \hd_{\Sigma}\h_{\Sigma^\prime} + \hc \r + \sum_\Sigma  V_{\text{sh}}(|\Sigma|) \hd_\Sigma \h_\Sigma
\end{eqnarray}
where it is understood that $\hd_\Sigma$ creates a chargon at a distance $\Sigma  \in \mathbb{Z} $ from the spinon, and $\sum_{\langle\Sigma,\Sigma^\prime \rangle}$ denotes a sum over nearest neighbors, $\Sigma^\prime = \Sigma +1$. In the following we will measure all energies relative to the ground state energy of the classical N\'eel state, $-L\varepsilon_0$. 

In the SC approximation the meson spectrum is obtained by calculating the chargon eigenenergies $E_h^{(n,\xi)}$ defined by $\H_h^{\eff}\ket{\psi_h^{(n\xi)}} = E_h^{(n,\xi)} \ket{\psi_h^{(n\xi)}}$. Here $n=1,2,\dots$ denotes the principal quantum number. The effective chargon Hamiltonian \eqref{HchEff} is inversion symmetric around the spinon position, with 
\begin{equation}
    \hat{I}\ket{\psi_h^{(n\xi)}} = \xi \ket{\psi_h^{(n\xi)}}.
\end{equation}
Here $\hat{I}$ is the inversion operator which maps $\Sigma \to - \Sigma$, and the corresponding eigenvalue is $\xi=\pm1$.

One can use the inversion symmetry to map the spinon-chargon problem to a single-particle problem on a semi-infinite 1D lattice. To this end the chargon wavefunction defined in the spinon frame is written as 
\begin{equation}
    \label{StringWaveFunc}
    \psi_h^{(n\xi)}(\Sigma) = (-1)^{\Sigma} \times \begin{cases}
                                    \phi_0^{(n,\xi)}, \quad \Sigma=0\\
                                    \dfrac{\xi}{\sqrt{2}}\phi_{|\Sigma|}^{(n\xi)}, \quad \Sigma <0\\
                                    \dfrac{1}{\sqrt{2}}\phi_{|\Sigma|}^{(n\xi)}, \quad \Sigma > 0.
                                                    \end{cases}
\end{equation}
The normalization condition, $\sum_\Sigma \left|\psi_h^{(n\xi)} (\Sigma)\right|² = 1$ now becomes $\sum_{\ell \geq 0} \left|\phi_\ell^{(n\xi)}\right|² = 1$. The wavefunction $\phi_h^{(n\xi)}$ can be understood as the string wavefunction which depends only on the length $\ell = |\Sigma|\ge 0 $ of the string. 

The inversion symmetry requires $\psi_h^{(n\xi)} (-\Sigma) = \xi \psi_h^{(n\xi)} (\Sigma)$, i.e. odd-parity string wavefunctions have a node in the center at $\ell=0$ with $\phi_0^{(n,-1)}=0$. This node is equivalent to a strong repulsive potential localized at $\ell =0$. As a consequence, the eigenstates with $\xi=-1$ and radial quantum number $n$ generally have a higher energy than their partners at the same $n$ but with $\xi =+1$. The repulsion from the central site for $\xi = -1$ states is a direct generalization of the centrifugal barrier discussed for magnetic polarons in the 2D $t-J_z$ model by a similar description \cite{Grusdt2018}. There it was argued that rotationally excited states, the 2D analog of the odd states with $\xi=-1$, are similar to mesonic resonances characterized by finite orbital angular momentum carried by a quark anti-quark pair observed in high-energy physics. In the same spirit, the excited states of the spinon-chargon mesons in our 1D setup with $\xi=-1$ can be understood as a set of resonances explained naturally by the parton theory.
\begin{figure}
    \centering
    \includegraphics[width=0.47\textwidth]{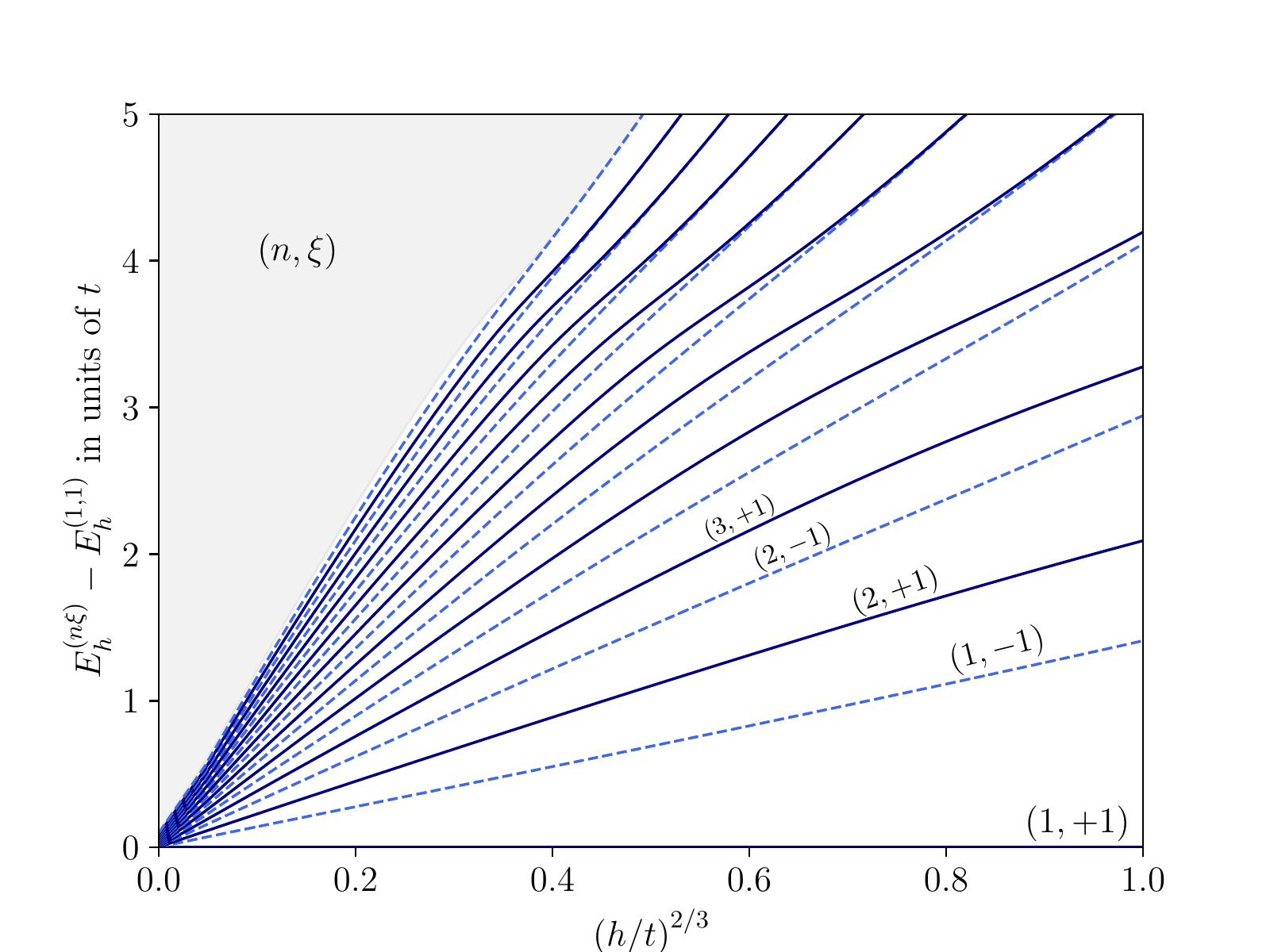}
    \caption{The chargon excitation energies above the ground state , $E_h^{(n\xi)}-E_h^{(1,1)}$ are shown for various values of $h/t$, assuming $J_z=J_\perp=J$. The eigenstates are labeled by their rotational and vibrational quantum numbers $(n,\xi)$. For small $h/t$ we observe a scaling of all excitation energies with the non-trivial power-law $(h/t)^{2/3}$. In the gray region we did not plot any states.}
    \label{ChargonEn}
\end{figure}

Now we discuss the effective Hamiltonians $\H_{\phi,\xi}$ which determine the string wavefunctions $\phi_{\ell}^{(n\xi)}$. They are defined in a Hilbert space $\{\ket{\ell}\}$ with positive string lengths $\ell=0,1,2,\dots$. For the states with even inversion symmetry, $\xi=+1$, the hoppings in the effective model are $t$ in the bulk and $\sqrt{2}t$ between $\ket{\ell=0}$ and $\ket{\ell=1}$. The factor of $\sqrt{2}$ arises because in the original Hamiltonian \eqref{HchEff} state $\hd_0\ket{0}$ is coupled to two states, $\hd_{\pm1}\ket{0}$. The even Hamiltonian $\H_{\phi,\xi=+1}$ thus reads
\begin{eqnarray}
    \label{Hphiplus}
    \nonumber\H_{\phi,+1} &=& - \left[\sqrt{2}t \ket{0}\bra{1} +\sum_{\ell > 0} \ket{\ell+1}\bra{\ell}\right] + \hc\\ 
    &+& \sum_{\ell\ge 0} V_{\rm sh}(\ell) \ket{\ell}\bra{\ell}.
\end{eqnarray}
For the states with odd inversion symmetry, $\xi=-1$, the hopping amplitude between the central site and the first site in the effective Hamiltonian is zero,
\begin{eqnarray}
    \label{Hphiminus}
    \H_{\phi,-1}= -t \sum_{\ell>0} \ket{\ell+1}\bra{\ell}+\hc + \sum_{\ell> 0} V_{\rm sh}(\ell) \ket{\ell}\bra{\ell}.~~~~
\end{eqnarray}

In the strong coupling regime a mapping to a continuum model shows that the radially excited states have energies given by \cite{Bulaevski1968}
\begin{equation}
    \label{Eairy}
    E_h^{(n\xi)}= E_0 - 2t + a_{\rm sh}^{(n\xi)} t^{1/3}h^{2/3} + \mathcal{O}(J_z,h),
\end{equation}
with numerical coefficients $a_{\rm sh}^{(n\xi)}$ related to the Airy-function\cite{Bulaevski1968}. The contributions of order $\mathcal{O}(J_z,h)$ can be easily calculated numerically by solving the single-particle problems Eqs. \eqref{Hphiplus} and \eqref{Hphiminus}. The coefficients $a_{\rm sh}^{(n\xi)}$ increase with $n$, and $a_{\rm sh}^{(n,+1)}< a_{\rm sh}^{(n,-1)}$.

\begin{figure}
    \centering
    \includegraphics[width=0.5\textwidth]{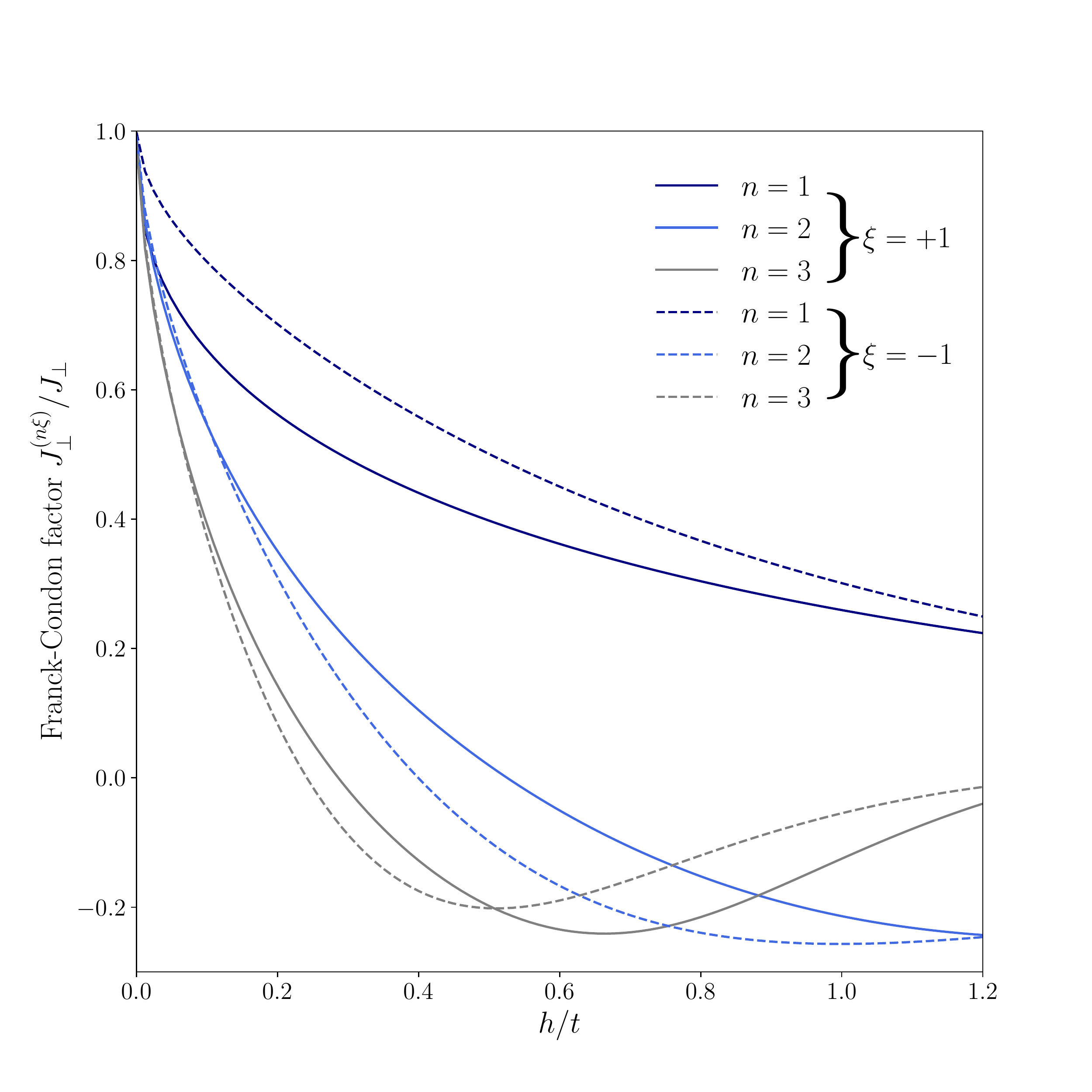}
    \caption{Franck-Condon factor renormalizing the spinon dispersion by dressing of the chargon in the strong coupling approach. We performed calculations for different values of $h/t$ and $n,\xi$.}
    \label{FC}
\end{figure}

In Fig \ref{ChargonEn} we calculate the SC meson excitation energies relative to the ground state energy $E_0^{(1,+1)}$ at $n=1,\: \xi=+1$ and assuming $J_z=J_\perp=J$. We find that all excitation energies scale as $J^{2/3}t^{1/3}$, confirming Eq. \eqref{Eairy}. Close inspection shows that $a_{\rm sh}^{(n,-1)} \approx a_{\rm sh}^{(n+1,+1)}$ and this approximation becomes more accurate for increasing values of the principle quantum number $n$ and larger values of $h/t$.

For the calculation of the chargon wavefunction $\ket{\psi_h^{(n\xi)}(j^s)}$ in Eq. \eqref{HchEff} we fixed the position of the spinon at $j^s$. Now, in a second step, we treat the spinon dynamics perturbatively and assume that the light chargon instantly follows the heavy spinon. This allows us to work with the following set of orthogonal low-energy basis states, $\{\ket{j^s,n,\xi}\}$, where
\begin{equation}
    \label{AppendixLowEnergyBasis}
    \ket{j^s,n,\xi} = \sd_{j^s}\ket{0} \otimes \ket{\psi_h^{(n\xi)}(j^s)}.
\end{equation}
The effective Hamiltonian $\H^\eff_s$ of the spinon is obtained by projecting the parton Hamiltonian \eqref{PartonHnoMag} to the new low-energy basis. Note that the basis in \eqref{AppendixLowEnergyBasis} formally corresponds to the introduced meson operators \eqref{mesonop} in the main text.

The non-trivial matrix elements are associated with spinon dynamics, see second line of Eq. \eqref{PartonHnoMag}, and lead to \begin{equation}
    \H_s^\eff = \dfrac{J_\perp^{(n\xi)}}{2} \sum_{j}  \ket{j^s+2,n,\xi}\bra{j^s,n,\xi} + \hc 
\end{equation}
As already mentioned in the main text, the quantum numbers $n$ and $\xi$ describe the internal state of the meson and can be treated as band indices. 

The expression for the Franck-Condon overlap \eqref{FCoverlap} given in the main text results from the mapping of the spinon part in \eqref{PartonHnoMag} to the basis \eqref{AppendixLowEnergyBasis}. It can be calculated directly from the string wavefunction $\phi_\ell^{(n\xi)}$ defined in the semi-infinite 1D geometry, see Eq. \eqref{StringWaveFunc}. We obtain 
\begin{multline}
    \dfrac{J_\perp^{(n\xi)}}{J_\perp} = \left[\dfrac{1}{\sqrt{2}} \l\phi_2^{(n\xi)} \r^* \phi_0^{(n\xi)}+ \dfrac{1}{2}\sum_{\ell=1}^{\infty} \l \phi_{\ell+2}^{(n\xi)}\r^*\phi_\ell^{(n\xi)}\right] + \\
    + \text{c.c.},
\end{multline}
where $\ell$ denotes the string length. In Fig. \ref{FC} we plot the Franck-Condon factor as a function of $h/t$ and for different values of $(n,\xi)$. In the limit $h\to 0$ we find no renormalization of the meson dispersion. This is expected since the string tension vanishes in this limit where free spinon and chargon excitations exist. 

\begin{figure}
    \centering
    \includegraphics[width=0.5\textwidth]{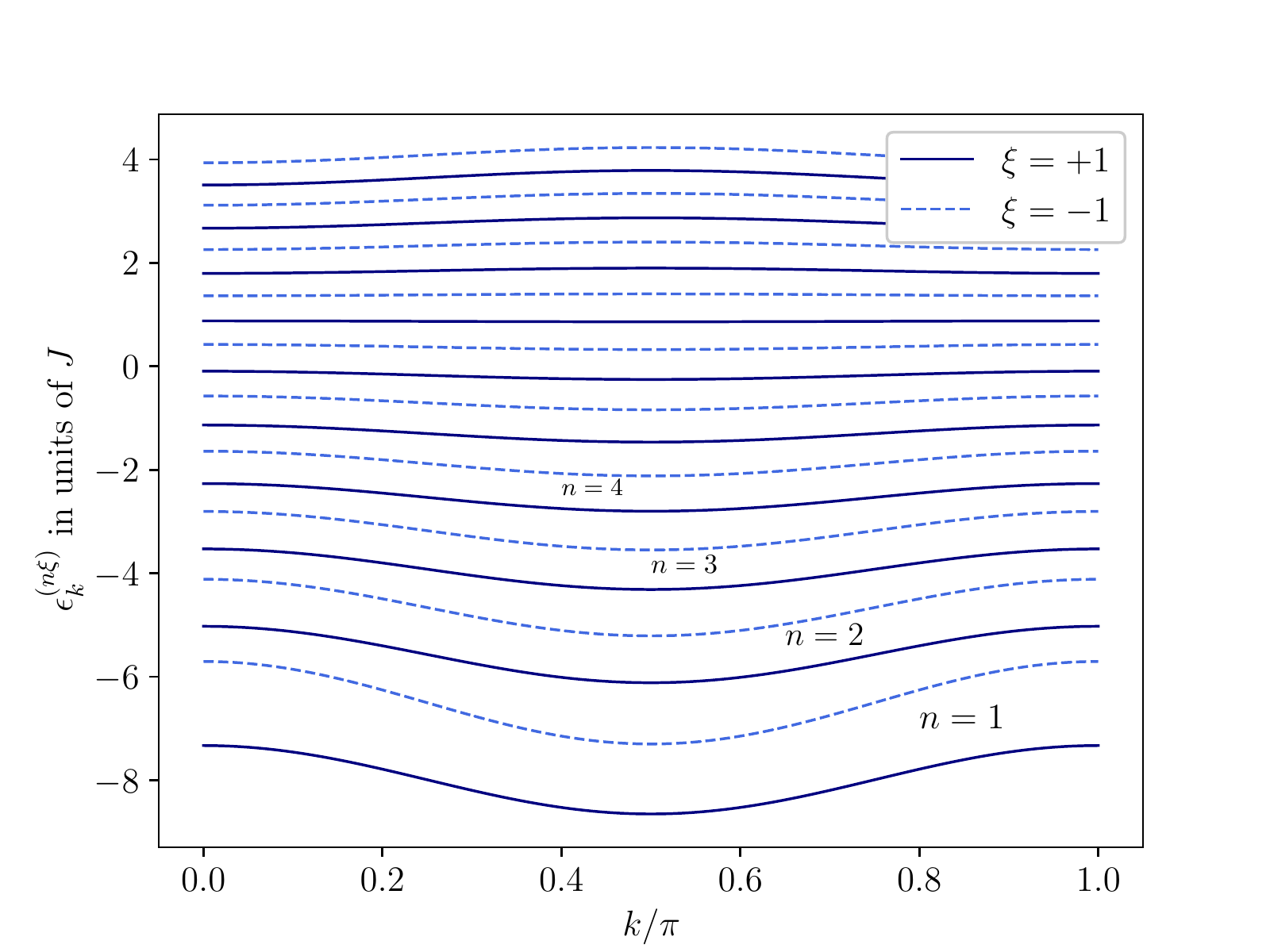}
    \caption{Momentum resolved excitation spectrum of the mesonic bound state at strong couplings. We assumed $t=5 J $ and $h=0.5 J$.}
    \label{MesonSpectrum}
\end{figure}

The SC approximation allows to calculate the excitation spectrum of the meson for arbitrary values of the total momentum $k$. Our result for the meson spectrum is 
\begin{equation}
    \varepsilon_k^{(n\xi)} = E_h^{(n\xi)} + J_\perp^{(n\xi)}\cos(2k),
\end{equation}
which we show in Fig. \ref{MesonSpectrum} for $t = 5 J$ and $h = 0.5 J$ well in the strong coupling regime. The curves correspond to approximate eigenenergies of the system, characterized by the quantum numbers $n$ and $\xi$ of the chargon wavefunction. In Fig. \ref{MesonSpectrum} we observe a series of resonances, alternating between even and odd parity states. 

\section{Meson-Magnon Interaction}
\label{ApdxMesMagInt}
Here, we derive the terms contributing to the polaron Hamiltonian \eqref{PolHamProjection}. This is achieved by computing the overlap matrix elements $\bra{\psi_{\rm sh}^{(n^\prime\xi^\prime)}(j^{s\prime})}{\hat{\mathcal{H}}}\ket{\psi_{\rm sh}^{(n\xi)}(j^s)}$ in Eq.~\eqref{PolHamProjection}.
In the previous section of Appendix \ref{ApdxMesonStatesSC} we found approximate eigenstates and -energies of the mesonic bound state of the heavy spinon and light chargon, neglecting magnon contributions. Namely the eigenstates and -energies are
\begin{multline}
    \ket{\psi_{\rm sh}^{(n\xi)}(k)} = \dfrac{1}{\sqrt{L}} \sum_{j^s} e^{-ikj^s}\sd_{j^s}\ket{0} \otimes \ket{\psi_{\rm h}^{(n\xi)}(j^s)} \\
    \equiv \dfrac{1}{\sqrt{L}} \sum_{j^s} e^{-ik j^s} \ket{\psi_{\rm sh}^{(n\xi)}(j^s)}
\end{multline}
and 
\begin{equation}
    \varepsilon_k^{(n\xi)} = E_h^{(n\xi)} + J_\perp^{(n\xi)} \cos(2k).
\end{equation}
Thus, the parton part of the Hamiltonian \eqref{partonHam} without magnon contributions is already diagonal in the basis $\ket{\psi_{\rm sh}^{(n\xi)}(k)}$ and yields the free meson Hamiltonian,
\begin{eqnarray}
    \nonumber\H_{\rm mes}^{(0)} &=& \sum_{j^s,n\xi} E_h^{(n\xi)} \fd_{j^s,n\xi}\f_{j^s,n\xi} \\
    &+& \sum_{j^s,n\xi} \dfrac{J_\perp^{(n\xi)}}{2} \l \fd_{j^s+2,n\xi}\f_{j^s,n\xi} + \hc \r.
\end{eqnarray}

The free magnon Hamiltonian $\H_{\rm mag}^{(0)} = \sum_q \omega_q \bd_q \b_q$ does not affect the SC wavefunction, $\ket{\psi_{\rm sh}^{(n\xi)}(k)}$, and thus the overlap is also trivial to compute in this case, $\bra{\psi_{\rm sh}^{(n^\prime\xi^\prime)}(j^{s\prime})}{\H_{\rm mag}^{(0)}}\ket{\psi_{\rm sh}^{(n\xi)}(j^s)} = \H_{\rm mag}^{(0)} \delta_{j^{s\prime},j^s}\delta_{n^\prime ,n}\delta_{\xi^\prime,\xi}$. 
Using that we work in a sub-space with only one meson, $\sum_{j^s,n\xi}\fd_{j^s,n\xi}\f_{j^s,n\xi}=1$, the free magnon term is the same in the new Hamiltonian.

The non-trivial part is to compute the overlap for the spinon-magnon and chargon-magnon interactions derived in Appendix \ref{ApdxMagnonCont},
\begin{equation}
    \H_{\rm mag}^{\rm int} = \H_{\rm mag}^{J} + \H_{\rm mag}^{\rm kin}.
\end{equation}
These interactions are further decomposed in spinon, chargon and string contributions, respectively, which we will treat separately. Because the above interaction terms are functions of the chargon operators $\hd_j, \h_i$, the desired overlap matrix elements of the corresponding parton-magnon interactions will involve new Franck-Condon factors similarly to the one that we already found in our SC treatment of the mesonic bound state, see Sec.~\ref{secStrongCplngThy}. These will lead to renormalized couplings between magnons and the meson in our effective polaron Hamiltonian shown in the main text in section \ref{SecPolHam}.

\subsection{Static Distortion of N\'eel Background}
We start our discussion with the Hamiltonian describing the exclusion of bonds occupied by the chargon. It can be written as,
\begin{widetext}
\begin{equation}
    \label{eqHJhmag}
     \H_{\rm mag}^{J,\rm h} = -\dfrac{1}{2}\sum_j \hd_j\h_j \left[\l J_z +h \r \l\ad_{j+1}\a_{j+1}+2\ad_j\a_j + \ad_{j-1}\a_{j-1}  \r 
    + J_\perp \l \ad_j\ad_{j+1}+ \ad_j\ad_{j-1} + \hc\r\right],
\end{equation}
and after applying a Fourier transformation to the HP boson operators, $\ad_j = \dfrac{1}{\sqrt{L}}\sum_p e^{-ipj}\ad_p$, we get,
\begin{equation}
    \nonumber \H_{\rm mag}^{J,\rm h} = -\dfrac{1}{L} \sum_{pq}\sum_j \hd_j \h_j \left[\l J_z + h \r \ad_p\a_q e^{-i(p-q)j} \l 1+ \cos(p-q)\r 
   + J_\perp \l \ad_p\ad_q e^{-i(p-q)j}\cos(q) + \hc\r\right].
\end{equation}
Sandwiching this Hamiltonian in the SC wavefunction \eqref{Born-Oppenheimer-ansatz}, leads to an overlap matrix element of the form 
\begin{equation}
    \sum_j \bra{\psi_h^{(n^\prime \xi^\prime)}(j^s)} \hd_j\h_j \ket{\psi_h^{(n\xi)}(j^s)} e^{-(p\pm q)j}
\end{equation}
which describes the renormalization of the couplings in the above Hamiltonian.
Its computation leads us the following effective meson-magnon interaction,
\begin{multline}
    \label{effPolHJh}
    \H_{\rm pol}^{J,\rm h} = -\dfrac{1}{L} \hspace{-0.15cm}\sum_{\substack{j^s,n\xi \\n^\prime\xi^\prime}} \hspace{-0.15cm}\sum_{pq}\fd_{j^s,n^\prime\xi^\prime}\f_{j^s,n,\xi} \Bigl[ \l J_z + h \r S_{J,p-q}^{(n^\prime\xi^\prime,n\xi)} \ad_p\a_q \l 1+ \cos(p-q) \r e^{-i(p-q)j^s} + \\
    +  J_\perp \cos(q)S_{J,p+q}^{(n^\prime\xi^\prime,n\xi)} \l \ad_p\ad_q e^{-i(p+q)j^s} + \hc \r\Bigr],
\end{multline}
where we defined the following Franck-Condon overlap, 
\begin{equation}
    S_{J,k}^{(n^\prime\xi^\prime,n\xi)}= \phi_0^{(n^\prime\xi^\prime)}\phi_0^{(n\xi)} + \sum_{j>0} \phi_j^{(n^\prime\xi^\prime)} \phi_j^{(n\xi)}\cos(k j),
\end{equation}
where $\phi_{|\ell|}^{(n\xi)}$ are the string functions introduced in appendix \ref{ApdxMesonStatesSC}. We used here that the string wavefunctions are real valued, $\phi_{|\ell|}^{(n\xi)}\in \mathbb{R}, \forall \ell,n,\xi $.  
The Franck-Condon factor $S_{J,k}^{(n^\prime\xi^\prime,n\xi)}$ explicitly depends on the momentum transfer of the involved magnon excitations and is thus site-dependent in real space. Namely, it depends on the distribution of the smeared-out chargon cloud, see Fig. \ref{Fig1}(b). 
This is expected since the chargon distorts the spin background in a certain distance - depending on the string tension - around the spinon, the meson center. 

It turns out that in the Hamiltonian \eqref{eqHJhmag} a zero-energy magnon mode is included which destabilizes the resulting polaron spectrum for small values of the staggered magnetic field. Physically, the magnon zero-mode results because the magnon has a zero energy cost to occupy the same lattice site as the chargon. In order to lift the energy of the magnon zero-mode we include the following phenomenological term into our effective Hamiltonian,   
\begin{equation}
    + \l J_z +h \r \sum_j \hd_j\h_j \ad_j \a_j. 
\end{equation}
This extra term changes the energy of the zero-mode to $\omega_0 \to J_z +h$, treating the site occupied by the hole like other sites. We emphasize that, by construction, magnon occupation on the site of the chargon can only arise if multiple magnons are present. Hence, the addition of the extra term should not modify the physics, but rather stabilize our approximate semi-analytical approach.

Effectively, the Hamiltonian \eqref{effPolHJh} describes a polaronic coupling of the (extended) mesonic impurity in the lattice to the bath of low-energy magnon excitations which results due to the suppression of the magnetization around the spinon by the fluctuating string.\\

We proceed with the discussion of the term~\eqref{MagStringInt}, resulting due to the presence of the geometric string in the chain along which the spins are displaced by one lattice site. Along the string the energy cost to create local spin flips becomes $-2h$, measured relative to the usual $+h$ cost for spin-flips without spinons or chargons. The corresponding Hamiltonian which adds this contribution to the free magnon Hamiltonian can be written as,
\begin{equation}
    \H^{\Sigma}_{\rm mag} = -2h \sum_{j^s,\ell>1} \sum_{\mu=\pm} \sd_{j^s}\s_{j^s} \hd_{j^s+\mu \ell}\h_{j^s+\mu \ell} \sum_{i=1}^{\ell-1} \ad_{j^s +\mu i}\a_{j^s + \mu i }
\end{equation}
As previously, we are interested in calculating the effective meson magnon interaction resulting from this string-magnon interaction term: $\H_{\rm pol}^{\Sigma}= \sum_{\substack{j^s, n\xi \\ n^\prime \xi^\prime}} \bra{\psi_{\rm sh}^{(n^\prime\xi^\prime)}(j^{s\prime})}{\H_{\rm mag}^{\Sigma}}\ket{\psi_{\rm sh}^{(n\xi)}(j^s)}\fd_{j^s,n^\prime\xi^\prime}\f_{j^s,n\xi}$.
A straightforward calculation yields, 
\begin{equation}
    \H_{\rm \rm pol}^{\Sigma} = -\dfrac{2h}{L} \sum_{\substack{j^s, n\xi \\ n^\prime \xi^\prime}} \sum_{pq} S_{\Sigma,p-q}^{(n^\prime\xi^\prime,n\xi)}e^{-i(p-q)j^s}\fd_{j^s,n^\prime \xi^\prime}\f_{j^s, n\xi}\ad_p \a_q ,
\end{equation}
where we defined the momentum dependent coupling 
\begin{equation}
    S_{\Sigma,k}^{(n^\prime\xi^\prime,n\xi)} = \sum_{\ell >1} \phi_\ell^{(n^\prime\xi^\prime)} \phi_\ell^{(n\xi)} \sum_{j=1}^{\ell-1} \cos(kj).
\end{equation}
The latter describes the average contribution of the fluctuating string to the energy cost for creating spin flips along the string $\Sigma$.

For large values of $h/t \gg 1$, the factor $S_{\Sigma,k}$ goes to zero. This is expected because the chargon gets more and more localized in this limit, thus reducing the probability to create spin flips along $\Sigma$. In the opposite limit $h/t \ll 1$ we find $S_{\Sigma,k} \to (1/L) \sum_{j=1}^{L-1} \cos(k j) \to \delta_{k,0}$, which depends strongly on the momentum transfer $k$. In this limit the string $\Sigma$ becomes very long and many magnon excitations can be excited along $\Sigma$.

The last term entering due to the distortion of the spin background consists of terms which subtract bonds from the free magnon Hamiltonian which have been used to describe the spinon tunneling instead of creating HP boson pairs at these bond.
The corresponding Hamiltonian is given by,
\begin{equation}
    \H_{\rm pol}^{J,\rm s} = -\dfrac{1}{L}\hspace{-0.15cm} \sum_{\substack{j^s,n^\prime\xi^\prime\\n\xi}} \hspace{-0.15cm}\sum_{pq} \fd_{j^s,n^\prime,\xi^\prime} \f_{j^s,n\xi} \left[ \l J_z +h \r X^{(n^\prime\xi^{\prime},n\xi)}_{p-q} e^{-i(p-q)j^s}\ad_p\a_q + J_\perp Y^{(n^\prime\xi^{\prime},n\xi)}_{pq} \l e^{-i(p+q)j^s} \ad_p\ad_q + \hc\r\right],
\end{equation}
with
\begin{eqnarray}
    \nonumber X^{(n^\prime\xi^{\prime},n\xi)}_{p-q} &=& \l 1-\phi_0^{(n^\prime\xi^\prime)} \phi_0^{(n\xi)} - \dfrac{\phi_1^{(n^\prime\xi^\prime)} \phi_1^{(n\xi)}}{2} \r + 
    \frac{1}{2} \cos(p-q)\l 3-\phi_0^{(n^\prime\xi^\prime)} \phi_0^{(n\xi)} - \phi_1^{(n^\prime\xi^\prime)} \phi_1^{(n\xi)} \r + \\  
    & \quad & \qquad\qquad\qquad\qquad\qquad\qquad\qquad\qquad\qquad\qquad \qquad + \frac{1}{2} \cos(2p-2q) \l 1+\phi_0^{(n^\prime\xi^\prime)} \phi_0^{(n\xi)} \r,\\
     Y^{(n^\prime\xi^{\prime},n\xi)}_{pq} &=& \cos(q)\l 1-\phi_0^{(n^\prime\xi^\prime)} \phi_0^{(n\xi)} - \dfrac{\phi_1^{(n^\prime\xi^\prime)} \phi_1^{(n\xi)}}{2} \r + 
     \frac{1}{2} \cos(p+2q) \l 1+\phi_0^{(n^\prime\xi^\prime)} \phi_0^{(n\xi)} \r.
\end{eqnarray}
These two renormalization factors depend only on the string wavefunction $\phi_\ell^{(n\xi)}$ at string lengths $\ell =0, 1$ where the close distance to the chargon further suppresses spinon tunneling.

\subsection{Couplings following from kinetic parton magnon interactions}
In this subsection we derive the meson-magnon couplings following from the influence of local spin flips on the dynamics of the spinon and chargon, which have been discussed in \ref{ApdxMagPartonDyn}. They consist of two terms, one where HP bosons block possible tunneling processes of the spinon and the other where the chargon and magnon perform correlated tunneling. 

First we consider the kinetic chargon-magnon coupling. In Fourier space it is given by
\begin{equation}
    \H^{\rm kin, h}_{\rm mag} = -\dfrac{1}{2L} \sum_{pq} \sum_j \l \hd_{j+1}\h_j + \hc\r \tilde{T}_{pq}\ad_p\a_q e^{-i(p-q)j}
\end{equation}
where for shortness of the expression we defined $\tilde{T}_{pq} = 2t \left[1+ e^{-i(p-q)} -e^{-ip}-e^{iq}\right]$.
As previously we sandwich this Hamiltonian in the SC wavefunction \eqref{Born-Oppenheimer-ansatz}. The appearing overlap matrix element to compute is 
\begin{equation*}
    \sum_j \bra{\psi_h^{(n^\prime \xi^\prime)}(j^s)} \l\hd_{j+1}\h_j +\hc \r \ket{\psi_j^{(n\xi)}(j^s)} e^{-(p\pm q)j} 
\end{equation*}
which results in the following meson-magnon interaction Hamiltonian:
\begin{equation}
    \label{MesMagChDyn}
    \H_{\rm pol}^{\rm kin,h} = -\dfrac{1}{L}\hspace{-0.15cm} \sum_{\substack{j^s,n^\prime \xi^\prime\\n\xi}} \hspace{-0.15cm}\sum_{pq} T_{pq} S_{t,p-q}^{(n^\prime\xi^\prime,n\xi)} e^{-(p-q)j^s}\fd_{j^s,n^\prime,\xi^\prime} \f_{j^s,n\xi} \ad_p\a_q   
\end{equation}
with $T_{pq} = 8t\sin(p/2)\sin(q/2)$.
Here the Franck-Condon factor results from the above overlap matrix element and is defined as,
\begin{equation}
     S_{t,k}^{(n^\prime\xi^\prime,n\xi)}= \sqrt{2} \phi_0^{(n^\prime\xi^\prime)}\phi_1^{(n\xi)}\cos(k/2) 
    + \sum_{j>0} \phi_{j+1}^{(n^\prime\xi^\prime)} \phi_{j}^{(n\xi)}\cos(k (j+1/2)).
\end{equation}
The resulting interaction term \eqref{MesMagChDyn} effectively describes a chargon-induced tunneling for the magnon excitations in the region of the chargon cloud, see Fig.~\ref{Fig1}.

The last term which has to be projected onto the SC wavefunction \eqref{Born-Oppenheimer-ansatz} is the one which describes the suppression of the spinon tunneling by local spin flips in the vicinity of the spinon \eqref{KinSpMagInt}. This is the only term which also suppresses the motion of the meson in our effective description. We just state here the final effective meson-magnon interaction because its derivation is similar as the previous ones; it follows by sandwiching the kinetic spinon-magnon interaction in the SC wavefunction. This procedure yields the interaction
\begin{equation}
    \H_{\rm pol}^{\rm kin, s} = -\dfrac{1}{4L} \sum_{j^s,n\xi}\sum_{pq} J_\perp^{(n\xi)}\l 1+ e^{-i(p-q)} \r^2 e^{-i(p-q)j^s}\ad_p\a_q \l  \fd_{j^s+2,n\xi}\f_{j^s,n\xi} + \hc\r,
\end{equation}
where $J_\perp^{(n\xi)}$ is the Franck-Condon overlap already introduced in section \ref{ApdxMesonStatesSC}.

Having derived all contributions entering the effective meson-magnon interaction there is one step left to arrive at the stated interaction~\eqref{PolInt} of the main text. We diagonalized the free magnon Hamiltonian by introducing Bogoliubov operators 
\begin{equation}
    \a_p = u_p \b_p -v_p \bd_{-p}
\end{equation}
with Bogolibov coefficients 
\begin{equation}
    v_p^2 = \dfrac{1}{2} \l \dfrac{J_z+h-\omega_p}{\omega_p}\r, \qquad u_p^2 = 1+ v_p^2.
\end{equation}
We introduce these Bogoliubov operators in the interaction terms, derived in this section of the appendix, and sum them all together to finally arrive at Eq. \eqref{PolInt}. We note that the Bogoliubov operators describe the elementary low-energy spin-wave excitations of the undoped system which then interact with the mesonic bound state - represented by the operators $\fd_{j^s,n\xi}$.

Because it becomes of importance in section \ref{SubSecChevy2}, we state here the form of the effective meson tunneling:
\begin{equation}
    J_{\perp,\eff}^{(n\xi)}=J_{\perp}^{(n\xi)} \l 1- \dfrac{2}{L} \sum_q v^2_q \r,
\end{equation}
The meson tunneling gets further reduced by the influence of magnon vacuum fluctuations.
\end{widetext}

\section{Meson-magnon binding at large staggered field}
\label{ApdxBoundStateLargeh}
\begin{figure*}
    \centering
    \includegraphics[width = .8\textwidth]{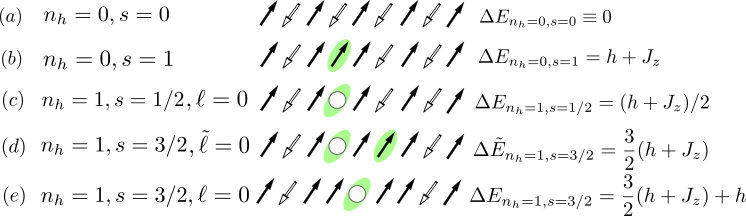}
    \caption{Overview of the classical spin-charge configurations and their corresponding zero-point energies, from which the relevant parton states can be constructed by allowing hole hopping. All energies are measured relative to the perfect N\'eel state shown in (a). (b) Flipping one spin creates an excitation with spin $s=1$ which costs an amount of energy $h+J_z$. Doping a hole into the spin chain changes spin by $s=1/2$ and costs an energy $\Delta E=(h+J_z)/2$, see (c). In (d) we show the configuration where an additional spin-flip excitation is created two sites right from the hole, which corresponds to having a spin of $s=3/2$ relative to the N\'eel state. This state lacks inversion symmetry around the hole, and we denote its string length as $\tilde{\ell} = \ell-1$. In (e) we allow one hole-hopping process to construct a new reference state with the same spin $s=3/2$. This state is inversion symmetric around the hole, and its string length is $\ell=0$ (corresponding to $\tilde{\ell}=+1$).}
    \label{FigClassConf}
\end{figure*}
In this appendix we provide an asymptotic description of the meson-magnon bound state in the large-$h$ limit. Specifically, we consider the regime $h\gg J_z,J_\perp$. In this limit quantum spin fluctuations $\propto J_\perp$ are strongly suppressed and we can restrict ourselves to studying the $t-J_z$ Hamiltonian: 
\begin{equation}
    \H \to \H_{t-Jz}, \qquad |h| \gg J_\perp.
    \label{eqHtoHtJz}
\end{equation}

In order to obtain an expression for the bound state energy, we consider the four relevant sectors independently: no hole $n_h=0$ and spin $S^z_{\rm tot}\equiv s=0,1$, and one-hole states, $n_h=1$, with spin $S^z_{\rm tot}\equiv s=1/2, 3/2$. In each sector we calculate the corresponding ground state energy $E_{n_h,s}$ semi-analytically using the parton picture. Note that the latter is exact within our approximation \eqref{eqHtoHtJz}, since magnon fluctuations can be entirely neglected in the $t-J_z$ Hamiltonian we consider here.

The cases with $n_h=0$ are trivial to solve, since $\H_{J_z}$ is diagonal in the $\hat{S}^z$ basis. For the cases with $n_h=1$ we will now construct an effective string potential describing the long-range force binding the chargon to one (for $s=1/2$) or three tightly bound (for $s=3/2$) spinons. Note that spinons in the $t-J_z$ model correspond to localized domain-walls of the surrounding N\'eel AFM. In the $\mathbb{Z}_2$ LGT formulation equivalent to the $t-J_z$ Hamiltonian, the long-ranged string potential we derive can be viewed as being mediated by the $\mathbb{Z}_2$ gauge field.

Once the string potential $V^{(s)}_\ell$ in the sector with spin $S^z_{\rm tot}\equiv s$ is known, the parton theory reduces to an effective hopping problem of the form
\begin{equation}
    \H_{\rm eff}^{(s)} = t \sum_{\Sigma} \l \ket{\Sigma+1}\bra{\Sigma} + \hc \r + \sum_\Sigma \ket{\Sigma}\bra{\Sigma} ~ V^{(s)}_{|\Sigma|},
\end{equation}
with $\Sigma \in \mathbb{Z}$. Using simple ED the ground state of $\H_{\rm eff}^{(s)}$ can be solved to obtain the energies $\Delta E_{1{\rm h},3/2} = E_{1,3/2} - E_{0,0}$ and $\Delta E_{1{\rm h},1/2}=E_{1,1/2}-E_{0,0}$ measured relative to the undoped ground state $n_h=S^z_{\rm tot}=0$, in the limit 
\begin{equation}
    \label{LargehCond}
    h \gg J_z, J_\perp
\end{equation}
but independent of $t$, which we have not specified here. I.e. this result is valid both for $t\geq h$ or $t\leq h$, as long as \eqref{LargehCond} is satisfied.

Finally, the binding energy of the meson-magnon pair is obtained as
\begin{equation}
    E_{\rm bind} = \Delta E_{1{\rm h},3/2}-\Delta E_{1{\rm h},1/2}- \Delta E_{0{\rm h},1}
    \label{ApdxEbind}
\end{equation}
with $\Delta E_{0{\rm h},1} = E_{0,1}-E_{0,0}$. For $E_{\rm bind} < 0$, the meson-magnon bound state exists below the meson-magnon scattering continuum.

\subsection{String potentials}
Before we construct the string potentials, we define reference parton states $\ket{\Sigma=0}_{n_h,s}$ in each sector $(n_h,s)$ with zero string length $\ell=0$. These are shown in Fig.~\ref{FigClassConf} (a)-(c) and (e) together with their zero-point energies relative to the classical N\'eel state. Note that all reference states are inversion symmetric around the hole. 

Next, we define states $\ket{\Sigma}_{n_h=1,s}$ for arbitrary $\Sigma \in \mathbb{Z}$ by starting from $\Sigma=0$ and applying hole-hopping terms to the right ($\Sigma > 0$) or left ($\Sigma<0$). The string potential is then given by the energy of the respective states: $V^{(s)}_{|\Sigma|} = ~_{1,s}\bra{\Sigma} \H_{J_z} \ket{\Sigma}_{1.s}$. Note that the inversion symmetry around the original hole position in the $\Sigma=0$ state guarantees that $V^{(s)}$ depends on $|\Sigma|$ only. 

In the case where only the hole is present and no additional spin-flips, $n_h=1, s=1/2$, the string potential is the same as already discussed in the main text, see Eq. \eqref{linear-pot},
\begin{equation}
    V_\ell^{(1/2)} = |\ell|h + \dfrac{1}{2} \l J_z + h \r + \dfrac{J_z}{4} - \delta_{\ell,0} \dfrac{J_z}{4}.
\end{equation}

\begin{figure}
    \centering
    \includegraphics[width=0.5\textwidth]{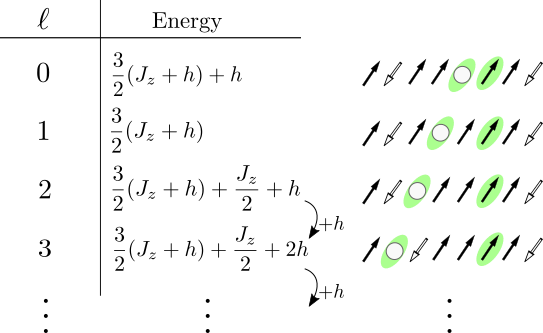}
    \caption{Spin configurations with one doped hole and an additional spin-flip, $n_h=1, s=3/2$, for different string lengths $\ell$. Their corresponding energies relative to the N\'eel state are also shown.}
    \label{FigSpin3_2Pot}
\end{figure}

Next we turn to the case $n_h=1$ with total spin $s=3/2$, with reference state shown in Fig.~\ref{FigClassConf} (e).  In order to derive the form of the string potential, we construct longer string configurations and their energies, see Fig.~\ref{FigSpin3_2Pot}. For $\ell \geq 2$ we find that each additional step $\ell \to \ell+1$ leads to the same increase in energy $h$. This leads to the following string potential,
\begin{equation}
    V_\ell^{(3/2)} = |\ell|h + \dfrac{h}{2}+ 2 J_z + \delta_{\ell,0}2h- \l \delta_{\ell,0} +\delta_{|\ell|,1} \r \dfrac{J_z}{2}.
\end{equation}

\subsection{Perturbative limit: $h \gg t$}
In order to get further analytical insight we now calculate the binding energy perturbatively in the limit $h\gg t, J_z$, when the hole hopping is also weak compared to $h$. This strongly restricts the relevant parton states to the smallest string lengths. 

A simple second order perturbation theory in $t/h$, for the case $n_h=1, s=1/2$ gives,
\begin{equation}
    \Delta E_{n_h=1,s=1/2} = \dfrac{1}{2} \l J_z + h \r - 2 \dfrac{t^2}{h + J_z/4}.
\end{equation}
Here the unperturbed ground state is $\ket{\Sigma=0}_{1,1/2}$, and we obtain perturbative admixtures of the excited states $\ket{\Sigma=\pm 1}_{1,1/2}$. 

In the spin $s=3/2$ case the ground state manifold is degenerate and given by $\ket{\Sigma= \pm 1}_{1,3/2}$, see Fig. \ref{FigSpin3_2Pot}. These states couple perturbatively to $\ket{\Sigma=0}_{1,3/2}$ and $\ket{\Sigma=\pm 2}_{1,3/2}$. The effective Hamiltonian describing second-order processes in the low-energy sector can be derived by a Schrieffer-Wolff transformation: 
\begin{multline}
    \H_\eff^{3/2} = \dfrac{3}{2}\l J_z +h \r - \dfrac{t^2}{h} - \dfrac{t^2}{h+ J_z/2} \\
    - \dfrac{t^2}{h}\bigl( \ket{+1}\bra{-1} + \hc \bigr).
\end{multline}
Consequently, the ground state energy in this case is 
\begin{equation}
    \Delta E_{n_h=1,s=3/2} = \dfrac{3}{2}\l J_z +h \r - 2\dfrac{t^2}{h} - \dfrac{t^2}{h+ J_z/2}.
\end{equation}

Combining our results, we obtain the perturbative binding energy from Eq.~\eqref{ApdxEbind}
\begin{equation}
    E_{\rm bind} = - 2 \dfrac{t^2}{h} \left[ 1+  \frac{1}{2+J_z/h} - \dfrac{1}{1+ J_z/4h}\right].
\end{equation}
Expanding the bracket in powers of $x= J_z/h \ll 1$ yields 
\begin{equation}
    E_{\rm bind} \approx - \dfrac{t^2}{h} < 0.
\end{equation}
We conclude that in the limit, $h \gg J_z,t$ the magnon binds to the hole.

\end{document}